# UNIVERSIDAD NACIONAL DE ROSARIO

FACULTAD DE CIENCIAS EXACTAS, INGENIERÍA Y AGRIMENSURA

## Tesis presentada para optar al grado de Doctor en Física

## **"Caracterización de austenita expandida generada por cementación iónica de aceros inoxidables. Estudio de la estabilidad frente a la irradiación con haces de iones ligeros energéticos"**

### Autor: Lic. Javier García Molleja

### Director: Dr. Jorge Néstor Feugeas
### Consejera de Estudios: Dra. María Dolores Calzada Canalejo

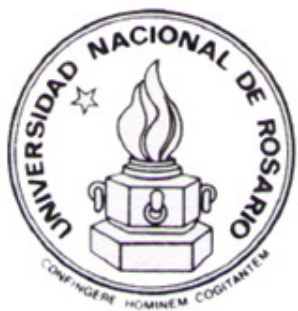

*Grupo de Física del Plasma*
*Instituto de Física de Rosario CONICET - UNR*
*2012*

Javier García Molleja                    Tesis presentada para optar al título de Doctor en Física

# Caracterización de austenita expandida generada por cementación iónica de aceros inoxidables. Estudio de la estabilidad frente a la irradiación con haces de iones ligeros energéticos

Javier García Molleja

Tesis presentada para optar al grado de Doctor en Física

 Tesis presentada para optar al título de Doctor en Física

*La ciencia es para conocer el mundo, la tecnología para cambiarlo.*
Jorge Wagensberg

# AGRADECIMIENTOS

También he de dar las gracias a todo el colectivo de hBirra, por haberme integrado en su círculo de amistad. Esos viajes y discusiones constructivas los llevaré siempre presentes. Cómo no, también a mis amigos de España, gracias a Melchor, Trini, Kike, Antonio, Paco, Rafa, José, Juan, David, Sara, Manolo, etc., por reclamar mi presencia allí y pasar el tiempo entre películas, tapas y juegos de rol.

Finalmente, deseo mencionar lleno de emoción a nuestro compañero y amigo Horacio, único a la hora de poder reparar cualquier aparato y localizar fugas de vacío. Una persona que no dudaba en agregarte a algún que otro almuerzo para pasar un rato entretenido. Sinceramente, todo nuestro grupo, y todo el IFIR también, lamentamos tu pérdida.

# Índice general

# Índice de figuras

# Índice de tablas

# Parte I

# Prefacio

# 0.1. Presentación

En el desarrollo actual de la industria se necesitan componentes para la fabricación de instrumentos que posean buenas propiedades, además de ser asequibles desde el punto de vista económico. La investigación actual presta una gran atención al campo de la *Nanotecnología*, ya que se están descubriendo de manera continuada nuevas propiedades y fenómenos inéditos cuando se trabaja con materiales, dispositivos y sistemas a escala nanométrica. Tales descubrimientos se están aplicando desde la construcción de ordenadores compactos de bajísima resistencia hasta la cirugía humana en el campo de la medicina celular [Tutor Sánchez, 2005]. Existen dos variantes para trabajar de manera nanométrica:

- **De arriba hacia abajo**. Se trabaja con un material hasta conseguir la precisión adecuada.

- **De abajo hacia arriba**. Está basado en el ensamblaje de átomos o moléculas entre sí hasta conseguir unos resultados adecuados a la aplicación ideada.

En este segundo caso, materiales ya conocidos empiezan a comportarse de manera diferente debido a efectos de confinamiento cuántico y correlación [Díez, 1999], altas tensiones residuales impuestas por los sustratos [Martin, 2004] o a un gran cociente superficie-volumen (provocando multitud de ligaduras colgantes) [Ibach, 2006].

La *ingeniería de superficies* trata de modificar las condiciones de la superficie de los materiales para protegerlos o mejorar su realización. Existen dos variantes: la deposición de una capa o la modificación de la zona superficial. Ambas técnicas han sido usadas profusamente durante las últimas décadas [Feugeas, 2003] pero aún quedan muchos interrogantes por descubrir, tales como la influencia de los parámetros de trabajo, la concentración de contaminantes y las tensiones residuales en la deposición de láminas delgadas, o como el proceso de optimización de la aplicabilidad, los fenómenos de difusión y la modificación de la estructura cristalina en el caso de la modificación de capas superficiales.

En cuanto al estudio de los procesos de modificación de superficies un candidato óptimo para ello es el acero inoxidable, el cual posee una gran calidad superficial, aparte de ser poco rugoso y muy resistente a la corrosión [Lo, 2009]. Sin embargo, también sería deseable que los aceros presentasen una gran dureza y una resistencia al desgaste apreciable. Como no lo poseen de manera natural es necesario otorgarle estas mejoras, siempre y cuando no desaparezcan sus cualidades inherentes.

Este tratamiento superficial de los materiales se lleva frecuentemente con el uso de *plasmas*, ya que no se producen contaminantes, el tratamiento es corto y fácilmente reproducible [Roth, 1995]. El uso de plasmas solo afecta a las capas más superficiales, por lo que los costes de producción se abaratan. Dentro del amplio campo del uso de plasmas dedicados a la modificación de superficies, nos centramos en la *difusión iónica* por la que se modificarán las capas cercanas a la superficie mediante el cumplimiento de la ley de Fick por parte de los iones presentes en el plasma. Esta técnica es muy utilizada en la industria de prótesis humanas y engranajes de motores por dar gran resistencia al desgaste y la corrosión, pero tiene el inconveniente de que aún no están bien comprendidos los procesos que desencadenan la difusión.

## 0.2. Planificación

Dentro de la difusión iónica existen dos técnicas principales: la *nitruración* y la *cementación.* La primera ha sido muy aplicada y estudiada, existiendo una enorme bibliografía sobre el tema. Menos atención ha recibido la segunda, quizás debido a la temprana época de desarrollo que hay hoy en torno al uso de plasmas en el tratamiento de superficies, o quizás por el poco interés demostrado por la industria a la hora de innovar y desechar técnicas de demostrado beneficio. Esto no ha sido un paso atrás a la hora de realizar un estudio de caracterización del acero austenítico tras la aplicación de un proceso de cementación iónica. Se utilizaron dos atmósferas diferentes (50% de argón, 45% de hidrógeno y 5% de metano y 80% de argón, 15% de hidrógeno y 5% de metano) para comparar cómo afectaban a los resultados y se realizaron tratamientos según un intervalo de tiempo (30, 60 y 120 minutos) evidenciando de este modo que el tiempo del proceso influenciaba drásticamente en los resultados. También, para determinar el comportamiento del acero tratado bajo irradiación iónica y tratamientos térmicos prolongados cementamos probetas con una atmósfera compuesta de 50% de argón, 45% de hidrógeno y 5% de metano, además de una nitruración sobre acero con una mezcla de 80% de hidrógeno y 20% de nitrógeno. En estos dos casos el tiempo de tratamiento se fija en 80 minutos.

- Mediante microscopía óptica se determina el espesor de las capas cementadas, realizándose mediciones más precisas mediante FIB/SEM a la hora de determinar la morfología superficial y transversal

- Con un tribómetro se determinará la resistencia al desgaste de las muestras tratadas frente al material base

- Mediante microindentaciones conoceremos la dureza de la capa superficial y su variación en función del espesor

- Con técnicas de difracción de rayos X, tales como GIXRD, se determinará la expansión de la estructura cristalina en función de la atmósfera, el tiempo de tratamiento, el gas empleado en las descargas (deuterio o helio) de plasma focus y la temperatura y tiempo del tratamiento térmico

- El análisis de corrosión se llevará a cabo con un baño en solución salina

- El perfil elemental con la profundidad se logrará con AES, que se basa en la eyección de electrones Auger

- Mediante la deposición a partir de sputtering por magnetrón de una capa de nitruro de aluminio observaremos si existen influencias de su presencia o ausencia tras sufrir tratamientos térmicos prolongados

La bibliografía sobre el mundo de la cementación no es exhaustiva, aunque es grande. Es de mencionar que los métodos hasta hoy día son de muy larga duración (desde cuatro horas hasta incluso días) y recurren a temperaturas muy elevadas en el proceso (más allá de 1100 ºC). Hay otros artículos que ni siquiera recurren al uso de plasmas. El avance que trae este trabajo al mundo de la física se fundamenta en que nuestros resultados ya son visibles a duraciones muy cortas (menores a dos horas) y temperaturas muy bajas (alrededor de 400 ºC). Esto demuestra la factibilidad de reproducir la cementación a gran escala y el consecuente ahorro económico. Como último dato, los resultados de este estudio podrían utilizarse para realizar comparaciones con el proceso de la nitruración iónica, llevando a ambas técnicas al grado de atención que se merecen. La comparativa entre tratamientos queda revelada en los experimentos de irradiación iónica mediante plasma focus, pudiendo determinar los efectos de los distintos tipos de iones sobre diferentes tratamientos superficiales, así como la influencia de una capa de nitruro de aluminio superficial en los procesos de oxidación a altas temperaturas.

## 0.3. Hallazgos

En el campo de la cementación iónica se ha descubierto que existe un umbral para la densidad de corriente que circula por el cátodo y las probetas a tratar. Por encima de cierto nivel se dan condiciones que favorecen la precipitación de carburos de cromo y hierro. También se ha descubierto que el bombardeo de argón en la probeta conlleva la creación de varias bandas de deslizamiento y fallos de apilamiento en la zona superficial del acero. Además, se destaca que el proceso de cementación conlleva el alojamiento del carbono dentro de la red austenítica de dos maneras diferentes: *grafítico* en una solución disuelta en la red que actúa como lubricante sólido cuando la superficie queda sometida a fricción intensa, y *carburo* en ligadura con los átomos que existen en la red cristalina y que pueden llegar a formar precipitados.

Frente a la irradiación iónica mediante plasma focus sobre probetas cementadas y nitruradas se observa una fuerte influencia en el tratamiento de modificación superficial previo, puesto que se observa una alta dependencia a este parámetro más que al tipo de gas que se utilice. También se descubre que un número creciente de disparos va reduciendo la expansión austenítica a la vez que hace aparecer un pico fijo en una posición angular, ocasionado posiblemente por las cristalitas tensionadas que se forman en la matriz amorfa superficial. Por último, se destaca que la capa superficial de nitruro de aluminio actúa como una buena barrera contra la oxidación cuando se somete a las probetas tratadas a entornos de alta temperatura durante tiempos prolongados.

## 0.4. Conclusiones

En el campo de la cementación iónica tenemos que:

- La austenita expandida sufre una rápida expansión (tiempos menores a treinta minutos) y se localiza incluso a una profundidad de ≈14 μm

- Ópticamente se observa la estructura típica de la austenita expandida y la presencia de bandas de deslizamiento

- Las observaciones FIB no indican la presencia de precipitados en forma de aguja. Sí se aprecian dominios de reorientación cristalina dentro de los granos de austenita

- En los tratamientos de 30 minutos, la capa más externa (≈150 nm) tiene todo el carbono ligado químicamente, independientemente de la atmósfera utilizada. Para tiempos de tratamiento mayor no se observan cambios significativos en las probetas tratadas con la atmósfera de menor porcentaje de argón, aunque sí para las de mayor porcentaje, identificando un incremento de la concentración del carbono puro a mayores tiempos de tratamiento

- El desarrollo de la austenita expandida es idéntico en todo el espesor de la capa de varios micrómetros, excepto en los primeros 150 nm debido a la llegada de átomos de Ar y C de alta energía y flujo. Esto hace que por sputtering haya modificaciones en la concentración de elementos

- La dureza, desde los 100 nm hasta 1 mm de profundidad, muestra valores semejantes entre las probetas conseguidas por ambas atmósferas. Hasta los ≈700 μm de profundidad la dureza es mayor que la del material base y alcanza un máximo de 12 y 11 GPa para las muestras con menor porcentaje de argón y mayor porcentaje, respectivamente. Este máximo se alcanza a los ≈150 nm de profundidad

- La resistencia al desgaste aumenta con el tiempo de tratamiento y se achaca a la mayor cantidad de carbono libre que hace las veces de lubricante sólido

- Aunque en principio la austenita expandida presenta unas buenas propiedades anticorrosivas, en nuestro estudio se comprobó una baja resistencia a la corrosión con respecto al material base. Esto se debe al desarrollo de carburos en la superficie que desencadenan el ataque corrosivo

- La creación de austenita expandida se logra a una velocidad cincuenta veces mayor que en los tratamientos en fase de gas a baja temperatura, con mejoras en la dureza y la resistencia al desgaste, pero el alto flujo iónico que se da en la superficie durante el proceso de cementación provoca carburos que desmejoran la resistencia a la corrosión

- Bajo irradiación iónica de deuterio o helio se aprecia una mayor resistencia al bombardeo por parte de las probetas cementadas. Las probetas nitruradas llegan a fundirse en su superficie a los diez disparos

- Superficialmente, además de cráteres, puntos de eyección y bandas de deslizamiento cruzadas, aparecen en las nitruradas descamado de zonas y agrietamiento

- Para un tratamiento dado, el uso de deuterio o helio provoca un mismo resultado

- Aparecen dos picos (111) de austenita expandida. El primero va reduciendo su parámetro de red a cada pulso, fenómeno achacado a la difusión del nitrógeno o del carbono tras cada choque térmico. El segundo queda flujo en la posición angular de 43,3º y es causado por cristalitas de alta tensión residual englobadas en una matriz amorfa

- El pico que se desplaza a cada pulso tiende a la posición del pico localizado a 43,3º

- La dureza en estas probetas depende del tratamiento superficial previo y no del tipo de gas que se emplee

- El nitruro de aluminio depositado en la superficie actúa como barrera contra la entrada de oxígeno en los tratamientos térmicos prolongados

- El pico (111) se desplaza a mayores valores angulares, es decir, a menores parámetros de red, con el aumento de temperatura sin importar el tiempo de tratamiento

# Bibliografía

# Parte II

# Introducción

La tecnología e industria actuales necesitan de materiales con buenas propiedades, que sean fáciles de maquinar y que no entrañen altos costes, además de no ser agresivos al medio ambiente. Un material muy utilizado y ampliamente conocido es el acero inoxidable austenítico. Desde el campo de la medicina hasta la fabricación de herramientas y motores, pasando por la fabricación de tuberías y cisternas, el campo de aplicación de este tipo de acero es muy extenso. Sin embargo, no presenta buenas propiedades de dureza ni de resistencia al desgaste. Existe multitud de bibliografía atenta a este tipo de acero, en especial al AISI 316L y hay muchas maneras de abarcar el problema de su poca dureza.

Un tipo de abarcamiento es el tratar su superficie, de tal manera que se modifiquen sus propiedades de interés, sin alterar las buenas características que ya posee de por sí. Hay dos amplios campos para el *tratamiento superficial*, que son la modificación de las primeras capas y la deposición de una capa de buena adherencia y con las propiedades que necesitamos. Para modificar la capa más externa hay gran número de técnicas, quedando destacadas por la amplísima producción de artículos científicos la *nitruración* y la *cementación*. Estas dos técnicas se basan en el alojamiento de átomos de nitrógeno o carbono en la red cristalina del acero austenítico, de tal manera que esta quede deformada, con tensiones debidas al aumento del parámetro de red, dando lugar a una fase conocida como *austenita expandida*. Dicha fase es de alta dureza y con gran resistencia al desgaste, teniendo además las propiedades del acero base intactas. La adherencia en este caso es altísima, puesto que modificamos las capas superficiales hasta cierto grado, por lo que sigue siendo parte del material y no la adición de otra estructura.

De las técnicas empleadas para nitrurar y/o cementar se puede recurrir al uso de plasmas fríos mediante una descarga luminiscente entre dos electrodos al usar una fuente de potencia (que puede ser de corriente continua) y una atmósfera con los gases de interés a baja presión. El uso del plasma comporta ciertas ventajas, como la facilidad de reproducción, el corto tiempo de tratamiento, la práctica ausencia de contaminantes, el bajo coste y la posibilidad de control de los productos mediante la modificación de los parámetros de trabajo. Con el plasma no se necesitan altas temperaturas para provocar reacciones químicas entre los precursores (tal y como se usaban anteriormente en los procesos con baños de sales y tratamientos gaseosos) y la combinación de los átomos de interés en la superficie del material a tratar. La cinética del plasma ayuda a provocar, con su producción de electrones energéticos y átomos e iones relativamente fríos, distintas reacciones químicas o producción de radicales a temperaturas menores, evitando dañar así el material.

En esta parte de la tesis se hará una breve introducción sobre el material a utilizar, el acero inoxidable austenítico AISI 316L, junto con un resumen de las diferentes técnicas de tratamiento superficial que existen hoy en día para otorgarle las propiedades de interés, tanto en el campo de la modificación superficial como en el de la deposición de capa superficial. Definiremos y explicaremos conceptos sobre el plasma y cómo utilizarlo para llevar a cabo las dos corrientes principales de tratamiento superficial descritas.

Para llevar a cabo esta tesis doctoral ha sido necesario realizar una serie de cursos y trabajos para habituarse a utilizar ciertos conceptos y métodos de investigación. Se realizaron en la Universidad Nacional de Rosario distintos trabajos de formación, idiomas y cursos de doctorado. En cuanto a los *trabajos de formación* se realizaron los siguientes: Multicapas y Superredes creadas mediante Deposición Física en fase Vapor Asistida por Plasmas (trabajo número 1, 2011), Estudio de la deposición de monocapas de AlN y superredes de AlN/TiN mediante sputtering por magnetrón. Análisis de la cementación de aceros mediante plasmas (trabajo número 2, 2011) y Capacitación en el dominio de técnicas metalográficas y ensayo de microdureza para el estudio de filmes finos (trabajo número 3, 2011). En el apartado de *idiomas* se cursaron Inglés (2007) y Francés (2007). Finalmente, los *cursos de doctorado* realizados fueron: Espectroscopia de Superficies (2007), Introducción a la Física del Plasma (2007), Cristalografía de Rayos X (2007), Microscopía de Fuerza Atómica (2008), Técnicas Experimentales en Física del Plasma (2009), Introducción a los Materiales Cerámicos (2009) y Diagnósticas Ópticas en Plasmas (2010, cursada en el Grupo de Espectroscopia de Plasmas de la Universidad de Córdoba, en España).

# Capítulo 1

# El acero

El acero es básicamente una aleación o combinación de hierro altamente refinado (más de un 98 %) y carbono (entre casi 0,05 % hasta menos de un 2 %). Algunas veces se agrupan con propósitos determinados otros elementos de aleación específicos, tales como el Cr (cromo) o Ni (níquel) [Acero].

Para fabricar acero se parte entonces de la reducción de hierro (Fe) puro (en la denominada producción de arrabio). Esto en principio no es tan sencillo como parece, puesto que nunca se encuentra Fe libre en la naturaleza al ser más favorable reaccionar químicamente con el oxígeno del aire para formar óxido de hierro (herrumbre). Dicho óxido se encuentra en cantidades significativas en el mineral de hierro, predominando en su parte más externa. Sin embargo, en su interior no solo hay un tipo de átomos, sino que aparecen impurezas y materiales térreos.

Los diferentes tipos de acero se clasifican de acuerdo a los elementos de aleación que producen distintos efectos en él:

- Aceros al carbono

- Aceros aleados (estructurales, para herramientas y especiales)

- Aceros de baja aleación ultrarresistentes

- Aceros inoxidables

## 1.1. El sistema hierro-carbono

Como hemos mencionado, el acero es una aleación de una gran cantidad de hierro y una pequeña proporción de carbono (C), además de otros aleantes que se añaden para darle características específicas. Si lo consideramos por tanto un sistema binario es susceptible de ser estudiado mediante un diagrama de fases.

En un sistema binario hay tres variables primarias que deben ser consideradas: temperatura, presión y composición [Darken, 1953]. Solo se considera la composición de uno de sus componentes, ya que la otra se obtiene inmediatamente al restar la unidad. En el caso de los aceros (hierro-carbono) casi siempre se elige como referente el C. Aunque se necesita una representación tridimensional, se utilizan las proyecciones para analizar el equilibrio a presión constante de una atmósfera. La creación del diagrama de fases debe obedecer algunas reglas impuestas por las leyes de la termodinámica.

Si el diagrama no es ideal se podrá distinguir la temperatura *eutéctica*, que es la temperatura a la que están en equilibrio dos fases sólidas y el líquido. Los diagramas se denominan *peritécticos* si hay un punto de conexión entre el equilibrio entre el líquido y una fase sólida y entre el líquido y otra fase. El punto en el que coexisten dos líquidos y un sólido se llama *monotéctico* [Darken, 1953].

En cualquier sistema en equilibrio el potencial químico de cualquier componente es el mismo en todas las fases que coexisten e independiente de la cantidad de cada fase. Dicho potencial puede ser considerado como una función de la composición, temperatura y presión.

Tras estos conceptos introductorios podemos describir el diagrama hierro-carbono (figura (1.1)), que suele dividirse en dos partes: una que comprende las aleaciones con menos del 2% de carbono y que se llaman *aceros*, y otra integrada por las aleaciones con más de un 2% de carbono, las cuales se llaman *fundiciones*. A su vez, la región de los aceros se subdivide en otras dos: una formada por los aceros cuyo contenido en carbono es inferior al correspondiente a la composición eutectoide (0,83 %) los cuales se llaman aceros *hipoeutectoides*, y la otra compuesta por los aceros cuyo contenido se encuentra entre 0,83 y 2 %, y que se conocen por aceros *hipereutectoides*.

En la figura aparecen tres líneas horizontales, las cuales indican reacciones isotérmicas [Diagrama]. La parte del diagrama situada en el ángulo superior izquierdo de la figura se denomina *región delta*. En ella se reconocerá la horizontal correspondiente a la temperatura de 1492 ºC como la típica línea de una reacción peritéctica. Por debajo de dicha línea (enfriamiento) se creará la fase gamma, mientras que por encima (calentamiento) se tendrá la fase delta y la fase líquida.

La máxima solubilidad del carbono en el hierro delta (red cúbica centrada en el cuerpo) es 0,10% de C (carbono), mientras que el Fe gamma (red cúbica centrada en las caras) disuelve al carbono en una proporción mucho mayor. En cuanto al valor industrial de esta región es muy pequeño, ya que no se efectúa ningún tratamiento térmico en este intervalo de temperaturas.

La siguiente línea horizontal corresponde a una temperatura de 1130 ºC, esta temperatura es la de solidificación del eutéctico, y se verifica que enfriando al líquido por debajo de la mencionada línea aparecerá la fase gamma junto a cementita ($Fe_3C$), dándose el proceso inverso si se procede al calentamiento de ambas fases.

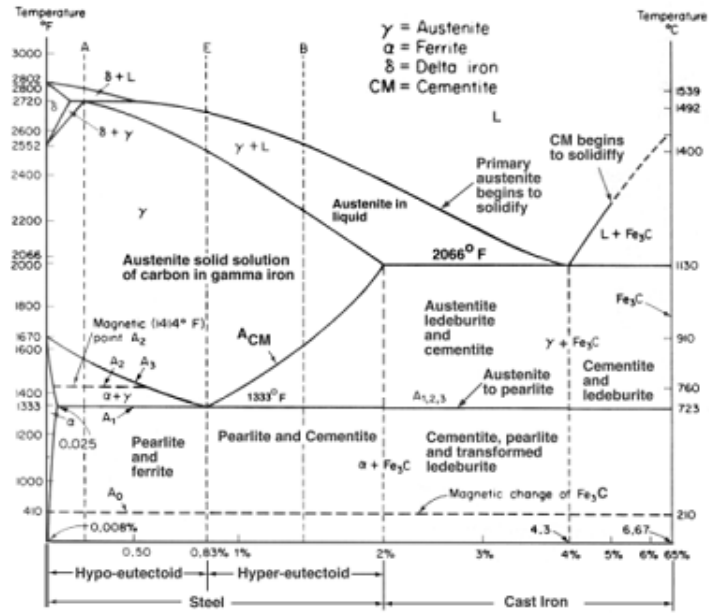

Figura 1.1: Diagrama de fases del hierro-carbono [Diagrama].

La mezcla eutéctica, por lo general, no se ve al microscopio, ya que a la temperatura ambiente la fase gamma no es estable y experimenta otra transformación durante el enfriamiento.

La última línea horizontal, se representa a los 723 ºC; esta línea corresponde a la temperatura de formación del eutectoide, y al alcanzarse en un enfriamiento lento la fase gamma debe desaparecer, dando origen a hierro alpha (de estructura cúbica centrada en el cuerpo de menor parámetro de red que la fase delta) y cementita. Evidentemente, calentando se dará el proceso inverso.

## 1.2. Propiedades físicas del acero

Cada elemento de aleación produce, en el acero, efectos específicos. Sin embargo, diversas variaciones en las propiedades generales se asocian con la formación de alguna fase de carburo (que aumenta la dureza del acero pero disminuye catastróficamente su resistencia a la corrosión) o de una solución sólida. Puesto que las partículas de carburo son duras, aumentan la resistencia al desgaste, hacen disminuir la maquinabilidad y aumentan, ligeramente, la dureza de la aleación. Evitan, además, el engrosamiento del grano durante el calentamiento. Por otra parte, un elemento aleante presente en solución sólida influye, muy activamente, en la mayoría de las propiedades del acero. Afecta a la dureza, la resistencia mecánica, la resistencia a la corrosión, a las propiedades eléctricas y a la templabilidad.

Conociendo las estructuras atómicas, cristalinas o de fase es posible estimar los valores de las propiedades. Existen pocos aspectos de estructura, mientras que el número de propiedades es casi ilimitado. Estas propiedades varían con la aplicación y como hay diferentes existe un gran número de estas: físicas, mecánicas, químicas y metalúrgicas adicionales.

- **Propiedades eléctricas**. Los fenómenos eléctricos constituyen un aspecto típico de las propiedades físicas que han sido estudiadas ampliamente [Guy, 1965]. La **resistencia eléctrica** nos indica la relación y rendimiento existente entre la conductividad eléctrica y el movimiento de los electrones. A partir de la definición de resistencia podemos distinguir entre semiconductores, conductores y superconductores. Los **efectos termoeléctricos** aparecen al aplicar una corriente a un conductor, provocando un aumento de temperatura.

- **Propiedades magnéticas**. Debido al interés industrial es necesario estudiar a fondo estas propiedades [Guy, 1965]. El **magnetismo atómico** se produce en átomos con niveles de energía parcialmente llenos, quedando espines desapareados, pudiendo entonces clasificarlos como ferromagnéticos, antiferromagnéticos, ferrimagnéticos y paramagnéticos. Los **dominios magnéticos** son unas regiones en las que los momentos magnéticos de los átomos están alineados y orientados en la misma dirección y dependiendo de su extensión los materiales pueden ser magnéticamente blandos o duros, llamándose imanes permanentes en este caso. La **magnetostricción** es la pequeña variación de longitud que experimenta un material ferromagnético cuando se altera su grado de imanación.

- **Propiedades térmicas**. Un cierto número de propiedades determina el efecto de añadir calor a una aleación [Guy, 1965]. La velocidad a la que el calor puede fluir a través de un material, bajo la influencia de un gradiente de temperatura dado, está determinada por la **conductividad térmica**. La dilatación térmica relaciona el cambio de longitud con el cambio de temperatura.

- **Propiedades nucleares**. Los problemas metalúrgicos que se presentan en la ingeniería nuclear [Bass, 2005] proceden de dos aspectos de las reacciones nucleares: los productos energéticos de la fisión y de la radiación y la necesidad de conservar tantos neutrones producidos como sea posible [Guy, 1965].

## 1.3. Aceros inoxidables austeníticos

Los aceros *inoxidables* son una gama de aleaciones que contienen un mínimo de 11% de cromo [Acero]. El cromo forma en la superficie del acero una película pasivante, extremadamente delgada, continua y estable. Esta película deja la superficie inerte a las reacciones químicas. Esta es la característica principal de resistencia a la corrosión de los aceros inoxidables.

El extenso rango de propiedades y características secundarias, presentes en los aceros inoxidables, hacen de ellos un grupo de aceros muy versátiles.

La selección de los aceros inoxidables puede realizarse de acuerdo con sus características [Aceros Inoxidables]: resistencia a la corrosión y a la oxidación a temperaturas elevadas; propiedades mecánicas del acero; características de los procesos de transformación a que será sometido; costo total (reposición y mantenimiento), y disponibilidad del acero.

Los *aceros inoxidables* tienen una resistencia a la corrosión natural que se forma automáticamente, es decir, no se adiciona. Tienen una gran resistencia mecánica, de al menos dos veces la del acero al carbono. Son resistentes a temperaturas elevadas y a temperaturas criogénicas. Son fáciles de transformar en gran variedad de productos y tiene una apariencia estética que puede variarse sometiendo el acero a diferentes tratamientos superficiales para obtener acabado a espejo, satinado, coloreado, texturizado, etc.

Aparte del cromo, existen otros elementos que mejoran las propiedades inoxidables [Lo, 2009] tales como el molibdeno (Mo), que se añade para aumentar la resistencia contra las picaduras.

Por otro lado, la *austenita* es una forma de ordenamiento distinta de los átomos de *hierro* y *carbono*. Esta es la forma estable del hierro puro a temperaturas que oscilan entre los 723 y 1492 ºC [Diagrama].

La estructura cristalina de la austenita es del tipo cúbica, de caras centradas, en donde se diluyen en solución sólida los átomos de carbono en los intersticios [Christiansen, 2004], hasta un máximo tal como lo indica el diagrama de fase de Fe-C, mostrado en la figura (1.1). Esta estructura permite una mejor difusión de los elementos intersticiales (N o C, por ejemplo), acelerando así el proceso de carburización (cementación) o nitruración del acero. Como se ha visto previamente, la solubilidad máxima del carbono en la red de hierro es solo del 2 %. Hay que recordar que por definición los *aceros* contienen menos de 2 % de carbono y pueden tener disuelto el carbono completamente a altas temperaturas.

La austenita no es estable a la temperatura ambiente excepto en aceros fuertemente aleados con níquel (Ni), aunque por encima de 500 ºC empiezan a precipitarse compuestos de Cr y Fe. La austenita no expandida es blanda y dúctil y, en general, la mayoría de las operaciones de forja y laminado de aceros se efectúa a aproximadamente 1100 ºC, cuando la fase austenítica es estable.

Finalmente, a diferencia de la *ferrita*, la austenita no es *ferromagnética* a ninguna temperatura. La ferrita, como se ha referido anteriormente, posee una estructura cristalina bcc (cúbica centrada en el cuerpo, por sus siglas en inglés) que puede llegar a distorsionarse en gran manera [Aleaciones] con la adición de carbono, cuya solubilidad máxima es de 0,025 % [Aceros Inoxidables].

Los *aceros austeníticos* inoxidables (serie 300) [Aceros Inoxidables] son los más utilizados por su amplia variedad de propiedades y se obtienen agregando níquel a la aleación, por lo que la estructura cristalina del material se transforma en austenita. El contenido de cromo varía de 16 a 28 %, el de níquel de 3,5 a 22 % y el de molibdeno de 1,5 a 6 %.

Los tipos más comunes son el AISI 304, 304L, 316, 316L, 310 y 317, mientras que las propiedades básicas son: excelente resistencia a la corrosión; excelente factor de higiene-limpieza; fáciles de transformar; excelente soldabilidad; no se endurecen por tratamiento térmico, y se pueden utilizar tanto a temperaturas criogénicas como a elevadas temperaturas. Las principales aplicaciones que podemos encontrar para estos aceros son: utensilios y equipo para uso doméstico, hospitalario y en la industria alimentaria, tanques, tuberías, etc.

Existe un caso especial para los aceros austeníticos, que es la denominada *fase S* [Lo, 2009], llamada también austenita expandida, austenita supersaturada, fase $m$, $S'$, $\varepsilon'$, $\gamma_N$ o $\gamma_C$, que resulta por la entrada de nitrógeno (N) o carbono (C) en una red cristalina austenítica más allá de la solubilidad máxima que indica el diagrama de fases. La obtención de austenita expandida se logra mediante diferentes técnicas de modificación superficial, tales como nitruración por plasma a baja temperatura, sputtering por magnetrón e implantación iónica por inmersión en plasma. Por consiguiente, la entrada de carbono o nitrógeno provoca una tensión residual compresiva [Lo, 2009] y una alta densidad de fallos de apilamiento que hace aumentar el parámetro de red de la austenita.

Algunos autores [Menthe, 1999] han sugerido que la austenita expandida realmente es una estructura tetragonal centrada en las caras, por la mayor expansión en el plano (200), aunque esta puede ser achacada a la mayor difusión del C y la mayor tensión residual que soporta de por sí este plano en una fcc (cúbica centrada en las caras, por sus siglas en inglés). También se ha propuesto una estructura tetragonal centrada en el cuerpo y una estructura triclínica [Fewell, 2000].

Esta estructura aumenta las propiedades mecánicas, tribológicas y corrosivas, pero por encima de 500 ºC se observa una degradación de la resistencia a la corrosión natural del acero debido a la precipitación de nitruros o carburos de cromo. En cambio, justo por debajo de 450 ºC no se aprecian signos de precipitación, estando en su lugar una capa delgada saturada de átomos de nitrógeno o carbono que ocuparán los huecos intersticiales de la red como una solución sólida.

La austenita expandida posee alta dureza y buena resistencia al desgaste. Su resistencia a la corrosión viene por el acero inoxidable, pero se puede mejorar en función de la cantidad de nitrógeno de la capa. También se aprecian interesantes propiedades magnéticas, ya que la cantidad de fallos de apilamiento y el porcentaje de átomos de N o C pueden hacer que sea paramagnético o ferromagnético.

 Tesis presentada para optar al título de Doctor en Física

# Capítulo 2

# Tratamiento de superficies

Los tratamientos de superficies se aplican para proteger a las piezas de los agentes externos, aumentando las propiedades de resistencia a la corrosión, al desgaste por fricción, confiriendo mayor dureza y evitando la rugosidad excesiva [Feugeas, 2003]. Una superficie puede ser cambiada de dos maneras:

- Mediante la modificación de las capas superficiales del material base

- Mediante la deposición de una capa de un determinado compuesto sobre la superficie

Es posible aplicar estos tratamientos bajo la presencia y ayuda de un plasma, eliminando así las limitaciones que presentan los otros tipos de tratamiento, mejorando además la reproducibilidad a diferentes escalas y la práctica ausencia de contaminantes.

A continuación iremos indicando de manera resumida los procesos que nos encontramos dentro de los dos grupos previamente descritos.

## 2.1. Modificación de las capas superficiales

Podemos distinguir dos procesos fundamentales: el *físico* en el que el proceso está fuera del equilibrio termodinámico, y el *químico* que al contrario que el anterior, está dentro.

- **Implantación iónica**. Los iones de una especie son acelerados al aplicar un campo eléctrico [Biller, 1999]. Este haz iónico es colimado y queda enfrentado con el sustrato a tratar. Debido a la gran energía cinética, los iones penetran como proyectiles y van interactuando con la red de átomos, frenándose por ello y alterándose por tanto la red [Möller, 1999]. Esta descripción para haces monoenergéticos vale para los polienergéticos, obteniendo así de esta manera la modificación uniforme de varias capas.

  Existen otros conceptos de implantación iónica que resuelven varios problemas: con *fuente* de plasma (PSII) o con *inmersión* en plasma (PIII) [Larisch, 1999]. La pieza se trata como cátodo de una descarga glow de un plasma formado por las especies a implantar y con impulsos de alta tensión se logran acelerar las cargas [Richter, 2000].

- **Difusión atómica**. Este tipo de tratamiento se basa en la exposición del sustrato a especies activas, de manera que estas sean absorbidas por la superficie [Möller, 2001]. Una vez integradas aparece la difusión gobernada por la *ley de Fick*:

$$\Gamma = -D \cdot \nabla n, \tag{2.1}$$

donde $\Gamma$ es el flujo de partículas, D es el coeficiente de difusión y $\nabla n$ es el gradiente de la densidad de partículas. En este caso sí trabajaremos en el equilibrio termodinámico [Fermi, 1968].

En estos tipos de tratamiento la muestra se introduce en un medio activo y se calienta, por lo que la especie de interés se liberará debido a las reacciones químicas y será absorbida por la superficie del sustrato [Philibert, 1991]. El inconveniente es que el proceso necesita de temperaturas muy estrictas.

Este problema puede ser eliminado si utilizamos como medio activo un plasma generado con descarga glow [Czerwiec, 2000]. El plasma estará entonces fuera del equilibrio termodinámico y los electrones tendrán una energía cinética elevada. Al colisionar excitarán a las moléculas, pudiendo entonces provocar una reacción química

$$e + X \rightarrow X^*$$
$$Y + X^* \rightarrow W + Z. \tag{2.2}$$

Para que esto sea posible debe darse que los caminos libres medios[1] sean lo suficientemente cortos como para que haya multitud de colisiones y poder mantener la cadena de reacciones.

## 2.2. Deposición de capas sobre la superficie

Otra forma de proteger la superficie de una pieza es mediante la deposición de otro material en forma de capa de un determinado espesor [Feugeas, 2003]. La ventaja de esta técnica sobre las anteriores es la libertad de elegir el material a depositar sobre la superficie. Sin embargo, ahora entran en juego los problemas de adherencia [Teixeira, 2001], por lo que es preciso una buena elección del material para el sustrato que debemos tratar [Evans, 1999]. Los procesos tradicionales se basan en la química y en la electroquímica, aunque la inclusión de plasmas da más versatilidad al asunto.

Los sistemas de deposición de recubrimientos basados en la generación de plasmas podemos dividirlos en los siguientes conceptos básicos:

- **PAPVD**. Es la deposición física en fase vapor asistida por plasmas [Schneider, 2000], nombrada así por sus siglas en inglés. Se basa en la generación de un plasma de un gas reactivo o de un haz de electrones [Zhu, 2005] y la emisión de átomos de un metal mediante un proceso de evaporación o de *sputtering*, de manera que permite la combinación de las especies para dar lugar a un determinado compuesto, que se dirige hacia la pieza a recubrir. El proceso permite, además de elegir el compuesto a desarrollar, el diseño de la interfase, permitiendo mejorar la adherencia [Vitiello, 2005] modificando los valores de microdureza y los gradientes de tensiones residuales. Tiene la ventaja de que se pueden modificar fácilmente las variables del proceso [Cheng, 2003], como cambiar de reactivos, la presión de llenado, la temperatura [Liu, 2005]. . . Con el fin de mejorar la adherencia se han desarrollado procesos en donde previo a la deposición de las capas duras, la superficie es sometida a difusión iónica [De las Heras, 2008].

---

[1] Es la distancia media en la que viaja una partícula entre colisión y colisión.

- **PACVD**. Es la deposición química en fase vapor asistida por plasmas, por sus siglas en inglés. Es un proceso en donde determinados compuestos son depositados en la superficie de un material a partir de reacciones químicas mantenidas entre moléculas precursoras en estado gaseoso [Olszyna, 1997]. Una alternativa al uso de elevadas temperaturas para crear las reacciones químicas es la utilización de plasmas [Sánchez Mathon, 2009] como medio para la producción de estas. Esto se debe al hecho de que en los plasmas las moléculas son excitadas por los electrones que poseen temperaturas mucho mayores que aquellas, reduciéndose en consecuencia la temperatura del sustrato [Corbella Roca, 2005]. Eligiendo adecuadamente los precursores es posible la generación de recubrimientos de variadas características.

- **Plasma Spray**. También es conocida esta técnica como termo-rociado por plasmas. Es un proceso en el cual un polvo es calentado rápidamente por encima de su punto de fusión, y proyectado en forma de pequeñas gotas a gran velocidad sobre la superficie de una pieza [Kang, 2007]. Las gotas se adhieren al sustrato y entre sí, formando tras su solidificación una capa uniforme. Este proceso es ayudado por un desplazamiento entre el flujo de partículas y la superficie para controlar el espesor de las capas.

  Para producir la fusión de las partículas es necesario disponer de un medio a gran temperatura y que produzca su arrastre a altas velocidades. Este medio puede ser un plasma en la región de *arco térmico* [Yamada, 2007] por el que se hace circular al polvo con el que se desea recubrir la muestra.

# Capítulo 3

# El plasma

Un *plasma* es una colección eléctricamente neutra (de manera aproximada) de cargas positivas y negativas que, en la mayoría de aplicaciones de interés industrial, interactúa fuertemente con un gas neutro de fondo [Roth, 1995]. La presencia de partículas cargadas en el plasma hace que también responda fuertemente con los campos eléctricos y magnéticos.

El régimen colisional del plasma en el que nos centraremos para nuestro posterior estudio es la aproximación lorentziana, que es una teoría del plasma basada en el *gas lorentziano*. Esta teoría argumenta que dentro del gas los electrones no interaccionan entre sí y se considera que los iones positivos permanecen en reposo. Los electrones sufren colisiones binarias con un gas neutro de fondo que actúa como un absorbente infinito de energía y momento proveniente de la población de electrones colisionales. Un refinamiento del gas lorentziano es el *modelo de Krook*, en el cual el tiempo efectivo de colisión es independiente del momento y energía de la partícula. Esto es a veces una buena aproximación de las interacciones de los electrones con gases nobles.

Existen otros modelos que describen otros regímenes en los que se puede encontrar el plasma. Estos modelos son el de *Boltzmann-Vlasov* encargado de analizar plasmas completamente ionizados, y el modelo de *Fokker-Planck* que estudia plasmas altamente turbulentos [Galeev, 1979].

En condiciones de densidades muy bajas no hay muchas colisiones entre partículas individuales y el comportamiento del plasma se estudia de acuerdo a las trayectorias de las partículas cargadas. Este es el análisis microscópico del plasma en donde se analiza el movimiento de cada carga inmersa en un campo eléctrico (ya sea uniforme, no uniforme o variable con el tiempo) o en un campo magnético (uniforme, no uniforme o variable con el tiempo), incluso ambos campos pueden actuar a la vez formando sus direcciones cualquier ángulo [Chen, 2006]. Al ir aumentando la densidad el plasma se empezará a comportar como fluido. En este caso se parte de la función de distribución de velocidades $f(r, v, t)$ de cada especie y se aplica en la *llamada ecuación de Boltzmann* para obtener ecuaciones macroscópicas necesarias para comprender el comportamiento del plasma en este régimen colectivo [Delcroix, 1968]. Las ecuaciones serán: la de conservación de las partículas; la de transporte de la cantidad de movimiento, y la de transporte de la presión cinética.

## 3.1. Parámetros del plasma

La propiedad por la que el plasma posee de manera aproximada el mismo número de partículas positivas que de negativas se denomina *cuasi-neutralidad*. Para que el plasma sea cuasi-neutro en el estado estacionario, es necesario tener casi el mismo número de cargas de un signo y de otro por elemento de volumen. Tal elemento de volumen debe ser lo suficientemente grande para contener el suficiente número de partículas, y lo suficientemente pequeño respecto a las longitudes características de las variaciones de los parámetros macroscópicos tales como la densidad y la temperatura. En cada elemento de volumen los campos de la distribución espacial de carga microscópica de los portadores de carga individuales deben cancelarse con los otros para conseguir la neutralidad de carga macroscópica.

La escala característica de longitud, $\lambda_D$, se denomina *longitud de Debye* y es la distancia sobre la cual se obtiene un balance entre la energía térmica de la partícula [Baumjohann, 1996], que tiende a perturbar la neutralidad eléctrica, y la energía potencial electrostática resultante de cualquier separación de carga, que tiende a restaurar la neutralidad de carga. Para que el plasma sea cuasi-neutro, la dimensión física del sistema, *L*, debe ser grande comparada con $\lambda_D$:

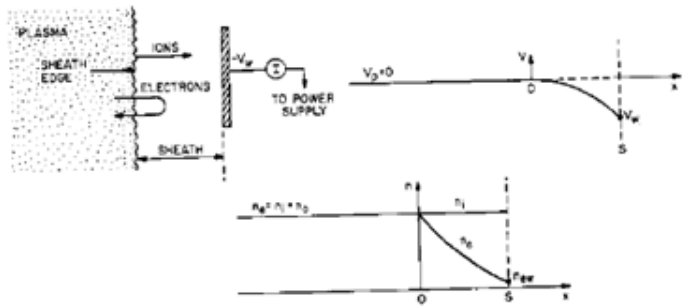

Figura 3.1: La conducta de electrones y parámetros del plasma en una vaina entre una pared polarizada negativamente y un plasma a potencial cero. Se indica un esquema de los perfiles de potencial y densidad numérica de partículas cargadas en la vaina [Roth, 1995].

Un electrodo o pared en contacto con el plasma normalmente afectará solo a sus alrededores más inmediatos del plasma. A menos que hayan grandes flujos de corriente en el plasma, o que sea altamente turbulento, un plasma tenderá a formar una vaina superficial para apantallarse él mismo de los campos eléctricos aplicados [Roth, 1995]. La *distancia de apantallamiento* característica es aproximadamente igual al espesor de la vaina que se forma entre el plasma y la pared envolvente. En las vainas no se verifica la cuasi-neutralidad. Según esta descripción el plasma siempre estará separado de la pared mediante vainas.

Ya que el efecto de apantallamiento es el resultado del comportamiento colectivo dentro de la esfera de Debye de radio $\lambda_D$, es necesario que esta esfera contenga las suficientes partículas [Baumjohann, 1996]. El número de partículas dentro de la esfera de Debye es $(4\pi/3)n_e\lambda^3_D$. El término $n_e\lambda^3_D$ se denomina normalmente *parámetro del plasma*, $\Lambda$. La densidad de electrones por unidad de volumen se denota como $n_e$.

La frecuencia de oscilación típica en un plasma completamente ionizado es la *frecuencia del plasma* (electrónica), $\omega_{pe}$. Si la cuasi-neutralidad del plasma está distorsionada por alguna fuerza externa, los electrones, teniendo más movilidad que la mayoría de los iones pesados, son acelerados en un intento de restaurar la neutralidad de la carga [Baumjohann, 1996]. Debido a su inercia se moverán más allá de su posición de equilibrio, resultando una oscilación colectiva rápida alrededor de los iones más masivos [Lieberman, 1994]. Una oscilación cuantizada del plasma de un conductor, con una energía típica de $E_P = \hbar\omega_{pe}$; se conoce como *plasmón* [Mochán], con $\hbar$ la constante de Planck reducida.

La frecuencia electrónica del plasma es crítica para la propagación de radiación electromagnética en plasmas [Milántiev, 1987], ya que bajo ciertas condiciones la radiación atravesará el plasma y en otras será reflejada.

## 3.2. Descarga eléctrica a baja presión

Consideremos para mayor claridad un tubo de vidrio evacuado, con electrodos con forma de discos circulares en cada extremo y conectados a una fuente de potencia DC de alta tensión. Ajustando un reóstato $R$, se puede barrer la curva característica de tensión-corriente, la cual es altamente no lineal. En el plasma, los electrones (y los iones negativos también) migran hacia el ánodo y los iones positivos hacia el cátodo, ambos colisionando frecuentemente con el gas neutro de fondo.

Si se considera un tubo de descarga como el anteriormente descrito y va aumentando la tensión *V*, mientras se mide la corriente *I* que fluye a través del tubo, la descarga trazará una curva que podremos medir y se esboza en la figura (3.2). La primera zona es la *descarga oscura* que es invisible al ojo y consta de varios regímenes: ionización de fondo, de saturación, de Townsend [Von Hippel, 1954] y de corona. Si la tensión se incrementa más allá del valor $V_B$ ocurrirá la ruptura eléctrica [Braithwait, 2000].

Una vez que la ruptura eléctrica se da, la descarga realiza una transición al régimen de *descarga glow (luminiscente)*, en la cual la corriente es bastante alta, al igual que la cantidad de excitación del gas neutro de fondo, por lo que el plasma es visible al ojo humano [Roth, 1995].

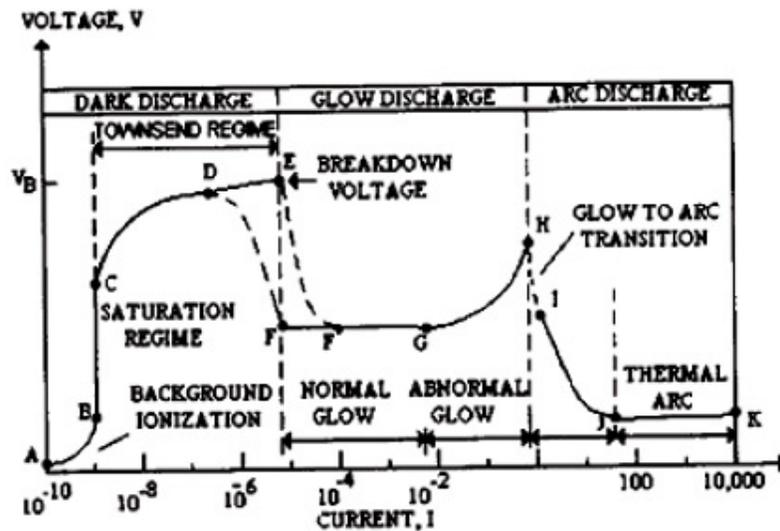

Figura 3.2: Característica tensión-corriente de un tubo de descarga eléctrica DC a baja presión. Se puede observar el ciclo de histéresis para la transición de descarga glow a oscura [Roth, 1995].

Después de una transición discontinua nos encontramos entonces en la región de *glow normal* de la curva tensión-corriente, en la que la tensión a través de la descarga es casi independiente de la corriente de descarga en varios órdenes de magnitud. Si se incrementa la corriente, la fracción del cátodo ocupada por el plasma incrementa, hasta que el plasma cubre toda la superficie del cátodo. En este punto, la descarga entra en el régimen *glow anormal*, en la que la tensión vuelve a incrementar en función de la corriente.

Más allá del glow anormal, la densidad de corriente en el cátodo puede llegar a ser suficientemente grande como para calentar el cátodo hasta la incandescencia [Roth, 2001], así pues se desencadena una *transición glow-a-arco* discontinua. Tras esta transición, la descarga se establece en algún punto de la región de arco, que dependerá de la resistencia interna de la fuente de energía DC. El *régimen de arco no térmico* es el régimen en el que la tensión de descarga decrece mientras aumenta la corriente. Tras este, la tensión aumenta lentamente con un incremento de corriente, entrando en el *régimen de arco térmico*.

A continuación se van a describir las zonas más importantes de la descarga glow en la que se van a llevar la mayoría de procesos de interés y que son totalmente necesarios para lograr el tratamiento superficial del acero.

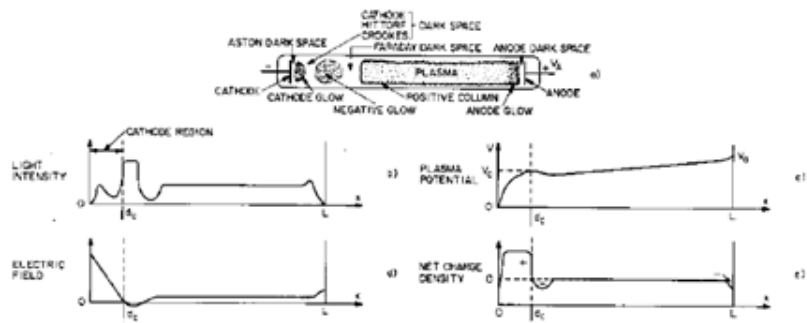

Figura 3.3: Variación axial de las características de la descarga glow normal
[Roth, 1995].

La primera es la *región del cátodo*. La mayoría de la caída de tensión a través del tubo de descarga aparece entre el cátodo y el contorno entre el espacio oscuro catódico y el glow negativo. Esta región se llama región del cátodo. Es de longitud $d_c$, desde la superficie del cátodo ($x = 0$) hasta la frontera del glow negativo ($x = d_c$). La caída de tensión se conoce como *caída catódica*, y es de $V_c$ voltios. La mayor parte de la potencia disipada en una descarga glow se da en la región del cátodo. En esta región, los electrones son acelerados a energías suficientemente altas como para producir ionización y avalanchas en el glow negativo.

Una descarga glow a baja presión ajustará la longitud axial de su región del cátodo, $d_c$, hasta que se establezca un valor mínimo del producto $d_c p$:
$$d_c p \approx (d_c p)_{min}, \tag{3.1}$$
donde este producto es el *mínimo de Paschen* [Raizer, 1991]. En este mínimo, la descarga es autosostenida bajo condiciones de una caída mínima de tensión y disipación de potencia en el cátodo. En la descarga glow normal, la densidad de corriente fluyendo al cátodo permanece aproximadamente constante conforme la corriente total varía, así como que el área de la descarga del plasma en contacto con el cátodo incrementa con la corriente total. Valores típicos de caída catódica están entre 100 y 300 V.

La segunda es el *glow negativo*. Inmediatamente a la derecha del espacio oscuro catódico (figura (3.3)) está el glow negativo, con la intensidad de luz más brillante de toda la descarga. El glow negativo posee un campo eléctrico relativamente bajo, es normalmente largo en comparación con el glow catódico y es más intenso por el lado del cátodo. Los electrones portan casi toda la corriente en la región del glow negativo [Francis, 1956]. Los electrones que han sido acelerados en la región catódica producen ionización e intensa excitación en el glow negativo, así pues se observa entonces la luz brillante. Conforme son frenados los electrones, la energía de excitación no vuelve a estar disponible y comienza por consiguiente el espacio oscuro de Faraday. La densidad numérica de electrones en el glow negativo es característicamente de unos $10^{16}$ electrones/m$^3$.

## 3.3. Cementación y nitruración iónicas mediante plasmas

Las conocidas *cementación y nitruración iónicas*, se fundamentan en la modificación de la superficie de nuestro acero austenítico, mediante el principio de *difusión iónica*. Técnicas similares de este grupo serán la boronización y la carbo-nitruración. Explicaremos con detenimiento el proceso de cementación y el de nitruración, cuyo conocimiento y estudio será de vital importancia para llevar a cabo el experimento de una manera satisfactoria, y ya asentamos que la cinética de cada proceso es casi igual, en cuanto a los procesos que se desencadenan. La cementación debe entenderse como el proceso de carburización, o sea, la entrada del carbono dentro de la red, no como el concepto clásico de cementación en el que se forma cementita ($Fe_3C$) en la superficie.

El proceso de *difusión iónica* se emplea para endurecer la superficie de piezas de metal. La pieza de metal es el cátodo de una descarga glow que se produce dentro de un dispositivo de calentamiento a temperaturas hasta 1000 ºC. La presión de trabajo será muy inferior a la atmosférica. Un gas típico que se utiliza para llevar a cabo el proceso de cementación es el metano, $CH_4$, aunque también es posible realizar el tratamiento usando otros gases de hidrocarburos, tales como el acetileno ($C_2H_2$). En cuanto a la nitruración iónica uno de los gases más empleados es el nitrógeno molecular, $N_2$, aunque también puede recurrirse al uso del amoniaco ($NH_3$) por su menor energía de disociación molecular. En el proceso de difusión iónica, las especies activas neutras e iónicas se producen en el plasma. Los iones son acelerados en la vaina del cátodo (la pieza a ser tratada) y entonces se calienta por impacto iónico [Feugeas, 2003]. La clave del proceso de cementación, al igual que el de nitruración, es el control de la producción de especies activas y la temperatura del plasma. A partir de la espectroscopia de emisión se puede determinar la temperatura del gas y compararla con la temperatura del sustrato, además de determinar la concentración de carbono o nitrógeno en las cercanías del sustrato, su nivel de excitación, etc.

Además, puede emplearse un reactor de post-descarga fluida para separar la producción de especies activas neutras por el plasma y el calentamiento de la pieza que opera en la parte de abajo de la descarga [Ricard, 1996]. Los reactores se pueden construir en base a una descarga glow DC o por una descarga de microondas. El interés industrial de estos reactores se centra en el tratamiento de piezas de diferentes formas superficiales con ranuras, huecos y tubos. Es una ventaja sobre el proceso iónico clásico donde las piezas a tensión negativa se rodean de una vaina catódica de espesor entre 1 y 5 mm. Este cátodo entonces apantalla la acción iónica dentro de pequeñas ranuras, huecos o tubos.

Los experimentos de cementación y nitruración mediante plasmas tienen parámetros de trabajo muy parecidos entre sí [Ricard, 1996]. La tensión que genera la descarga glow se sitúa entre 400-600 V para la cementación y entre 600-800 V para la nitruración a una presión de ≈3,750 Torr (≈5 mbar). Esto hace que el camino libre medio sea lo suficientemente pequeño como para que haya multitud de colisiones y lo suficientemente grande como para que la partícula llegue antes de colisionar con una energía mayor que la de ionización. Durante el proceso, los iones calientan por colisión al sustrato y contribuyen al proceso de nitruración o cementación, ya que favorece la difusión.

En un proceso industrial la nitruración y la cementación se realizarán en una cámara de vacío con un sistema de evacuación de gases que permita llegar a una presión ≤1,0·10$^{-3}$ mbar. Al mismo tiempo debe existir otra válvula que ingrese en la cámara los gases reactivos hasta la presión de trabajo óptima. La idea principal es que al llegar el nitrógeno o el carbono a la superficie, estos entren en el interior del material por un proceso de difusión [Sudha, 2010]. Experimentalmente, para la cementación usando metano, se observa que a la superficie llegan neutros e iones, ambos excitados, tanto de C, como de $C_2$ y de CH [Rie, 1998], mientras que para la nitruración con nitrógeno molecular se observa en la superficie N y $N_2$ [Ricard, 1996] tanto en su forma neutra como ionizados y casi siempre en un estado excitado. Como la velocidad de las partículas es elevada penetrarán en el material favoreciendo así el proceso de difusión. Se tiene que una vez dentro del material quedan eléctricamente neutralizados si la energía por nucleón no es de varias decenas de keV [Bürgi, 1990]. Esta es una ventaja frente a los métodos que solo se basan en la intervención de una reacción química.

Experimentalmente, se tiene que la difusión de N o C en la fase γ (austenita) incrementa la resistencia al desgaste profundo de la pieza [Fernandes, 2010]. A partir de las micrografías pueden estudiarse *a posteriori* las capas presentes en el acero tratado y mediante los espectros de difracción de rayos X se caracterizan las capas que se obtienen en un tratamiento bajo ciertas condiciones, ayudando así a reproducir las sesiones con idénticos resultados.

Las piezas de acero son muy sensibles a la oxidación que se produce por impurezas de aire o agua. También se tiene que la capa de óxido desaparece cuando se introduce una cantidad de metano [Ricard, 1996]. Por otra parte, el hidrógeno aumenta la eficiencia del proceso de difusión iónica, luego en principio (aunque no está totalmente demostrado) la presencia de hidrógeno (H) elimina el oxígeno (O) del acero, tanto de su capa pasivante como de las impurezas de vapor de agua que existan [Kumar, 2000]. Por todo esto, podemos suponer entonces que para evitar estos óxidos ha de añadirse gas hidrógeno en el proceso iónico estándar.

En muchos casos es esencial que el tratamiento superficial evite que en la misma se desarrollen otros compuestos. Se puede solucionar utilizando una configuración de los electrodos denominada *triodo* que permite trabajar con presiones de llenado inferiores y tensión de 200 V [Feugeas, 2003]. Con este tipo de configuración se pueden conseguir capas de difusión de elevada dureza y gran espesor. Es posible la obtención de este tipo de estado superficial final utilizando reactores de configuración bipolar.

La cementación y nitruración iónicas en estas condiciones permiten lograr en tiempos de proceso muy cortos y a bajas temperaturas [Feugeas, 2003b] capas de una fase conocida como austenita expandida que aumenta la resistencia al desgaste y le otorga una microdureza alta [Wang, 2009].

## 3.4. Sputtering por magnetrón

Es una importante técnica de tratamiento superficial [Kelly, 1984] de la que se hará uso en esta tesis y tiene una amplísima proyección en el campo tecnológico y experimental [Clement, 2004]. Se basa en la deposición de una capa sobre la superficie a tratar. Mediante el uso del magnetrón[2] se origina un plasma por el que se logran depositar de manera física partículas en un sustrato [Mahieu, 2006]. Las descargas generalmente son tipo glow DC, tratándose en consecuencia de plasmas fríos fuera del equilibrio termodinámico [Drüsedau, 2002]. Existen variantes en función de la configuración del magnetrón y el tipo de alimentación que tiene, así como la polarización o no del sustrato, aunque en todo caso la presión de trabajo es muy baja.

Este concepto se basa en la generación de un plasma de un gas reactivo (por ejemplo, $N_2$) y la emisión de átomos de un metal (por ejemplo aluminio, Al) mediante un proceso de evaporación o de *sputtering*, de manera que permite la combinación de las especies para dar lugar a un determinado compuesto (por ejemplo, AlN). Este compuesto es incorporado al plasma pudiendo ser ionizado por colisión electrónica y dirigiéndose preferentemente hacia a recubrir [Feugeas, 2003], que puede estar *polarizada* con tensión negativa. Polarizar la muestra no es necesario en este proceso aunque ayuda a mejorar la tasa de deposición si esta muestra es conductora [Cheng, 2003b]. En caso de que tengamos la muestra sin polarizar esta estará a un *potencial flotante*. Una gran cantidad de compuestos pueden generarse siguiendo este mecanismo [Chan, 2009], como por ejemplo el WC (mediante la generación de C en descargas tipo glow en $CH_4$ y el *sputtering* de tungsteno, W), el CrN (plasma de $N_2$ y evaporación de Cr), TiCN (evaporación de Ti en un plasma de $N_2$ y $CH_4$), el TiN (plasma de nitrógeno y evaporación de titanio, Ti) [Ingason, 2009], etc. Este tipo de proceso permite, además de la elección del compuesto a desarrollar según las propiedades deseadas para la superficie a recubrir, el diseño de la interfase, posibilitando optimizar la adherencia mediante la reducción de los gradientes de tensiones residuales y de microdureza, la compatibilidad química, etc. La versatilidad de este concepto se puede inferir considerando que sin la necesidad de la apertura a una atmósfera de la cámara de reacción, es posible modificar fácilmente las variables del proceso, obteniendo recubrimientos de estructuras complejas [Xu, 2001]. Con solo cambiar la naturaleza de los gases reactivos [Cheng, 2003c] que ingresan en la cámara, o la naturaleza de los materiales a evaporar, además de otros parámetros auxiliares como la presión de llenado [Cheng, 2003d], la temperatura del proceso, corrientes de descarga, etc., es posible cambiar totalmente la naturaleza de los compuestos a depositar. Se logra, por ejemplo, la deposición de recubrimientos multicapas consistentes en una sucesión de láminas delgadas [Auger, 2003] de diferentes compuestos. Con el fin de mejorar la adherencia al sustrato por ejemplo [Dauskardt, 1998], se han desarrollado procesos en donde previo a la deposición de las capas duras, la superficie del sustrato es sometida a un proceso de difusión iónica para generar una capa superficial con el compuesto en solución sólida [De las Heras, 2008]. Este tipo de tratamiento se conoce como *dúplex*.

---

[2] Dispositivo refrigerado con agua en el que se coloca en su cátodo el material a eyectar. Posee un campo magnético que provoca el confinamiento de un altísimo número de electrones al iniciar la descarga, ionizando a multitud de átomos haciendo que, por consiguiente, sus iones sean atraídos hacia el cátodo.

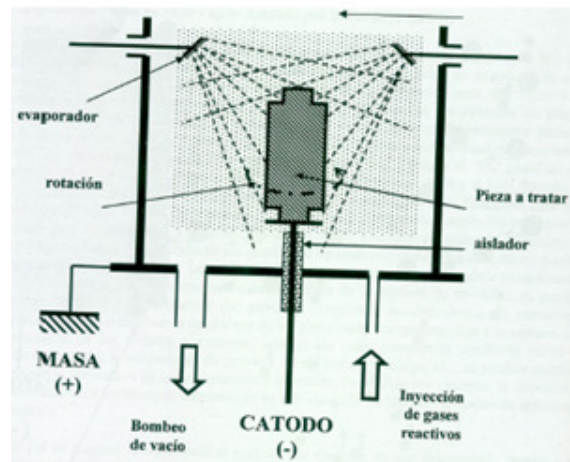

Figura 3.4: Esquema de un reactor PAPVD empleado para recubrir superficies de aceros. Es un sistema equivalente a los utilizados en procesos de nitruración iónica, solo que las presiones de llenado de gases son 2 o 3 órdenes de magnitud inferiores [Feugeas, 2003].

Los átomos metálicos también pueden originarse mediante la evaporación por *haces de electrones* [Zhu, 2005], por *cátodo hueco* [Pessoa, 2007], por *descarga catódica*, etc. El proceso de deposición mediante *sputtering* usando la configuración de magnetrones [Smentkowski, 2000] se emplea para endurecer la superficie de piezas de metal, así como la fabricación de compuestos de propiedades piezoeléctricas [Loebl, 2003] y ópticas [Venkataraj, 2006].

# Bibliografía

# Parte III

# Experimento

En esta parte del trabajo, describiremos la realización experimental llevada a cabo. Enunciaremos los aparatos con los que contamos y los pasos que realizamos para fabricar los sustratos que modificaremos. A continuación, indicaremos todos y cada uno de los instrumentos que poseemos para el conveniente estudio y caracterización de las probetas conseguidas. En este paso abordaremos la marca y características de cada aparato usado, los principios físicos en los que se sustenta y las condiciones de trabajo a los que fueron sometidos para conseguir la caracterización.

Los procesos de Cementación y Nitruración Iónica, así como la deposición mediante sputtering por magnetrón y el estudio de Tribología se llevaron a cabo en el *Grupo de Física del Plasma*, una rama del Instituto de Física de Rosario (IFIR), localizado en Rosario (Argentina) y a cargo del Dr. Jorge N. Feugeas. Las sesiones de Microscopía Óptica, cálculo de Dureza mediante Indentación, Tratamientos Térmicos mediante horno con control de temperatura y tiempo y estudios de la Resistencia a la Corrosión se realizaron en el *Laboratorio de Metalurgia*, dependiente del Instituto de Mecánica Aplicada y Estructuras (IMAE) de la Facultad de Ciencias Exactas, Ingeniería y Agrimensura de Rosario (Argentina) y a cargo de la Dra. B. Liliana Nosei. Parte de los análisis de corrosión se hicieron en colaboración con el *Institut Universitaire de Technologie* de la Université des Sciences et des Technologies de Lille, situada en Lille (Francia) a cargo de los Doctores D. Chicot y J. Lesage. Los análisis de Nanoindentación y las pruebas de FIB se llevaron a cabo en el *Dipartimento di Ingegneria Meccanica e Industriale* de la Università "Roma Tre" de Roma (Italia) a cargo del Dr. E. Bemporad. El estudio de las probetas mediante rayos X se realizó en el *Laboratorio de Rayos X*, parte del Instituto de Física de Rosario (IFIR), localizado en Rosario (Argentina) y a cargo del Dr. Raúl E. Bolmaro. La caracterización concerniente a la Espectroscopia Electrónica de Auger se realizó en el *Laboratorio de Superficies e Interfaces* del Instituto de Desarrollo Tecnológico para la Industria Química del INTEC, localizado en Santa Fe (Argentina) y a cargo del Dr. Julio Ferrón. Las sesiones de Plasma Focus se implementaron en el *Instituto de Física de Arroyo Seco*, laboratorio de Plasmas Densos, de la Universidad Nacional del Centro de la provincia de Buenos Aires, de Tandil (Argentina), a cargo de la Dra. María M. Milanese.

# Capítulo 4

# Montaje experimental

## 4.1. El reactor de plasma

Las sesiones de cementación iónica se llevaron a cabo en un reactor de plasma con una fuente de corriente continua, donde se controlaba la presión de trabajo, la tensión y la temperatura [Isola, 2007].

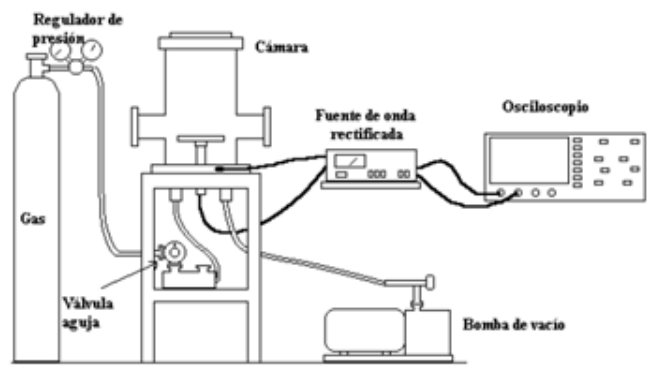

Figura 4.1: Dispositivo experimental utilizado para el trabajo [Isola, 2007].

A continuación, iremos describiendo cada una de las partes de las que consta el montaje:

- **Reactor**. Tiene las paredes construidas de acero inoxidable, con un espesor de 6 mm, encerrando un volumen de 8,8 litros. La cámara posee dos ventanas de vidrio *Pirex*, las cuales no dejarán pasar longitudes de onda menores a 300 nm [Nosei, 2004]. Por último, posee un cátodo interior aislado eléctricamente de la cámara (que actuará como ánodo) por un soporte de vidrio *Pirex*. Los conductos de entrada y salida de gases están colocados en la parte inferior de la cámara. Los cierres de la cámara, aparte de tener tuercas, poseen un *o-ring* engrasado para evitar pérdidas de vacío [Chambers, 1989]. Debido a las temperaturas de trabajo se coloca exteriormente una tubería de cobre por la que circulará en todo momento agua, para refrigerar así las zonas de la cámara que poseen *o-ring*, o sea, las ventanas y la tapa superior.

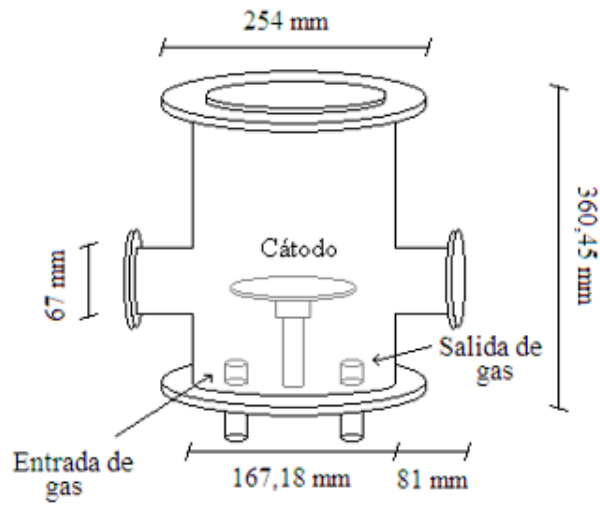

Figura 4.2: Reactor de plasma, indicando sus dimensiones y las partes más destacadas [Isola, 2007].

Se usó además un nuevo reactor de cementación en el que se suprimen las ventanas laterales, evitando así posibles riesgos de degradación de los o-ring. Está fabricada con acero AISI 304 de 7,2 mm de espesor y el volumen total es de 5,1 L [Nelson, 2002]. Para mejorar el llenado de gas de la cámara se usa una tubería de vidrio que lleva la entrada de gases (en la parte inferior) hacia la parte superior, justo a 9 cm del cátodo. De esta manera la renovación de los gases reactivos y el efecto de barrido harán mejorar los resultados.

Además, el cátodo también sufrió modificaciones, de tal forma que en los últimos experimentos se suprimió el efecto de borde sobre las probetas mediante el encastre de estas en el propio cátodo, fabricado con acero AISI 304. Esto se llevó a cabo haciendo un dispositivo de dos placas: una conectada a la alimentación donde se asientan las probetas y otra con agujeros para que queden al mismo nivel del cátodo, obligando al efecto de borde a colocarse en la parte más externa, lugar donde no están situadas las probetas. Esto mejora el proceso de cementación al no haber variaciones de campo eléctrico en ellas. Los agujeros son un poco más grandes que el diámetro de las probetas para una fácil colocación.

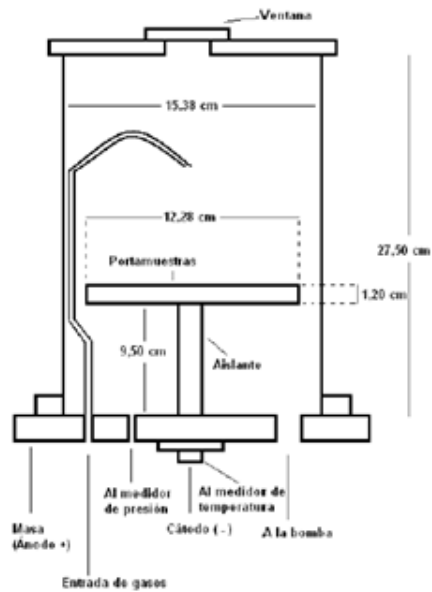

Figura 4.3: Esquema del segundo reactor de cementación.

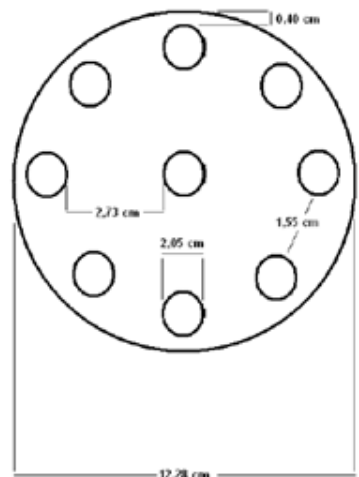

Figura 4.4: Diseño del cátodo.

- **Bomba de vacío**. Es de tipo rotatoria, de la marca VARIAN SD-200, equipada con un motor monofásico de 220 V y 60 Hz. Posee una velocidad de desplazamiento del aire de 12 m$^3$/h [O'Hanlon, 1989] y la presión mínima que puede alcanzar (según datos del fabricante) es de 7,5·10$^{-5}$ Torr (1,0·10$^{-4}$ mbar). La contracorriente de aceite intenta minimizarse con codos y filtros.

- **Válvulas agujas**. Se utilizan para regular la entrada de gases en la cámara. Son correspondientes al modelo LVIOK, producido por *Edwards (Britain)*. El proceso de cementación (o nitruración) es dinámico: en todo momento la bomba evacúa gases y entran mediante las válvulas que llevan a un reservorio y de ahí a la cámara.

- **Medidor de presión absoluta**. El modelo que utilizamos es el MKS Baratrón tipo 112A. Este aparato mide la presión por la deformación de un diafragma que está en contacto con el gas, lo cual produce una variación en la capacidad del sistema, lo que se traduce a una señal eléctrica. El rango de medición de presiones es de 0-26,6 mbar con un error de $1,3 \cdot 10^{-2}$ mbar.

- **Osciloscopio**. Con este aparato podremos medir la tensión y la intensidad de la señal que está recorriendo al plasma. Por su pantalla observaremos la forma de la señal del plasma. Es necesario advertir que no se mide realmente la intensidad del plasma, sino la caída de tensión sobre la resistencia de carga (u otra colocada en la fuente para este propósito), aunque aplicando la ley de Ohm conoceremos la intensidad de manera inmediata. El modelo es LECROY 9361 con dos canales. Admite una frecuencia máxima de 600 MHz y un barrido de 5 Gs/s.

- **Termopar**. Mide la temperatura del cátodo por *efecto Seebeck*. Es de tipo K (un filamento de Ni-Cr y otro de Ni-Al). El rango de temperaturas de este modelo oscila entre -200 y 1250 ºC, con un error del 0,75 %.

- **Fuente de energía**. Da origen a una señal que se enviará al plasma. Esta señal es de la clase DC rectificada, es decir, emite una señal senoidal de periodo fijo (la frecuencia de la línea de tensión alterna comercial, 50-60 Hz) a la que se le elimina la parte negativa de la oscilación y se minimiza la permanencia en el valor 0. La resistencia interna (de carga) que posee la fuente es de 39 Ω y la tensión de salida es regulable mediante un autotransformador VARIOSTAT.

  Se utilizó otra fuente de potencia que entregó una señal rectificada. La tensión máxima para este dispositivo es de 900 V y la corriente se mide sobre una resistencia de 1 Ω. Esta fuente permite mayor paso de corriente, siendo conveniente para acortar el tiempo de calentamiento al tener una resistencia de carga de 16,8 Ω.

- **Tubos de gas**. Son provistos por *Gases Rosario*. Son de gran volumen y la pureza de cada uno de ellos es la siguiente: para el argón, 4,5; para el hidrógeno molecular, 4,8; y para el metano, 4,5: Para recurrir a la nitruración emplearemos un tubo de nitrógeno molecular de pureza 5,0. Los controladores de flujo que conectan los tubos con las válvulas aguja son de la marca COLE-PALMER. Se eligieron estos gases por su facilidad de reabastecimiento y de uso, en combinación con la literatura existente que los utiliza.

# 4.2. Proceso previo a la sesión

En primer lugar se obtuvo de ROLDÁN S.A.-Aceros Inoxidables una barra de acero austenítico AISI 316L de una pulgada de diámetro (≈2,54 cm). Según la hoja de características aportada por el vendedor la barra poseía la siguiente composición porcentual de elementos:

| Elemento | Porcentaje |
|---|---|
| Carbono | 0,0300 |
| Fósforo | 0,0400 |
| Azufre | 0,0260 |
| Silicio | 0,3190 |
| Manganeso | 1,3440 |
| Cromo | 17,0180 |
| Níquel | 10,6590 |
| Molibdeno | 2,1840 |
| Titanio | 0,0040 |
| Cobre | 0,3800 |
| Hierro | 67,9960 |

Tabla 4.1: Porcentaje atómico de los elementos presentes en la barra de acero austenítico AISI 316L adquirida.

La barra se cortó en probetas de 7 mm de espesor y fueron rectificadas por ambas caras para eliminar posibles imperfecciones y rugosidades y así minimizar las diferencias del paso de corriente (que alteraría el valor de energía con la que llegan los iones) por diferentes valores de resistencia que pueden tener unas probetas con respecto a otras. Para la colocación en el nuevo cátodo, esquematizado en la figura (4.4), algunas de las probetas se rebajaron para conseguir un diámetro de 2 cm y una altura de 6 mm. Tras esto se comenzó el pulido, el cual consta de tres fases:

1.  Pulido en correa motorizada usando lijas de tamaño de grano 80, 180 y 400. Como refrigeración se recurrió al uso de agua. El pulido con cada lija se hizo en una única dirección hasta que las marcas que estaban en otras direcciones desaparecieron.

2.  Se aplicaron lijas de tamaño de grano 600, 1000 y 1200, repitiendo de forma manual (sin usar la correa motorizada) el proceso anterior.

3.  Se intentó dar un acabado espejado recurriendo a un torno con un paño empapado con una solución acuosa de alúmina en suspensión, cuyo tamaño es de 1 μm. Tras el acabado, las probetas se lavaron con agua, se limpiaron de restos de grasa mediante alcohol y se pusieron bajo un secador durante medio minuto. En el caso de observar rayaduras se repetirá el pulido a partir de la lija de tamaño de grano 600.

Tras este trabajo se guardaron en lugar oscuro y seco envueltas en algodón. Para determinar la rugosidad superficial de estas probetas, una selección de estas se sometió a un análisis con un dispositivo Surface Roughness Tester, modelo TR200, que consta de una punta de diamante con su vértice formando un ángulo de 90º con las paredes y aplicando sobre la superficie una carga de 4 mN. La resolución del dispositivo está alrededor de 10-40 nm. Los resultados determinaron que el valor de $R_a$ (rugosidad media obtenida al realizar la diferencia entre crestas y valles) dan un valor de ≈40 nm.

Ya construidas y pulidas las probetas, se indica el proceso que se lleva a cabo para cementarlas (proceso que es idéntico al de nitruración). En primer lugar se limpian con acetona para eliminar así cualquier polución existente, por lo que se debe limpiar toda su superficie, no solo la parte a cementar. Inmediatamente, se las limpia también con papel secante y se colocan en el cátodo de la cámara. Durante este proceso deben llevarse guantes protectores para impedir marcas de huellas dactilares en la probeta, ya que esto influiría negativamente sobre el experimento.

Como convenio se colocan en el cátodo tres probetas lo más posiblemente alejadas entre sí, ya sea tanto sobre una placa como en el nuevo cátodo donde van encastradas (en este caso el resto de huecos se rellenarán de probetas testigo que no serán estudiadas). La razón de colocar varias probetas se debe a que así hay un número suficiente de ellas para aplicar todos los procesos de caracterización, ya que algunos de los que se van a aplicar serán destructivos y solo podrán utilizarse una vez. La explicación de por qué deben estar separadas entre sí en el caso de no utilizar el cátodo que permite el encastre es para evitar la creación de un cátodo hueco[3] y la consecuente elevación desmesurada de la temperatura en esa zona.

Colocadas las probetas en el cátodo se cierra la cámara y se abre la tubería que porta el agua de refrigeración. A continuación se activa la bomba de vacío y las válvulas de evacuación se abren al máximo. La presión base de trabajo es $2{,}67 \cdot 10^{-2}$ mbar.

Al alcanzar la presión base se abre el tubo de hidrógeno y se llena el reactor de este gas. Para ello, es necesario cerrar la válvula de mayor evacuación de la bomba y se deja abierta la válvula de menor evacuación por la que saldrá un caudal fijo y minimizando la cantidad de gas necesario para llevar a cabo el proceso. Abriendo y cerrando periódicamente la válvula aguja que controla el ingreso del hidrógeno se realizan algunas purgas para asegurar que el vacío base consta mayoritariamente de hidrógeno molecular. La presión de $H_2$ en la cámara será (tomando valores promediados de las sesiones) de $1{,}484 \pm 0{,}004$ mbar. Se conecta entonces la fuente de potencia. Con el Autotransformador se aumenta la tensión que se entrega al plasma lentamente, recorriendo toda la zona oscura de manera uniforme y llegando así a la ruptura. Si el aumento de tensión se realizase rápido es probable que se formasen arcos peligrosos en la cámara. Una vez que se consigue la descarga glow es necesario dejar este plasma durante 15 minutos a una tensión de 446±10 V, intensidad de 150±20 mA y temperatura de 40±3 ºC, cuyos valores y errores también están promediados sobre los datos de las sesiones. Este plasma limpiará la superficie de posibles óxidos y ayudará al posterior proceso de cementación (o nitruración).

---

[3] El cátodo hueco es una fuente de electrones y los crea en gran cantidad.

Tras la limpieza con hidrógeno la cámara se llenará con los gases restantes (argón y metano) a las proporciones establecidas para cementar, llegando así a la presión de trabajo. Se eligen estas mezclas para determinar cómo influyen diferentes proporciones de argón e hidrógeno en las características de las muestras cementadas, así como el intento de adaptar a la cementación el uso de atmósferas de tres gases. En esta tesis la presión a la que se cementa será de 3,750 Torr, es decir, 5,000 mbar (caracterizaciones previas demostraron que es una presión óptima para las cámaras a utilizar y para los tratamientos que realizaremos), a una temperatura que se situará siempre entre 400-410 ºC, dejando libre la tensión que recorrerá el plasma. Por tanto, los únicos parámetros de ajuste con los que se cuenta serán la presión (mediante el ajuste de cada una de sus válvulas) y la tensión del plasma. El trabajo se realizará con dos mezclas diferentes y cada una durante sesiones de 30, 60 y 120 minutos. Los gases que se utilizan para crear nuestra atmósfera serán el Ar, el $H_2$ y el $CH_4$ en las siguientes proporciones:

- Las probetas indicadas como C50- poseerán una atmósfera de 50 % Ar – 45 % $H_2$ – 5 % $CH_4$: Como la presión de trabajo será de 5,000 mbar esto se traduce en los siguientes valores de presiones parciales: 2,500 mbar de Ar, 2,250 mbar de $H_2$ y 0,250 mbar de $CH_4$: La nomenclatura para esta mezcla irá seguida de 030, 060 o 120, según el tiempo de tratamiento.

- Las probetas indicadas como C80- poseerán una atmósfera de 80 % Ar – 15 % $H_2$ – 5 % $CH_4$: Como la presión de trabajo será de 5,000 mbar esto se traduce en los siguientes valores de presiones parciales: 4,000 mbar de Ar, 0,750 mbar de $H_2$ y 0,250 mbar de $CH_4$: La nomenclatura para esta mezcla irá seguida de 030, 060 o 120, según el tiempo de tratamiento.

El hidrógeno es fácilmente ionizable y la producción de electrones servirá para mantener un valor de corriente adecuado. Por su lado, el argón tendrá la suficiente masa para calentar el cátodo por transferencia de momento tras colisionar con este. Sus estados metaestables además servirán como reserva energética para mantener la descarga. Una vez conseguida la presión de trabajo con las proporciones adecuadas se va aumentando la tensión del plasma poco a poco hasta llegar a un valor de temperatura por encima de 400 ºC, momento en el que se empieza a tomar el tiempo, ya que es cuando comienza el proceso de cementación. La subida progresiva de 40 hasta 400 ºC durará normalmente entre 35-45 minutos. En las sesiones con nuevo cátodo, el calentamiento se lleva a cabo de una manera diferente: en vez de suministrar la mezcla de gases de trabajo tras la limpieza con hidrógeno, se introduce una mezcla de hidrógeno (4,000 mbar) y de argón (1,000 mbar) para dejar la entrada del gas de interés hasta el momento de alcanzar la temperatura adecuada, eliminando así posibles malinterpretaciones. Además, el tiempo de tratamiento será en este caso de 80 minutos.

Para el caso de la nitruración iónica la mezcla de gases que se utiliza es de 80 % de $H_2$ y 20 % de $N_2$, con una presión de trabajo de 5,000 mbar y una temperatura entre 400 y 410 ºC, dejando libres los valores de tensión y corriente para que se alcancen las condiciones anteriores. Esta atmósfera de nitruración ya fue empleada en el *Grupo de Física del Plasma* en otros experimentos. Por esta causa, y por motivos únicamente comparativos entre tratamientos, no se usa argón en esta atmósfera.

Durante el proceso de cementación (o nitruración), se controlará en todo momento la presión y la temperatura (esta ajustada mediante la tensión), tomándose estos valores, además de la tensión y de la corriente cada cierto tiempo. Finalizado el tiempo de cementación se irá reduciendo progresivamente la tensión hasta que la fuente no entregue más. Se cerrarán las válvulas de entrada de gases a la cámara y se abrirá la válvula de máxima evacuación de la bomba de vacío. De esta manera se enfriarán de manera natural las probetas bajo un ambiente de vacío y solo podremos abrir la cámara una vez que el termopar indique que se llegó a la temperatura de 70 ºC. De esta manera se evita que se produzca una oxidación u otras reacciones al abrir por encima de este valor. La presencia del vacío durante el enfriamiento se debe a evitar en lo posible cementación residual y no aparezcan compuestos no deseados. Desactivada la bomba y cerrada la refrigeración de los *o-ring* se sacarán las probetas con guantes.

Antes de realizar cualquier estudio de caracterización debemos dejar que las probetas lleguen a temperatura ambiente y tras eso limpiarlas con alcohol para eliminar la capa de grafito formada en la superficie (si se realizó la cementación, puesto que la nitruración, evidentemente, no presentaba dicha capa). Finalmente, se realiza una nueva medición de la rugosidad superficial, obteniendo un valor de $R_a$ de ≈130 nm.

# Capítulo 5

# Instrumentos y principios físicos para la caracterización

Tras conseguir las probetas cementadas es necesario estudiarlas para conocer las propiedades y caracterizarlas. De esta manera determinaremos los diferentes efectos de una u otra mezcla de gases y qué ocurre con un aumento del tiempo de tratamiento. Lo que interesa conocer de cada una de las muestras es:

- El espesor de la capa cementada

- La expansión de la red

- La dureza de la superficie

- La resistencia al desgaste

- La resistencia a la corrosión

- La composición elemental de la superficie

Por consiguiente, se necesita de aparatos cualificados para llevar a cabo la medición de manera fidedigna. Cada aparato y cada técnica estarán orientados en descubrir cada una de las propiedades. A continuación, se enumeran los aparatos a los que se recurre para caracterizar las muestras, indicando además los procesos físicos en los que se basan.

## 5.1. Microscopio Óptico

Gracias a este instrumento se realiza un estudio visual de las muestras. El microscopio que se utiliza es de marca OLYMPUS modelo MG, con posibilidad de observar la muestra con 100 e incluso 400 aumentos al cambiar de objetivo y contando con un buen ocular (para más aumentos se recurre a otro microscopio que permite observar muestras a 500X). Posee una fuente propia de luz y una serie de tornillos micrométricos para desplazar la superficie de observación [Microscopios]. Con la ayuda de una Cámara Digital se puede fotografiar la zona de estudio.

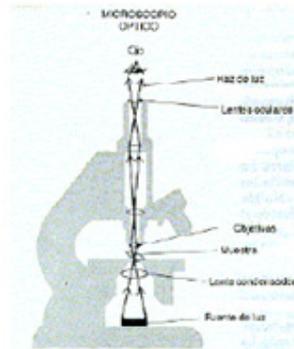

Figura 5.1: Esquema de un microscopio óptico [Óptica].

Los estudios ópticos se centrarán en la superficie y en la sección transversal. Para este segundo caso es necesario cortar la probeta y realizar un pulido hasta obtener un acabado espejado. Para revelar la estructura se acude a un ataque electroquímico durante 30 segundos, recurriendo a una solución de 10 % de ácido oxálico (cuya fórmula semidesarrollada es HOOC-COOH) en un aparato LECTROPOL-STRUERS.

## 5.2. Difracción de Rayos X en Ángulo Rasante

Es necesario recurrir al uso de rayos X para identificar la estructura cristalina y determinar si existe alteración de fases. Mediante el proceso de difracción que sufren los fotones de cierta longitud de onda al interactuar con la red cristalina se pueden conocer estos parámetros [Alonso, 1987]. Para este estudio se usa un Difractómetro de Rayos X cercado de la marca PHILIPS X'Pert con montaje de geometría de Bragg-Brentano ($\theta$-$2\theta$). El cercado afecta a todo el contorno y se dispone de una ventana de vidrios con plomo para visualizar el proceso y poder manejar las muestras o colocar los filtros antes de cada sesión. El software que utilizaremos para crear los difractogramas es *X'Pert Data Collection*.

La difracción es el proceso de interferencia constructiva de este haz de rayos X en el momento que interactúa con el material a evaluar [López]. La condición de interferencia constructiva viene dada por la *ley de Bragg*.

$$2d_{hkl}\,\mathrm{sen}\,\theta = \lambda,  \qquad (5.1)$$

donde $d_{hkl}$ es el espaciado interplanar, $\theta$ es la mitad del ángulo formado entre el rayo difractado y la proyección del incidente y $\lambda$ es la longitud de onda de la radiación empleada.

Para el análisis de la pequeña capa cementada o nitrurada es necesario recurrir a una configuración diferente del difractómetro. Esta técnica es la *Difracción de Rayos X con Incidencia Rasante* (GIXRD, siglas del inglés Grazing Incidence X-Ray Diffraction) en la que la incidencia del haz primario se hará de manera rasante, o sea, a bajos ángulos [Birkholz, 2006]. Esto provoca que el trayecto de los rayos X se dé mayoritariamente cerca de la superficie y obteniendo por tanto la información estructural de esta zona con mayor detalle.

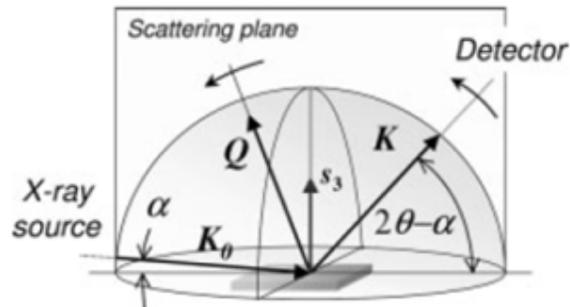

Figura 5.2: La geometría en la difracción de incidencia rasante está caracterizada por un bajo ángulo de incidencia $\alpha$, que se mantiene constante durante la medida [Birkholz, 2006].

Denotaremos al ángulo de incidencia como $\alpha$. El trayecto que recorren los rayos X será entonces $l=t/\text{sen}\alpha$ con $t$ el espesor de la capa cementada. La elección de $\alpha$ debe ser motivada por el valor del coeficiente de atenuación lineal de la muestra $\mu$, es decir, $l\approx 1/\mu$ [Bubert, 2002].

La generación de rayos X se lleva a cabo mediante un tubo con ánodo de cobre en el que escogeremos la línea Cu K$\alpha$ [Cullity, 1956], correspondiente a una longitud de onda de 1,54 Å. El tubo opera a una tensión de 40 kV y a una intensidad de 30 mA. Para colimar el haz paralelizado se deben colocar unas ranuras *Soller* a la salida del tubo de 4x4 mm$^2$. Los ángulos de incidencia serán 2 y 10º. El ángulo $2\theta$ varía entre 30 y 80º. El recorrido del detector se realiza en pasos de 0,03º, demorándose 1 segundo por cada punto de estudio. El detector es de la clase de *centelleo*[4].

## 5.3. Dureza por Indentación

Para determinar cómo influye la cementación en el aumento de dureza se debe recurrir al proceso denominado microindentación [Indentaciones]. Las pruebas se realizarán con un Indentador *Vickers* piramidal de base cuadrada de la marca SHIMADZU, modelo HMV-2.

El método de prueba Vickers consiste en indentar el material de prueba con un indentador de diamante, en la forma de una pirámide recta de base cuadrada y un ángulo de 136º entre caras opuestas sujetas a una carga de 1 a 100 kgf. La carga total se aplica normalmente durante 10 o 15 segundos [Müller, 1973]. Se miden mediante un microscopio las dos diagonales [Dureza, 2009] dejadas por la indentación en la superficie del material tras retirar la carga y se calcula su promedio. Se calcula entonces el área de la superficie inclinada. La dureza Vickers es por tanto el cociente obtenido al dividir la fuerza en kgf por el área cuadrada en mm$^2$ de la indentación.

---

[4] En los detectores de centelleo el fotón X dará lugar a un fotón lumínico que por efecto fotoeléctrico generará un electrón. Este electrón colisionará en los dínodos presentes, aumentando así el nivel de la señal, ya que cada vez que los electrones los golpean se liberan el doble de estos.

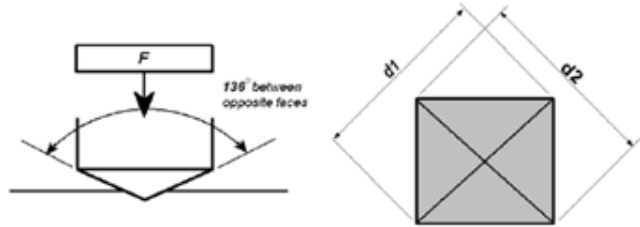

Figura 5.3: Esquema de la prueba de dureza Vickers [Indentaciones].

F es la carga en kfg; d es la media aritmética de las dos diagonales, $d_1$ y $d_2$ se expresan en mm, y HV es la dureza Vickers:
$$HV = 2F\,sen(136^\circ/2)/d^2 \approx 1,854F/d^2.$$

La indentación se basa en la combinación de deformaciones elásticas y plásticas que sufren los materiales al incidir la punta [Carvalho, 2001]. En las primeras instancias, la red cristalina se deformará a causa de la aparición de una carga normal al sustrato. Llegará un momento en el que sea tan grande que los átomos de la red no volverán a recobrar su posición original, por lo que esta parte del proceso se le denominará deformación plástica. Al retirar la carga quedará una impronta causada por las deformaciones plásticas, cuyas dimensiones están vinculadas fuertemente con la dureza que presenta el material.

## 5.4. Ensayos de Tribología

Para determinar la resistencia al desgaste de nuestras probetas cementadas se recurre al ensayo de tribología [Tribología]. Se dispone de un Tribómetro TRM 100 Dr.Ing. GEORG WAZAU, Mess-+Prüfsysteme con portamuestras giratorio gracias a un motor UNIMOTOR UM. Este tribómetro se clasifica como de tipo CSEM y el desgaste se realiza mediante el método *ball-on-disc* (bola sobre disco) a temperatura ambiente, sin lubricación. La contraparte es una bola de alúmina de 5 mm de radio, que contacta perpendicularmente sobre la probeta. Todo el proceso está controlado por el software *TriboControl V4* de Wazau.

La contraparte de alúmina tiene un largo brazo donde en cualquier punto de él se puede colocar un peso definido, ejerciendo una determinada fuerza sobre el metal a desgastar. El motor girará a una velocidad indicada durante un trayecto o un tiempo previamente programado. Es posible realizar un estudio visual de la huella que queda al finalizar la prueba y compararla con el material base [Mauricio].

Dicho estudio se realiza partiendo del hecho de que la probeta gira a velocidad angular constante con un periodo $\tau$ de giro y 0,1 m/s de velocidad lineal. La bola fija de radio r = 5 mm se coloca apoyando en la probeta a una distancia R = 9 mm de su centro (dicho valor se elige siempre y cuando sea posible), momento en el que se aplica una fuerza F = 10 N constante durante la rotación. Cuando ha transcurrido un tiempo t; que corresponde a un número de vueltas N = t/$\tau$, se ha recorrido una distancia l = 2$\pi$R·N. Llegada a una distancia en particular (por norma general, 500 m), indicada previamente al software de control, el dispositivo se detendrá, pudiendo medir con microscopio la profundidad de la huella $\delta$ y su ancho D. Gracias a todos estos datos se puede llegar a calcular el volumen por unidad de longitud, S, que se elimina de material tras el experimento con el tribómetro [García Molleja, 2010]. Mediante simples cálculos geométricos podemos conocer este valor, considerando que el perfil de desgaste se puede aproximar a una sección de un disco con radio igual al radio r de la bola de alúmina:

S = ½ r$^2$ ($\theta$-sen$\theta$),                    (5.2)

donde $\theta$ es el ángulo sólido de un cono con el vértice en el centro de la bola y la base es el círculo formado al intersecarse la superficie de la bola con la superficie de la muestra.

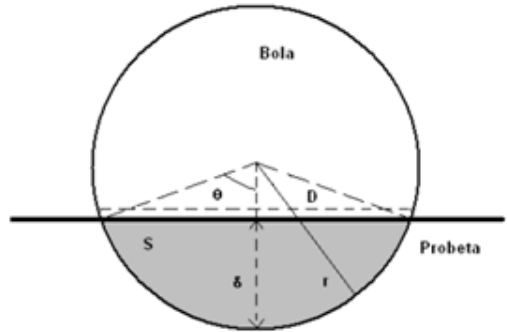

Figura 5.4: Esquema donde se localizan los parámetros necesarios para calcular el desgaste producido por una bola de alúmina sobre una probeta de acero. La zona gris muestra la sección S.

Dicho círculo posee un diámetro igual a D y $\theta$ es igual a

$\theta$ = arc cos (1-D$^2$/2r$^2$).                    (5.3)

## 5.5. Resistencia a la Corrosión

Es necesario que exista una fuerte resistencia a la corrosión de esta capa. Tras aplicarle el proceso corrosivo se estudian ópticamente y mediante SEM para localizar las zonas más afectadas.

• **Corrosión por picaduras**. Es la disolución localizada y acelerada de un metal como resultado de la ruptura de la película pasivante de óxido que producen de manera natural los aceros inoxidables [Corrosión a]. La corrosión por picaduras se desarrolla solo en presencia de especies aniónicas agresivas e iones de cloro.

- **Corrosión intergranular**. Cuando el acero inoxidable se somete a un tratamiento térmico sobre los 1000 ºC y un posterior templado, el carbono disuelto en la red no es estable y precipita como carburo de cromo, localizándose en los bordes de grano [Corrosión b], lo que hace que sea altamente propenso a la corrosión.

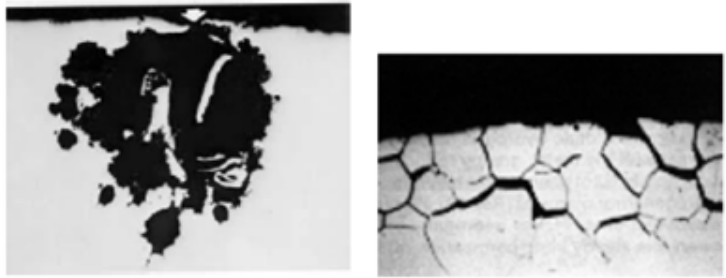

Figura 5.5: Izquierda: estudio de la corrosión generada por picaduras. Derecha: imagen del proceso de corrosión intergranular [Corrosión b].

Se colocan las seis probetas diferentes en un recipiente de plástico. Se les aplicó un esmalte en la base y en los laterales para que la solución solo atacase a la superficie cementada. También se colocó en el recipiente una probeta del material base sin tratar, a efectos comparativos. Tras esto, se llenó el recipiente con un litro de agua, con una concentración de NaCl de 5,85 % de peso en volumen. El recipiente se tapó y selló durante 60 días.

## 5.6. Espectroscopia Electrónica de Auger

Con el estudio de la muestra mediante la *Espectroscopia Electrónica de Auger* (AES por sus siglas en inglés, Auger Electron Spectroscopy) se analiza la composición elemental en los primeros nanómetros de superficie. Recurriremos entonces a un aparato PERKIN-ELMER de ultra-alto vacío.

El concepto físico de la técnica a utilizar se fundamenta en las transiciones de Auger, que son no radiativas. En este caso existirá una vacancia en un orbital interno que será ocupada por un electrón de una capa superior. En esta transición no se obedecerán las reglas de selección dipolar, por lo que es susceptible que cualquier electrón lleve a cabo este proceso. La energía liberada se dará entonces a otro electrón, el cual quedará eyectado del átomo [Feldman, 1986]. La energía de este está determinada por las diferencias en las energías de enlace asociadas a la desexcitación de un átomo conforme reagrupa sus capas electrónicas por la emisión de este mismo electrón de Auger de energía característica.

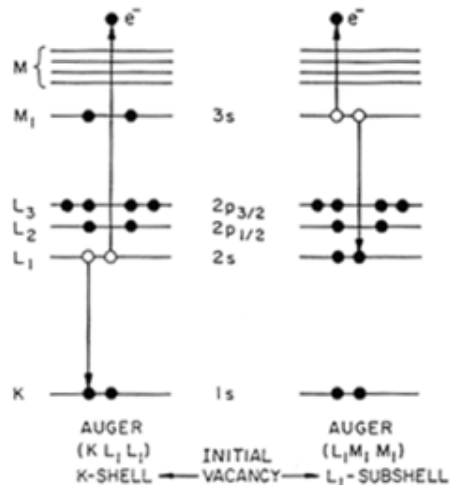

Figura 5.6: Diagrama de dos procesos de desexcitación mediante dos electrones. La transición Auger $KL_1L_1$ corresponde a un hueco inicial K rellenado con un electrón $L_1$ y otro electrón $L_1$ eyectado. La transición $LM_1M_1$ indica entonces un hueco inicial en la capa L que es rellenado con un electrón $M_1$ y otro $M_1$ eyectado [Roth, 1995].

Como el enlace químico entre átomos afecta la distribución electrónica, AES será susceptible a este hecho, dando un valor de la energía cinética del electrón eyectado ligeramente diferente a su valor con el átomo aislado. De todas formas es necesario advertir que esta alteración del valor es muy pequeña. La aplicación de AES deberá darse bajo condiciones de alto vacío para que los electrones lleguen con su velocidad de partida. Estos serán recolectados por un CMA (Analizador de Espejo Cilíndrico) que contendrá axialmente una pistola electrónica que será la encargada de originar las vacancias internas. Para poder analizar la composición elemental con la profundidad deberemos interrumpir de vez en cuando este proceso y realizar una sesión de sputtering para eliminar las capas superiores [Bubert, 2002]. Las transiciones electrónicas de Auger generalmente aparecen como pequeños trazos superpuestos al alto ruido de los electrones secundarios, por lo que para resaltarlos es necesario recurrir a técnicas derivativas [Davis, 1978]. La energía con la que llegan estos electrones Auger será característica del átomo donde se llevó a cabo el proceso, y dentro de cada átomo particular, de la transición acaecida (los niveles involucrados).

     

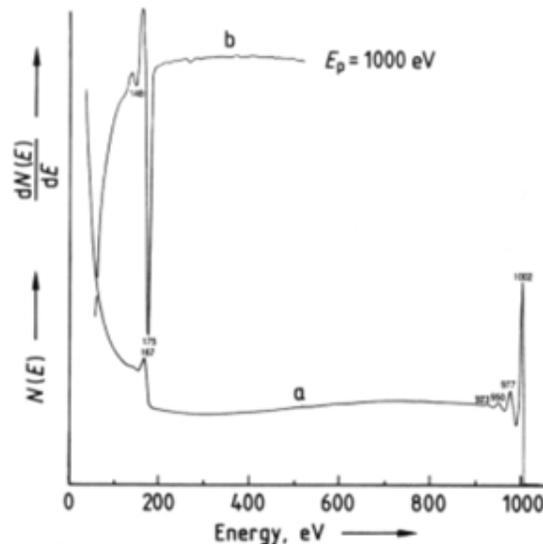

Figura 5.7: Distribución de electrones secundarios, (a) el modo *N(E)* y (b) el modo *dN(E)/dE*, donde la diferenciación revela los picos de Auger y los satélites achacados al plasmón [Bubert, 2002].

Para llevar a cabo estas sesiones fue necesario cortar las probetas para conseguir un tamaño de 12 mm de longitud, 8 mm de anchura y 1,5 mm de espesor. Se limpiaron previamente con un baño de acetona mediante ultrasonido. La presión base a la que se llegó antes del estudio fue de $4,67\cdot10^{-8}$ mbar y se mantuvo una presión de trabajo de $5,33\cdot10^{-8}$ mbar. Para analizar la composición elemental con la profundidad se tuvo que aplicar un sputtering iónico con iones de $Ar^+$: La energía con la que colisionaban los iones fue de 4 keV a una corriente de 15 mA en el filamento de emisión, originando una corriente iónica de 3,5 nA. Este proceso de sputtering se realizaba cada ciertos intervalos de tiempo. La técnica propiamente dicha de AES se logra con un cañón electrónico inclinado 30º respecto a la normal de la superficie con una corriente de 2 µA y una energía de 2 keV. El barrido se realiza en tres conjuntos de valores de energía de los electrones Auger eyectados para localizar los elementos de interés: carbono, cromo, hierro, níquel y oxígeno. La corriente que recorre la muestra es de unos 215 nA. Gracias a un amplificador *lock-in* provocaremos la derivación de la señal captada al aplicar una pequeña tensión periódica. La resolución de este aparato será de un 0,6 % y una modulación en la amplitud de 2 $V_{p-p}$. Los espectros obtenidos tendremos que clasificarlos por elementos; para ello, mediremos la altura de los picos principales de cada elemento, se normalizarán al dividirlos por un factor y se calculará por consiguiente el porcentaje total de cada elemento a un tiempo de sputtering concreto.

## 5.7. Haz Iónico Focalizado

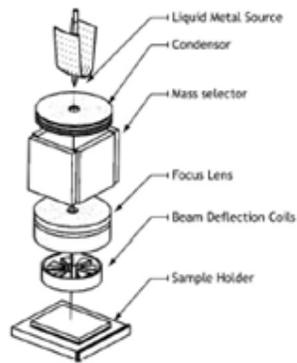

Figura 5.8: Dispositivo FIB [Fibics, 2009].

Es una técnica que puede ser utilizada para barrer material de localizaciones muy específicas de una muestra, por lo que se convierte en apropiado como paso de preparación de las muestras para XTEM, creando incluso estructuras de multicapas de parámetros elegidos por el científico [Bemporad, 2008]. Esto no es posible con la preparación ordinaria de las muestras debido a la falta de precisión de un barrido iónico. El dispositivo de Haz Iónico Focalizado (FIB, siglas en inglés de Focused Ion Beam) puede compararse con un microscopio de barrido electrónico (SEM), pero que en vez de emitir electrones [Goldstein, 2003] se recurre a un haz de iones de galio [Fibics, 2009]. Por consiguiente, las imágenes FIB se toman de una manera similar a como se hace mediante SEM. Esta técnica posee ventajas y desventajas por el simple hecho de utilizar Ga. Los iones pesados (en comparación con los electrones) logran barrer el material a una tasa considerable [Song, 2011]. Sin embargo, esto induce un daño a la muestra por el mero hecho de llevar a cabo la observación, por lo que en ciertas ocasiones se recurre a la deposición superficial de una capa de platino como protección [Bemporad, 2008].

El sistema FIB usa un haz de iones de galio finamente focalizado mediante lentes electromagnéticas, por lo que se puede operar tanto a corrientes bajas del haz para crear imágenes, como a altas corrientes del haz para sputtering específico o barridos.

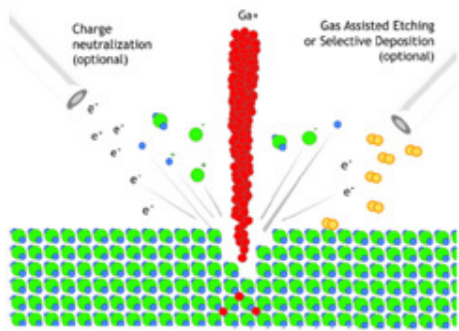

Figura 5.9: Detalle esquemático del proceso con FIB [Fibics, 2009].

Tal y como muestra la figura (5.9), el haz de iones primarios de galio ($Ga^+$) golpea la superficie de la muestra, eyectando entonces una pequeña cantidad de material que abandona la superficie como iones secundarios ($i_+$ o $i_-$) o átomos neutros ($n_0$). El haz primario también produce electrones secundarios ($e^-$). Conforme el haz primario se direcciona sobre la superficie de la muestra, se lleva a cabo la recolección de la señal de los iones eyectados o electrones secundarios para así crear una imagen. Estos procesos se llevan a cabo en condiciones de alto vacío para no afectar al haz primario ni al haz secundario.

En el caso de corrientes bajas del haz primario se constata que hay poca eyección del material; para sistemas FIB modernos se puede alcanzar la resolución de 5 nm en las imágenes. Para altas corrientes primarias se observa una gran eliminación del material mediante sputtering [Söderberg, 2004], permitiendo entonces realizar un barrido de precisión del espécimen a una escala submicrométrica [Bemporad, 2007]. Si la muestra no es conductora debe utilizarse un cañón de electrones de baja energía para inundar la superficie y así procurar la neutralización de la carga. De este modo, se pueden llevar a cabo imágenes con iones secundarios positivos mediante un haz primario de iones positivos.

Hasta hace poco, la técnica FIB era usada ampliamente en la industria de semiconductores [Fibics, 2009]. Se aplicaba para realizar análisis de defectos, modificación de circuitos, reparación de máscaras y como preparación de la muestra para TEM (siglas en inglés de Microscopía de Transmisión de Electrones) en localizaciones específicas de circuitos integrados. Actualmente, los sistemas FIB poseen una gran capacidad para la obtención de imágenes de alta resolución; esta capacidad, acoplada con la posibilidad de seccionar in situ las muestras, ha eliminado la necesidad de estudiar mediante SEM las muestras seccionadas con FIB [Song, 2011].

Las condiciones experimentales varían según el modo en el que se trabaje. El dispositivo utilizado fue de la marca FEI Helios NanoLab$^{TM}$ 600 (Dualbeam). Aplicando una tensión entre 2 y 5 kV se trabaja en el modo SEM, mientras que con una tensión de 30 kV y una corriente de ≈10 pA se trabaja en el modo FIB. El dispositivo FIB/SEM que se utilizó en esta tesis trabaja en un amplio rango de aumentos, englobados entre 2500X y 350000X. El portamuestras puede inclinarse 52º para observar la sección transversal de las probetas.

# Bibliografía

 Tesis presentada para optar al título de Doctor en Física

# Parte IV

# Resultados y Discusión

En esta parte de la tesis se estudian los resultados obtenidos tras realizar las sesiones. El primer caso consistirá en realizar estudios pertinentes en cuanto al proceso de cementación iónica, haciendo hincapié en cómo es la estructura cristalina austenítica y de qué manera los átomos de carbono pueden ocupar ciertas posiciones para provocar la expansión de la red. Tras eso caracterizaremos con diversas técnicas al acero AISI 316L, que actúa como material base. De esta manera se podrán comparar las propiedades del acero base con las del acero tratado.

Para caracterizar las muestras de acero tratado mediante atmósferas diferentes y en tiempos distintos recurriremos a Microscopía Óptica y FIB/SEM para analizar la capa tratada y la morfología superficial y transversal. Utilizando rayos X determinaremos la expansión austenítica. Recurriendo al uso de AES determinaremos la composición elemental de las primeras monocapas y, en especial, la cantidad de carbono presente en esta región, así como la fase en la que se encontrará disuelto en la red cristalina. También se determinarán la dureza, la resistencia al desgaste y la resistencia a la corrosión.

Finalizada la caracterización, recurriremos a la cementación y nitruración iónicas para comprobar cómo se comporta el acero AISI 316L tratado de diferentes maneras ante la irradiación energética mediante iones de deuterio y helio. La caracterización óptica y la de FIB/SEM serán útiles para analizar la morfología superficial tras el bombardeo iónico y para determinar los efectos del choque térmico en la sección transversal. La estabilidad de la estructura austenítica se analizará mediante rayos X. Se medirá la dureza superficial. Como último punto, se aplicarán tratamientos térmicos a alta temperatura y durante tiempos prolongados para determinar la estabilidad de la austenita expandida, recurriendo a la deposición de una capa superficial de nitruro de aluminio en vistas a comprobar si actúa como barrera contra la oxidación.

# Capítulo 6

# Estudio sobre el material base

## 6.1. Estructura cristalina de la austenita

Como ya se ha mencionado, el acero austenítico posee una estructura cristalina del tipo cúbica centrada en las caras (fcc, por sus siglas en inglés, face centered cubic) [Hinojosa, 2000]. Esta estructura es inestable a la temperatura ambiente, a menos que se forme parte de un acero inoxidable fuertemente aleado, tal y como es nuestro caso.

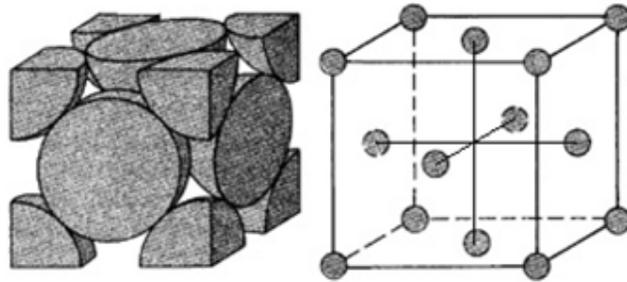

Figura 6.1: Representación de la red fcc en dos configuraciones, la de la izquierda es la configuración de átomos. La de la derecha es la configuración de puntos [Elgun, 1999].

En este tipo de estructura cristalina los metales ocuparán los puntos, que por cada celda tendremos cuatro [Sands, 1969]. Supondremos para los cálculos que en estos puntos se localizarán preferentemente los átomos de hierro, aunque abunden también los de Ni y Cr. A partir del radio atómico es posible conocer el volumen de la celda elemental fcc, que valdrá $V_{fcc} = 16R^3 \sqrt{2}$. Conociendo este dato sabemos que el factor de empaquetamiento atómico es de 0,74, por lo que determinamos que el 74 % del volumen de la celda estará ocupado por los átomos de Fe, además de átomos de Cr y Ni, puesto que en los aceros inoxidables están en un porcentaje suficientemente relevante, mientras que el 26 % está vacío [Hinojosa, 2000]. Este porcentaje vacío son los intersticios de la red, en la que el carbono que proviene de la descarga se disolverá en ella, provocando la distorsión de esta y la consecuente aparición de la denominada *austenita expandida*. Los intersticios propensos a ser ocupados por átomos se denominan huecos intersticiales y para una red fcc existen de dos tipos: los *tetraédricos* y los *octaédricos*.

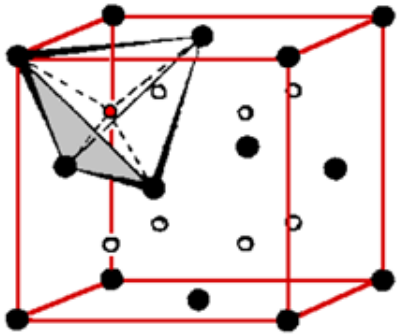

Figura 6.2: Red fcc que muestra la generación y localización de los huecos intersticiales tetraédricos [Föll].

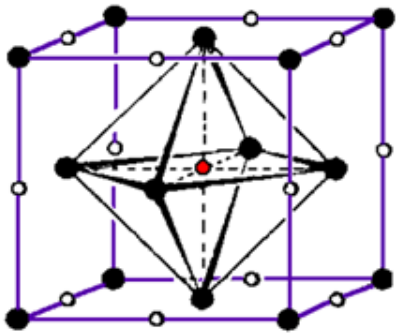

Figura 6.3: Red fcc que muestra la generación y localización de los huecos intersticiales octaédricos [Föll].

Es fácil comprobar, a partir de las anteriores figuras, que por cada celda elemental fcc existen cuatro huecos octaédricos y ocho huecos tetraédricos. Esto significa que por cada cuatro átomos de hierro por celda, son susceptibles de que existan doce átomos de carbono, contribuyendo a la expansión de la austenita [Ibach, 2006]. En el caso de que atengamos a los planos, en vez del volumen, es posible ver en la figura (6.4) cuáles son los que más se expandirán en función de los huecos que contengan. En el plano (111) solo hay dos puntos de posición de metal y no hay ningún tipo de hueco. En el caso del plano (200) vemos que existen dos puntos metálicos y dos huecos de tipo octaédrico, mientras que para el plano (202) existe un único punto metálico, un hueco octaédrico y dos tetraédricos. A la vista podemos observar que la mayor expansión se dará tanto en el plano (200) como en el (202). Elegimos estos planos pues son los que corresponden a los tres primeros de primer orden que aparecen en los difractogramas e identifican inequívocamente la estructura fcc.

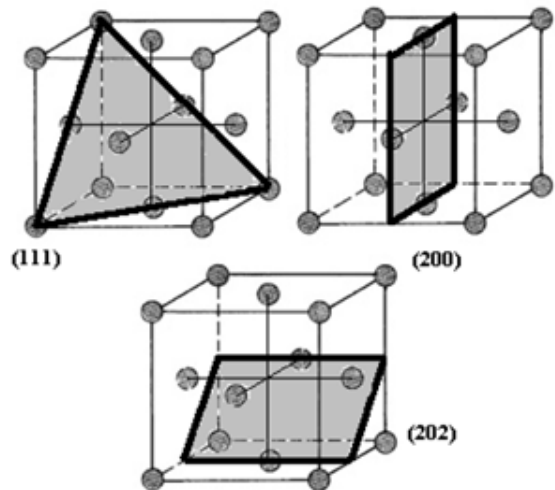

Figura 6.4: Planos de la red fcc, donde se indican según sus índices de Miller.

Para calcular el grado de expansión de la austenita, sin tomar en cuenta los cálculos arriba realizados, deberemos centrarnos en los huecos ocupados por los átomos de carbono [Bürgi, 1990], ya que cada uno de los tipos de huecos posee un tamaño determinado. Consultando la Tabla Periódica de los Elementos [Rolando, 1992] determinamos el radio iónico del hierro, $R_{Fe}$ = 69 pm y el radio atómico del carbono $R_C$ = 127 pm. Por otro lado, el parámetro de red de la austenita (que es la configuración fcc) es de a = 358,4 pm. Poseyendo todos y cada uno de estos datos determinaremos los radios de los huecos intersticiales:

- **Huecos Octaédricos**. Estos huecos se localizan entre dos puntos ocupados por hierro, situados en los vértices de la red o en el centro de cada cara. Como ya sabemos el valor de a (determinado, como se verá, mediante rayos X) solo debemos restar $2R_{Fe}$ (un valor del radio para cada uno de los dos átomos) a este valor para conocer el valor del radio del hueco: 110,2 pm. Aunque los lugares atómicos también estén ocupados por cromo y níquel no van a ser considerados en estos cálculos [Rammo, 2006]. Por consiguiente, al considerar el valor del radio atómico del carbono, la expansión de la austenita por cada hueco octaédrico ocupado es de 33,6 pm.

- **Huecos Tetraédricos**. Los huecos pertenecientes a este grupo ocupan las diagonales de la celda. El valor de la diagonal se conoce al saber el parámetro de red de la celda, $\sqrt{3}$ a = 620,8 pm. A la diagonal debe restársele dos veces $R_{Fe}$ que se identifican con los átomos ionizados de hierro (sus electrones de conducción llevan a cabo el enlace metálico) que están localizados en sendos vértices, junto al diámetro del hueco octaédrico existente en el centro de la celda fcc. De esta manera nos quedarán dos huecos tetraédricos, cada uno poseyendo un radio de 65,6 pm. Esto quiere decir que la austenita se expandirá en su diagonal 122,8 pm (es decir, 86,8 pm de expansión del parámetro de red) por cada átomo de C que penetre en esta clase de hueco.

Es necesario mencionar que el proceso de ocupación de los huecos depende de las condiciones de difusión que se dan en el experimento, por lo que estos parámetros de expansión calculados deben considerarse promediados con respecto a la totalidad de huecos a la hora de hacer coincidir el parámetro de red de la austenita expandida con el número y radio de los huecos, ya que existirán algunos de los huecos ocupados y otros estarán libres. Además, es necesario considerar los probables defectos que posee la red austenítica, así como la predominancia de ocupación de los huecos octaédricos frente a los tetraédricos [Rammo, 2006].

Se han estudiado los parámetros de red para la austenita expandida desarrollada bajo un proceso de nitro-carburización mediante la técnica PIII [Abd El-Rahman, 2009], donde se observa que existe una coexistencia entre dos estructuras fcc que corresponden a las soluciones sólidas de nitrógeno y carbono, presentando un máximo de expansión de red de 7,5 % y 2,1 %, respectivamente [Blawert, 2000]. Además, la mencionada estructura de doble capa también se observa en la nitruración asistida por plasma de $N_2$-$H_2$ a temperaturas menores de 420 ºC, con un proceso previo de limpieza mediante una atmósfera de Ar-$H_2$ para eliminar los óxidos superficiales [Czerwiec, 2006]. De todas maneras, se produce una contaminación de carbono en la superficie en el proceso de precalentamiento de la muestra. Entonces, cuando se da comienzo a la limpieza previa, el $O_2$ desaparece de la superficie, no así el carbono, que queda intacto y se difunde hacia el interior de la muestra gracias a la alta temperatura, desarrollando por consiguiente una capa $\gamma_C$. Cuando la nitruración comienza se desarrolla la capa de $\gamma_N$, que empuja a la de $\gamma_C$ hacia el interior, conformándose como una intercapa entre la zona nitrurada y el acero austenítico [Tsujikawa, 2005]. No obstante, dicha capa cementada en una sesión de nitruración no aparece en los análisis a menos que se lleve a cabo una limpieza previa in-situ.

Cuando la austenita expandida carece de tensiones, tal y como puede crearse mediante nitruración gaseosa, adopta una estructura cristalina fcc perfecta [Christiansen, 2004]. Esto da a entender que la variación de la medición de parámetros de red (y su asignación en diferentes tipos de redes cristalinas que no son fcc) que se encuentra en la literatura puede ser achacada a otros efectos secundarios, tales como las fuertes tensiones que se inducen durante la nitruración, cementación o nitro-carburización, desarrollando defectos (fallos de apilamiento, bandas de deslizamiento, maclas, etc.) dentro de los granos, los cuales generan esfuerzos asimétricos para cada plano cristalográfico [Czerwiec, 2009].

## 6.1.1. Difusión e implantación del carbono

El tratamiento realizado en laboratorio a las probetas de acero AISI 316L se basa en la difusión iónica. Por tanto, las condiciones físicas son propicias para que se cumpla la ley de Fick, por lo que el carbono llegará a la superficie y difundirá hacia las zonas de menor concentración, o sea, hacia el interior del acero [Philibert, 1991]. Sin embargo, como la muestra a tratar es el cátodo de nuestra descarga, los iones llegarán con una determinada energía cinética. Esta energía es otorgada por la diferencia de potencial a la que se somete el plasma entre sus electrodos y como consecuencia de esto los iones se implantarán dentro de la estructura cristalina del acero y al ser detenidos por la pérdida energética comenzarán el proceso de difusión. Esta implantación será beneficiosa para nuestro tratamiento, ya que fomenta la difusión hacia las capas más internas.

En los procesos de difusión se generan flujos provocados por los gradientes de concentración. Estos están provocados por gradientes de magnitudes [Philibert, 1991] y los mecanismos están gobernados por las leyes de Fick:

- **Primera ley de Fick**: el flujo es proporcional al gradiente, donde el gradiente simboliza la variación espacial de una magnitud en concreto, en nuestro caso, la concentración. Esta ley es aplicable a sistemas en régimen estacionario,
  $J = -D \, dC/dx$.
  La constante de proporcionalidad se denomina *coeficiente de difusión*, que se mide en $cm^2/s$. Este coeficiente depende de la temperatura, por lo que la difusión será un proceso térmicamente activado [Sun, 2005], conllevando entonces que ecuaciones de tipo Arrhenius guíen la difusión [Chiang, 1997].

- **Segunda ley de Fick**: considera el gradiente de flujo como consecuencia de la variación de concentración a través del tiempo,
  $\partial C(x, t)/\partial t = D \partial^2 C(x, t)/\partial x^2$.
  Esta ley es aplicable a sistemas en régimen transitorio.

Si existe una interfase definida entre un material A (los átomos implantados) y otro B (los átomos del material base), con el tiempo se producirá una *difusión atómica* del material A hacia el B formando entonces una *solución* (zona en donde A y B están mezclados) [Ibach, 2006]. Se tiene que la tensión interfásica depende del grosor de la interfase y este espesor tiende a aumentar con el tiempo.

Cuando finaliza la difusión se puede llegar a la formación de un *compuesto* (tras una reacción química) o a la formación de una *solución* (tras un mezclado aleatorio). A qué resultado se llegará dependerá en gran medida del valor de la energía libre de Gibbs, denotada como G.
$\Delta G = G - G_0 < 0$,
para que sea posible la mezcla de A y B (con el subíndice 0 denotando el instante inicial). Sigamos analizando más detenidamente [Ragone, 1995]:
$\Delta G = \Delta H - T \Delta S$,

el segundo miembro es siempre positivo (T es la temperatura absoluta) y la entropía de configuración se define como $\Delta S \equiv$ -R[x ln x+(1-x) ln(1-x)] > 0; x < 1, con x siendo la fracción de material A en la matriz B y R la constante universal de los gases. Ahora analicemos el término entálpico.

$\Delta H = H - H_0$,

donde estos valores se relacionan con las energías de *interacción*, $\varepsilon$. Si se verifica que $\varepsilon_{AB}$-($\varepsilon_{AA}$+$\varepsilon_{BB}$)/2 < 0 la mezcla de A y B es muy favorable (los términos entálpico y entrópico dan una energía libre altamente negativa). En cambio, si se cumple que $\varepsilon_{AB}$-($\varepsilon_{AA}$+$\varepsilon_{BB}$)/2 = 0 se dará el caso ideal ($\Delta H = 0$) y la mezcla será algo menos favorable. Con ambos casos se puede comprobar que $\Delta H =$ cte $\Delta \varepsilon$. Con un valor positivo la mezcla puede ser, o no, favorable, dependiendo del grado de solubilidad máxima de A dentro de B. Esto puede conocerse llevando a cabo la primera y segunda derivadas de g (la energía libre de Gibbs molar) respecto a la fracción x de la impureza en la matriz. Entre los puntos de inflexión está la *cúpula espinodal*, donde las fases coexistirán separadas una de otra y desde dichos puntos hasta los mínimos la mezcla será metaestable [Ragone, 1995].

Al variar T en los diagramas se recrearán los típicos diagramas de fase donde se pueden analizar las mezclas puras y las coexistencias entre fases.

Para continuar el proceso de difusión es necesario que los átomos vayan de un lugar a otro, pero para lograr esto se ha de superar una energía de activación. Este paso se denomina *complejo activado*.

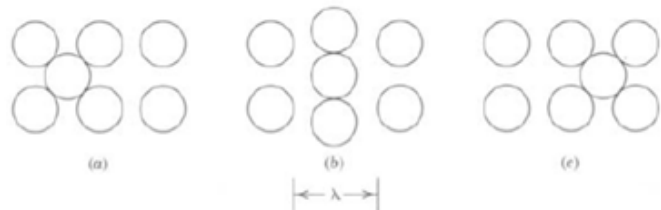

Figura 6.5: Representación del paso de impurezas intersticiales [Kingery, 1960].

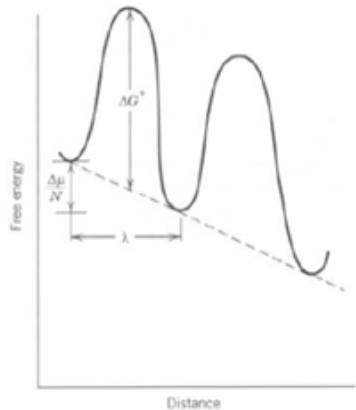

Figura 6.6: Paso de las impurezas intersticiales desde el punto de vista de la energía de activación [Kingery, 1960].

Si $\Delta G^*$ es la energía de activación para que tenga lugar la migración, se comprueba que la frecuencia de salto será de tipo Arrhenius al depender de la temperatura: $\Gamma = \nu \exp(-\Delta G^*/k_B T)$. Si sustituimos esto en el coeficiente de difusión observaremos que también es de tipo Arrhenius
$D = \gamma \lambda^2 \Gamma = D_0 \exp(-E_{aM}/k_B T)$,
con $D_0 = \gamma \lambda^2 \nu \exp(\Delta S^*/k_B T)$ y $E_{aM} = \Delta H^*$. Se tiene que $\nu$ es la frecuencia de salto de la impureza en la red y $\lambda$ la distancia de dicho salto, que media entre un intersticio y otro. Se tiene que $\gamma$ es el inverso del número de coordinación, que da la cantidad de primeros vecinos que rodean un hueco y a los que pueden migrar las impurezas [Chiang, 1997].

Hay que mencionar que los mecanismos de difusión dependen de la zona en que ocurra: en la masa del cristal, en la interfase o en el borde de grano. Además, influye la presencia de vacancias, intersticios y bordes de grano. También se da la *autodifusión*, en la que interviene la propia red. Generalmente se tiene que a gran cantidad de impurezas los defectos tienden a agruparse, pudiendo dificultar la difusión en toda la red al no estar distribuidos. La contribución de los fonones de la red a la mejora de la difusión [Gómez, 1997] no conlleva un aporte importante. A las temperaturas con las que hemos trabajado, la vibración del hierro alrededor de su posición de equilibrio no supera una amplitud cuadrática media [Cohen-Tannoudji, 1977] de un 1,03 % del parámetro de red. Se obtienen semejantes resultados de amplitud de vibración si consideramos el cromo o el níquel presentes en la red cristalina austenítica.

Existen varios estudios centrados en los procesos de difusión que ahondan más en el comportamiento de las partículas que llegan a la red cristalina y comienzan la difusión. Parece ser que el comportamiento difusivo del nitrógeno queda marcado por el atrapamiento de este en los sitios del cromo de la red cristalina [Williamson, 1994]. Esto induce un alto valor de supersaturación, a la vez que una reducción en su difusividad a la hora de comparar con la difusión del carbono. El carbono tiene una menor capacidad de interacción con el cromo, por lo que esto permite una mayor difusividad en el acero inoxidable austenítico, en detrimento de una menor tasa de saturación. De todas formas, ya sea la difusión del nitrógeno o del carbono, ambos ocupan probablemente los sitios intersticiales y permanecen como solución sólida.

Estudios mediante Espectroscopia Electrónica de Auger (AES) y Espectroscopia Electrónica de Dispersión (EDS, por sus siglas en inglés) sobre el acero AISI 316L sometido a un proceso de cementación, muestran que los niveles de carbono en la estructura sobrepasan el 12 % atómico sin la aparición de precipitados de carburos.

También se tiene que realizando la cementación en fase gas a 470 ºC durante ≤ 36 horas, los elementos intersticiales, tales como el carbono, difunden

 

rápidamente en el acero AISI 316L hasta lograr una supersaturación colosal de carbono con un porcentaje de concentración de carbono de ≈12 % ($\gamma_C$) sin precipitados de carburos [Ernst, 2004]. En este punto el grado de expansión de la austenita reduce el desajuste del volumen atómico metálico inicial entre $\gamma_C$ y $\chi$ ($Fe_5C_2$) a un nivel que logra disminuir en gran manera la contribución de la energía de esfuerzo por desajuste en la barrera de nucleación [Ernst, 2007]. Por consiguiente, si el proceso de cementación continúa más tiempo, los carburos de Hägg, o también llamados fase $\chi$, comienzan a precipitar. Existe además una propuesta que indica que la nucleación de los carburos y su crecimiento puede relacionarse con la orientación relativa de las redes cristalinas de $\gamma_C$ y $\chi$ en la interfase.

Desde que entran los iones de C (y quedan neutralizados) hasta que se detienen en el acero acaecen multitud de colisiones con la red. Estas colisiones son las responsables de la pérdida de energía y causan una desviación de la trayectoria inicial del átomo de carbono. Al considerar la gran cantidad de iones que llegan a la superficie del acero, obtendremos una distribución de trayectorias internas aleatoria, por lo que debemos recurrir al método de Monte-Carlo para estimar la distribución final. Gracias a este método podemos conocer el rango medio de profundidad en el que los átomos se detendrán, puesto que los iones se neutralizan dentro de la red.

Para que los iones lleguen con la máxima energía al cátodo es necesario que el camino libre medio no sea excesivamente pequeño [Lewin, 1965]. El cálculo de este parámetro se va a llevar a cabo de manera estimativa, puesto que el objetivo es determinar cuánto favorece la implantación del ión en la red austenítica el proceso de difusión y la formación de austenita expandida. Supondremos por tanto que un ión de carbono colisionará en todo el trayecto con átomos de argón, hidrógeno y carbono, por lo que únicamente se sopesarán las interacciones ión-neutro. Para el caso de una mezcla de 50 % de argón, 45 % de hidrógeno molecular y 5 % de metano las proporciones atómicas serán 30,3 % de Ar, 66,6 % de H y 3,1 % de C. Conociendo esto, el camino libre medio de una partícula que colisiona con varias especies diferentes se puede definir de la siguiente manera [Chen, 2005]

$$l_{Ci} = [\Sigma_{x=C,Ar,H}\, \pi n_x \sigma^2_{Cix}\, \sqrt{(1+m_{0Ci}/m_{0x})}]^{-1}, \tag{6.1}$$

donde $\sigma_{Cix} = \frac{1}{2}\,(\sigma_{Ci}+\sigma_x)$, con $\sigma_x$ el diámetro del átomo x. $C^i$ denota un ión de carbono; $m_0$ es la masa atómica en u, y $n_x$ es la densidad volumétrica de las partículas x y se obtiene mediante la aplicación de la ley de los gases ideales $p_x = n_x k_B T$; donde $k_B = 1,38 \cdot 10^{-23}$ J/K es la constante de Boltzmann, T = 406 ºC es la temperatura de trabajo del plasma, $p_x = \%at. \cdot 5,000$ mbar es la presión parcial que ejercen los átomos x.

Los datos necesarios para los cálculos aparecen en la siguiente tabla:

| Partícula | Diámetro (pm) | Masa (u) |
|-----------|---------------|----------|
| $C^i$ | 32 | 12,0107 |
| C | 127 | 12,0107 |
| Ar | 142 | 39,948 |
| H | 106 | 1,00794 |

Tabla 6.1: Datos experimentales para el cálculo del camino libre medio de un ión de carbono en un gas de átomos de carbono, argón e hidrógeno.

Con todo esto determinamos que el camino libre medio del ión de carbono es de l = 418 μm. En caso de repetir los cálculos para una mezcla de gas de 80 % Ar – 15 % $H_2$ – 5 % $CH_4$ el valor que obtenemos es de l = 506 μm.

A pesar de que la tensión aplicada es de 527 V es necesario calcular cuál es la energía con la que llegan a la superficie del cátodo, puesto que la caída de potencial se da en la región catódica, que va desde el mismo cátodo hasta la frontera del glow negativo. Por consiguiente, la energía (en eV) será la diferencia de potencial entre la superficie del cátodo y el punto que está a un camino libre medio de distancia. Qué tensión tiene mencionado punto es algo factible de calcular. Esto se logra suponiendo que la vaina obedece el modelo de la ley de Child, válida para casos de alta tensión en una zona donde no existen electrones en el medio (eliminados por su gran movilidad al aplicar una tensión negativa en el cátodo) y los iones comienzan a dirigirse hacia el electrodo negativo. Según este modelo [Roth, 1995] la ecuación de la variación de tensión eléctrica es
$V(x) = -V_0 (x/d)^{4/3}$,
con $V_0$ la tensión impuesta en el cátodo y d el espesor de la vaina, o sea, la región catódica donde se da la caída. Identifiquemos como x = 0 la frontera del glow negativo y x = d la localización del cátodo. Según esto, x = d - l: Para determinar el valor de d podemos recurrir a la ley de Paschen, donde se puede localizar un mínimo de tensión de chispa en un plasma según la distancia entre electrodos y la presión a la que está dicho plasma. Se tiene que la región catódica siempre tiene el mismo espesor para una presión dada, por lo que se verifica que pd = cte. Este valor mínimo se determina por [Roth, 1995]
$(pd)_{min} = 2{,}718/A \ln(1+1/\gamma)$ (Pa·m).
A es una constante experimental que se puede obtener de $1/l \approx Ap$, siempre que el valor del camino libre medio sea mayor que $x_l$ (la distancia mínima a recorrer para ganar la energía necesaria para producir ionizaciones). γ es el coeficiente de emisión de electrones secundarios [Petry, 1925] que para un cátodo metálico bombardeado por iones de energía no muy elevada presenta un valor $\approx 10^{-2}$ $e^-$/ión. Con todos estos valores es posible llevar a cabo cálculos:
$(pd)_{min} = 2{,}62 \rightarrow d_c = 2{,}62/p = 5{,}24$ mm,
para la mezcla con menor porcentaje de argón. En cambio, para el caso de mezclas utilizadas con el máximo porcentaje de Ar, los resultados dan un valor de d = 6,35 mm. Con esto determinamos que la última colisión se dará a x = 4,822 mm, por lo que el valor de la tensión de este punto es de 471,71 V. Si restamos este valor a 527 V es sencillo determinar que los átomos de carbono llegan al cátodo con una energía de aproximadamente 55 eV, resultado muy parecido si se repiten los cálculos para la mezcla de mayor proporción de argón.

Existen trabajos que indican que en la vaina catódica las colisiones predominantes son de intercambio de carga entre iones y átomos de carbono. Conociendo la sección eficaz de transferencia de carga [Suno, 2006] para la colisión $C^+(^{2S+1}L) + C(^2P) \rightarrow C + C^+$, se puede observar que en el intervalo energético entre 0,2 y 80 eV/u la sección eficaz se sitúa entre $(1-8) \cdot 10^{-16}$ cm$^2$. Si para bajas energías el intervalo se reduce entre $(3-8) \cdot 10^{-16}$ cm$^2$ se determina que el camino libre medio es de 7,50 mm y la vaina posee un espesor de 94,04 mm, resultado que no se observa experimentalmente, por lo que se concluye que este mecanismo no está de manera presente en nuestros experimentos.

Para calcular el rango de detención y la distribución utilizamos el software SRIM 2006 [Ziegler, 2006], cuyos resultados para 20000 iones de carbono incidiendo a ≈55 eV sobre un bloque de acero son los siguientes:

|  | Rango (Å) | Diseminación (Å) |
|---|---|---|
| Longitudinal | 5 | 3 |
| Proyección Lateral | 3 | 4 |
| Radial | 5 | 3 |

Tabla 6.2: Rango de penetración del carbono en la muestra de acero y diseminación bajo diferentes proyecciones.

Por consiguiente, esta penetración, alrededor de 5 Å, fomentará y ayudará al proceso posterior e inmediato de la difusión atómica gobernada por la Ley de Fick. Además, con este rango de penetración el carbono podrá atravesar la fina capa pasivante creada por el cromo sin problemas.

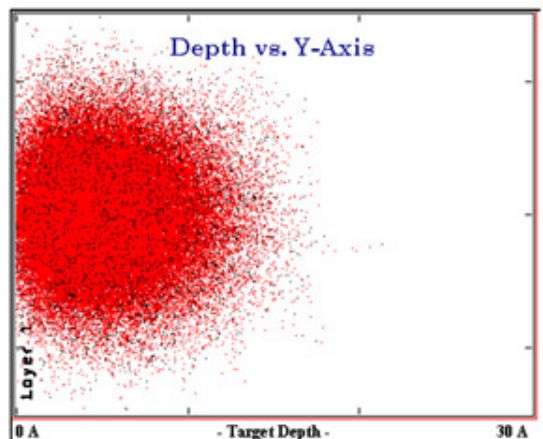

Figura 6.7: Distribución de los átomos de carbono en la matriz de acero según el método de Monte-Carlo, implementado en SRIM 2006.

## 6.2. Parámetro de red de la austenita

Ante la necesidad de estudiar la capa cementada y la consecuente expansión de la austenita necesitamos conocer, para comparar, los valores que están presentes en el material base. Por tanto, es necesario calcular el valor del parámetro de red de la austenita no expandida. Deberemos recurrir al estudio de los difractogramas de rayos X obtenidos mediante la técnica de GIXRD.

Es sabido que la austenita se organiza como una fcc, por lo que los planos que veremos serán los (111), (200) y (202). A partir de estos planos y la información que obtengamos del difractograma (los ángulos a los que se dan los máximos de intensidad) conoceremos el parámetro de red de la austenita y con la extrapolación del Nelson-Riley (cálculo de la intersección de la recta con el eje de ordenadas al representar parámetros de red contra $\cot(\theta)\cdot\cos(\theta)$) determinaremos un valor más exacto [Czerwiec, 2006]. El valor del parámetro de red se consigue a través de la aplicación de la Ley de Bragg [Cullity, 1956]
$2d_{hkl} \operatorname{sen} \theta = \lambda$,
como la red fcc es una cúbica conoceremos el valor de $d_{hkl}$ :
$$d_{hkl} = a/\sqrt{(h^2 + k^2 + l^2)}, \tag{6.2}$$
con a el parámetro de red y la terna (h; k; l) los índices de Miller. Sustituyendo (6.2) en (5.1) y despejando el valor de a tenemos por tanto que
$$a = \lambda \sqrt{(h^2 + l^2 + k^2)}/2 \operatorname{sen} \theta. \tag{6.3}$$

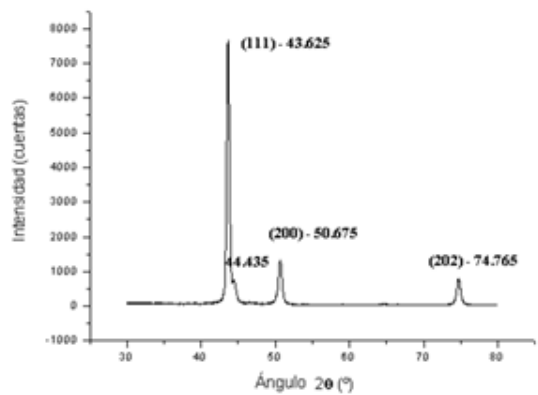

Figura 6.8: Difractograma del material base en el que se observan los picos característicos de la austenita y un pico satélite.

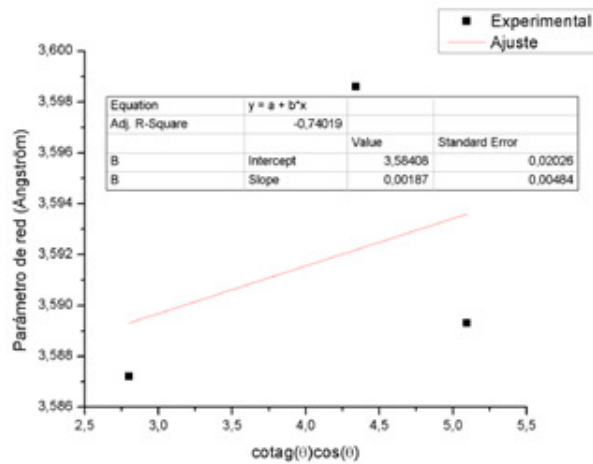

Figura 6.9: Obtención del parámetro de red por el método Nelson-Riley.

Al conocer (h, k, l) y el valor del ángulo (realizando la mitad del valor de 2θ) obtenemos que el parámetro de red del acero inoxidable austenítico AISI 316L es a = 3,584 Å. Aparte de los picos característicos de la austenita podemos observar un pico satélite. Mediante el uso de las tablas de PCPDFWIN podemos decir que este pico se debe a la presencia de un carburo de cromo o de hierro, posiblemente de este segundo, ya que la cantidad de hierro es mucho mayor que la de cromo y que de 4 posibles candidatos que muestran las tablas tres de ellos son carburos de hierro aunque de diferente estequiometría. Al tener este compuesto el material base, no sería de extrañar que vuelva a aparecer dicho pico en los difractogramas de las probetas cementadas.

## 6.3. Dureza del material base

La dureza del material base y de los materiales tratados se estudió utilizando indentación tipo Vickers. Es independiente dónde realicemos la prueba, ya que al no estar tratado debe tener el mismo valor de dureza en cualquier punto que escojamos. Normalmente, estas indentaciones se realizarán en el centro de la probeta cortada y lejos de la región modificada cuando se trata de una probeta tratada por cementación.

Realizamos las indentaciones aplicando dos cargas diferentes: 25 y 300 g. Realizamos una indentación por cada carga sobre cada una de las seis probetas que estudiamos, por lo tanto podemos calcular el valor medio de todos estos datos.

| Carga (g) | Dureza (HV) |
|-----------|-------------|
| 25 | 264 |
| 300 | 264 |

Tabla 6.3: Valores de dureza Vickers del acero AISI 316L obtenidas con dos tipos de carga diferentes.

Según esta tabla, la dureza del acero AISI 316L es de 264 HV, independientemente de la carga aplicada. Este resultado coincide con los valores conocidos para este material.

## 6.4. Resistencia al desgaste del material base

Según las condiciones descritas en un apartado anterior, realizamos las pruebas de desgaste con tribómetro sobre el material base. De esta manera determinaremos el desgaste (la cantidad de volumen eliminado por unidad de longitud, es decir, la superficie que tiene la sección transversal del surco creado) realizado por la bola de alúmina sobre la superficie del acero austenítico y podremos comparar la mejora de la resistencia al desgaste que presentan las probetas tratadas frente a la probeta sin tratar.

En primer lugar debemos conocer el valor del volumen por unidad de longitud eliminado, junto con la distancia recorrida y la carga aplicada, 500 m y 10 N, respectivamente, aplicando (5.2) tras determinar el ángulo sólido.

$$S = \frac{1}{2} 0,005^2 (0,2204 - \text{sen } 0,2204)$$
$$= \frac{1}{2} 0,000025 (0,2204 - 0,2186) \qquad (6.4)$$
$$= 2,23 \cdot 10^{-8} \text{ m}^3/\text{m}.$$

Este valor, por tanto, nos servirá para realizar comparaciones con el resto de probetas tratadas y así poder evaluar las posibles mejoras de las propiedades de resistencia al desgaste.

## 6.5. Resistencia a la corrosión del material base

Tras realizar las pruebas de corrosión con una solución de sal en agua a un 5,85 % de peso en volumen (equivalente a 100 g de NaCl en un litro de agua destilada), siguiendo las normas de un ensayo normalizado, podemos observar que el material base no presenta una capa marrón asociada al grafito, tal como se observa en la superficie de las probetas cementadas.

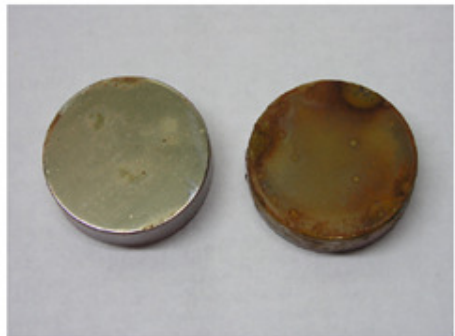

Figura 6.10: Fotografía comparativa que muestra la ausencia de una capa de grafito sobre la superficie del material base (izquierda) tras las sesiones de corrosión.

Una vez limpias las probetas se puede observar que no está atacada y la superficie es idéntica a la que se obtiene tras el proceso de pulido y espejado, lo que confirma las conocidas propiedades anticorrosivas de los aceros inoxidables.

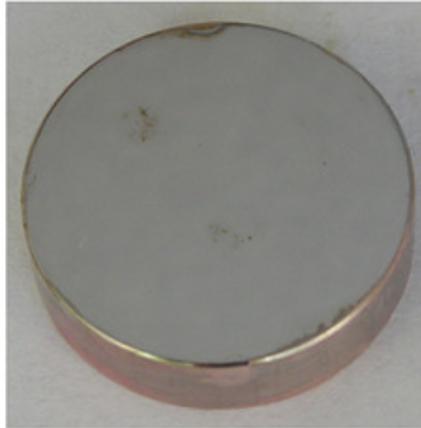

Figura 6.11: Fotografía del material base atacado por agentes corrosivos tras su limpieza. No se observan signos evidentes de corrosión.

# Capítulo 7

# Estudio sobre los aceros tratados superficialmente

En este capítulo nos dedicaremos al estudio de las probetas cementadas bajo ambas atmósferas de trabajo: 50 % de argón, 45 % de hidrógeno y 5 % de metano, en un caso y 80 % de argón, 15 % de hidrógeno y 5 % de metano en el otro, ambas en sesiones de 30, 60 y 120 minutos. Para el primer caso, identificado como C50-, las condiciones experimentales utilizadas durante el proceso están consignadas en la siguiente tabla.

| Probeta | Tiempo (min) | Presión (mbar) | Tensión (V) | Corriente (A) | Temperatura (°C) |
|---------|--------------|----------------|-------------|---------------|------------------|
| C50-030 | 30 | 4,985 | 572 | 1,14 | 401 |
| C50-060 | 60 | 5,001 | 545 | 1,08 | 406 |
| C50-120 | 120 | 4,982 | 549 | 1,35 | 411 |

Tabla 7.1: Valores experimentales durante el tratamiento de las probetas C50-.

Como puede observarse la presión que se da en cada sesión es próxima al valor fijado de 5,000 mbar que decidimos utilizar. La tensión de trabajo de cada una de las tres sesiones es parecida, fomentando así de igual manera el proceso de llegada de iones a la superficie del acero. En cuanto a la corriente de trabajo hemos de aceptar que son valores cercanos entre sí, variando solo en el orden de las décimas. En el campo de la temperatura intentamos mantenernos siempre en el intervalo que prefijamos: 400-410 ºC.

Como la superficie del cátodo es de ≈164 $cm^2$ (contando la superficie superior e inferior del cátodo, así como su altura y la de las tres probetas) tenemos que la densidad de corriente resultante sobre la superficie es de 7,0 mA/$cm^2$.

Tras realizarse la cementación de cada una de estas tres probetas y retirarlas del reactor tras el tiempo de espera indicado, observamos que la superficie presentaba un color oscuro en todas ellas.

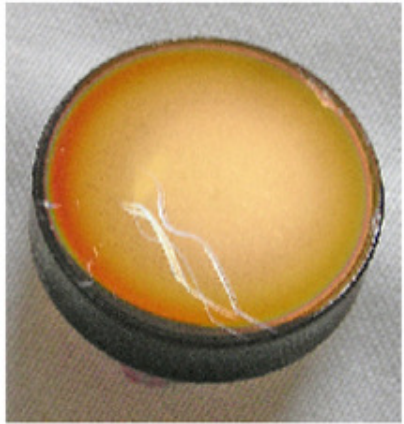

Figura 7.1: Fotografía en la que se observa la presencia de una fina capa de grafito que aparece por el proceso de cementación.

Tras una limpieza superficial se eliminaba esta capa oscura. La composición de esta podría ser seguramente de grafito, el cual proviene del carbono que no tuvo tiempo de difundirse hacia el interior del acero. Esto puede ser debido a varios factores, tales como la llegada de iones de carbono a baja energía que no pueden penetrar en el acero, un mecanismo de exodifusión que elimina de la estructura el carbono y la imposibilidad de entrar por difusión si no es a altas energías cinéticas al alcanzar el carbono el nivel de saturación máxima dentro de la red.

En cuanto a las sesiones C80- las condiciones experimentales obtenidas durante el proceso de cementación están en la siguiente tabla.

| Probeta | Tiempo (min) | Presión (mbar) | Tensión (V) | Corriente (A) | Temperatura (°C) |
|---------|-------------|----------------|-------------|---------------|------------------|
| C80-030 | 30 | 4,993 | 415 | 1,43 | 409 |
| C80-060 | 60 | 4,993 | 455 | 1,39 | 404 |
| C80-120 | 120 | 4,984 | 520 | 1,36 | 406 |

Tabla 7.2: Valores experimentales consignados durante el tratamiento de las probetas C80-.

Observamos que la presión de las tres sesiones está cerca de la presión prefijada de trabajo: 5,000 mbar, aunque siempre por debajo de este valor. Mencionamos que la presión para la sesión C80-120 es algo inferior a las otras dos, pero admitimos que aún está lo suficientemente cerca del valor prefijado. La tensión de trabajo fue parecida en las dos primeras sesiones, por lo que se dará un comportamiento de llegada de los iones con idéntica energía cinética, facilitando así la difusión tras la implantación previa. Para el caso de la sesión C80-120 tenemos una tensión más elevada, pero es algo lógico al tener un valor de presión más bajo (todo relacionado según la *ley de Paschen*[5] [Roth,

---

[5] Relaciona la tensión de ruptura de un plasma con el producto de la presión del gas y la distancia entre electrodos. Al representarla obtenemos una curva en donde se identifica

1995]). Referente a la intensidad de corriente observamos valores muy cercanos, los cuales solo varían en el orden de las décimas. La temperatura de trabajo estuvo siempre dentro del intervalo prefijado (400-410 ºC), por lo que podemos asegurar que la cementación se dio a bajas temperaturas.

En cuanto a la densidad de corriente en esta sesión alcanzamos un valor de 8,1 mA/cm$^2$.

Tras realizar el proceso de cementación y llevar a cabo el protocolo de espera y enfriamiento vemos que la capa de grafito variaba considerablemente según la sesión. En la sesión C80-030 esta capa era muy tenue, mientras que en la sesión C80-060 resultó inexistente. En cambio, en la C80-120 sí aparecía esta capa. Su origen, tal y como se debatió más arriba, se debe a la llegada de carbono al acero.

Las inspecciones realizadas con microscopio óptico sobre la superficie de la capa cementada muestran la existencia de granos con planos de deslizamiento, asociados estos a los planos (111) de la austenita tras sufrir un trabajado en frío. Los granos son observables en la superficie tanto cubierta con grafito como limpia, por lo que podemos asegurar en primera instancia que la estructura cristalina no ha sido alterada, o sea, sigue siendo austenita que no ha sufrido deformación visible en sus granos.

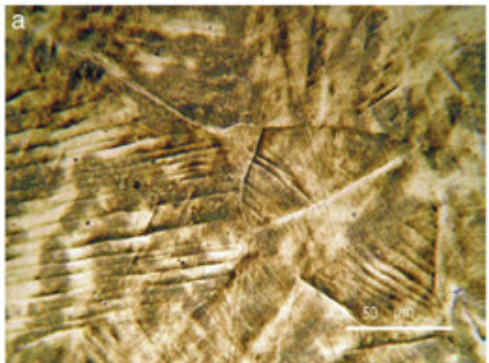

Figura 7.2: Microscopía Óptica de la muestra C50-030. La imagen superficial permite identificar las bandas de deslizamiento dentro del grano.

## 7.1. Espesor de la capa cementada

El primer paso necesario para caracterizar nuestro tratamiento de cementación es medir el espesor de la capa cementada. Esto se consigue cortando la probeta, tal y como se ha explicado anteriormente. La inspección mediante microscopio óptico se realiza con un aumento de 400X.

Para la sesión C50-030 obtenemos una capa que ocupa toda la superficie de la probeta y es bastante uniforme. Su espesor es de 14 μm. Se puede observar en la imagen (7.3) que la zona de austenita expandida se corta abruptamente, aunque hay que admitir que el carbono sigue difundiéndose hacia zonas más internas. Este corte es típico de la técnica de difusión empleada.

En cuanto a la sesión C50-060 también obtendremos una capa uniforme que cubre toda la superficie. El espesor medido es de 13 μm, es decir, inferior al tratamiento de media hora.

En el último caso, la sesión C50-120, obtenemos una capa que ocupa toda la superficie pero es rugosa en la interfase austenita expandida-acero. Esto provoca que el espesor varíe entre 11-22 μm. En el caso de no atender a los valores extremos que indicamos tenemos que el espesor medio está por encima de los obtenidos en las anteriores sesiones.

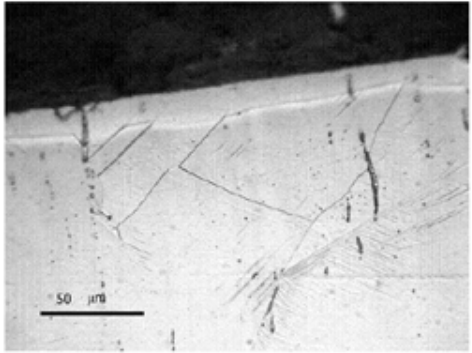

Figura 7.3: Fotografía mediante microscopio óptico que muestra la capa cementada de la muestra C50-030. Las bandas de deslizamiento en la capa cementada son más tenues, ya que esta es resistente al ataque con ácido oxálico (cuya fórmula molecular es $H_2C_2O_4$) realizado para revelar la estructura.

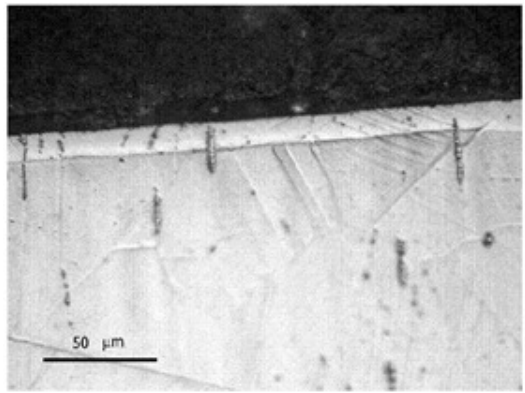

Figura 7.4: Imagen del microscopio óptico donde aparece la capa cementada de la sesión C50-060.

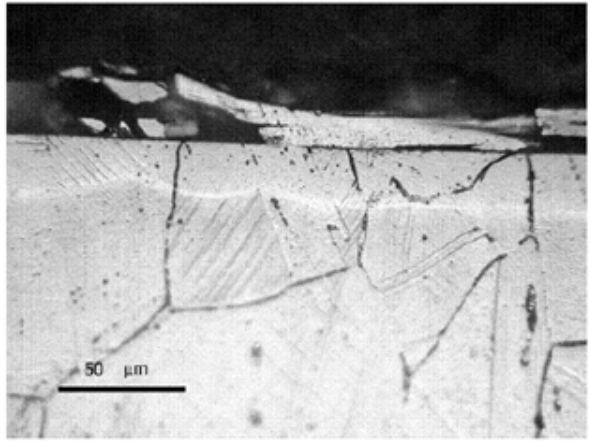

Figura 7.5: Fotografía de la sesión C50-120 obtenida al observarla con un microscopio óptico.

Se ve claramente que el espesor aumenta conforme aumenta el tiempo de cementación, resultado ya observado anteriormente por varios investigadores [Sun, 2005]. La estructura de la capa concuerda con resultados publicados [Tsujikawa, 2005b] sobre la austenita expandida, al igual que la presencia de austenita debajo de la mencionada capa. La continuidad de los bordes de grano a través de la interfase no se puede observar con total definición a causa de la resistencia de la capa cementada a la corrosión electroquímica que hemos empleado para revelar la estructura. De todas maneras, las inclusiones sí se pueden observar, demostrando que atraviesan la interfase.

Para el caso de las probetas C80- también determinaremos el espesor de cada capa cementada a partir de un análisis mediante Microscopía Óptica a 400X.

Para la sesión C80-030 tenemos una capa que ocupa toda la superficie, siendo además muy uniforme, de 17 μm de espesor. Tras el corte abrupto típico de la capa continúa la difusión del carbono hacia el interior, aunque no forma austenita expandida.

Considerando la probeta C80-060 comprobamos que la capa cementada no ocupa toda la superficie, sino la parte central. También observamos que la superficie es altamente rugosa. El espesor oscila entre 13-24 μm, aunque el espesor medio es superior al de la sesión anterior.

En el caso de la sesión C80-120, conseguimos una capa muy uniforme de 27 μm de espesor que ocupa toda la superficie sometida a tratamiento.

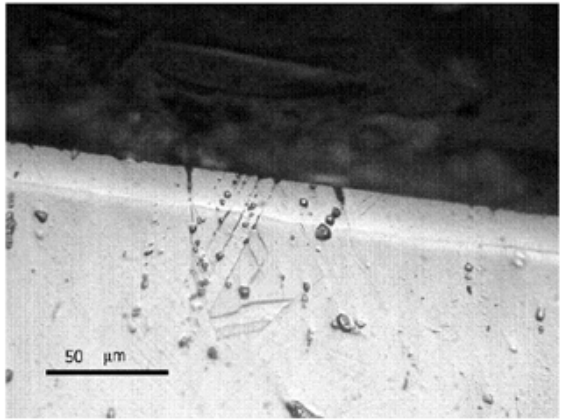

Figura 7.6: Imagen obtenida por Microscopía Óptica de la capa cementada correspondiente a la sesión C80-030. Se puede observar en la parte central un grano que atraviesa la interfase, demostrando que el proceso de cementación modifica la capa superficial, por lo que no se presenta la deposición de materiales sobre la superficie.

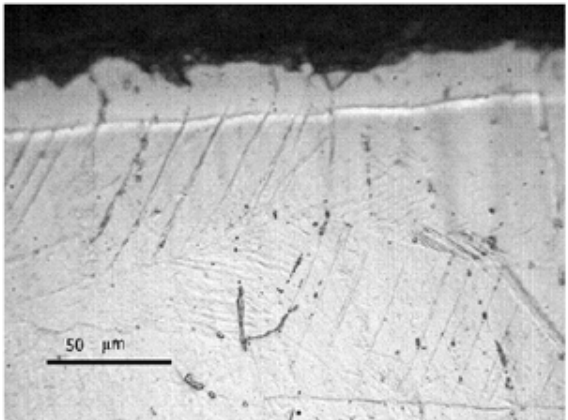

Figura 7.7: Fotografía de la capa cementada de la muestra C80-060.

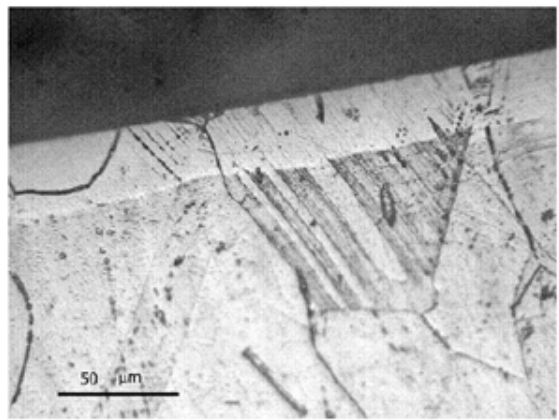

Figura 7.8: Imagen fotográfica de la capa cementada correspondiente a la sesión C80-120.

Como resultado global podemos admitir que el espesor aumenta con el tiempo de la sesión, debido a la mayor cantidad de carbono que llega al acero inoxidable, que, como ya se indicó, ingresa al reactor a la misma tasa en todas las sesiones, lo cual convierte al tiempo en la única variable que hace aumentar la cantidad de C.

La sección transversal de las muestras también se analizó mediante FIB. Para hacer un análisis en función de la profundidad se utiliza el haz de iones de galio a 5 keV, originando entonces una cavidad mediante sputtering donde las paredes son lisas y pueden observarse si inclinamos la probeta, que en nuestro caso fue de 52º. Si analizamos el material base (parte a de la figura (7.9)) es posible observar la diferente orientación cristalográfica de los granos a través de la gradación en grises que se presenta. Este fenómeno aparece debido a la eficiencia en la emisión de electrones secundarios al interaccionar el ión de galio con un átomo del cristal a una profundidad determinada. Estas bandas paralelas presentan una anchura de entre 2 y 5 µm y se pueden asociar a las maclas que componen el grano. En la parte más superficial se identifican pequeñas regiones recristalizadas originadas, probablemente, por el proceso de maquinado con el que se preparó la muestra. Esta región tiene un espesor cercano a los 0,5 µm.

Por otro lado, realizando un estudio con FIB para la muestra C50-030 (parte b de la figura (7.9)) se logra observar una estructura de mosaico caracterizada por los diferentes valores de grises y las pequeñas estructuras regulares. Esto puede atribuirse a los defectos, fallos de apilamiento y desarrollo de bandas de deslizamiento que se originan por el impacto de los iones de argón y la difusión hacia el interior del carbono. Estos defectos pueden atribuirse a la inclusión del carbono dentro de la red cristalina, proceso que provoca la aparición de tensiones residuales de carácter compresivo. Esta aparición de tensiones afecta de manera drástica a la dureza y a la resistencia al desgaste de la probeta. El tamaño de las estructuras varía entre 0,5 y 2 µm, excepto para la parte más superficial de la muestra, que varía entre 0,1 y 0,5 µm.

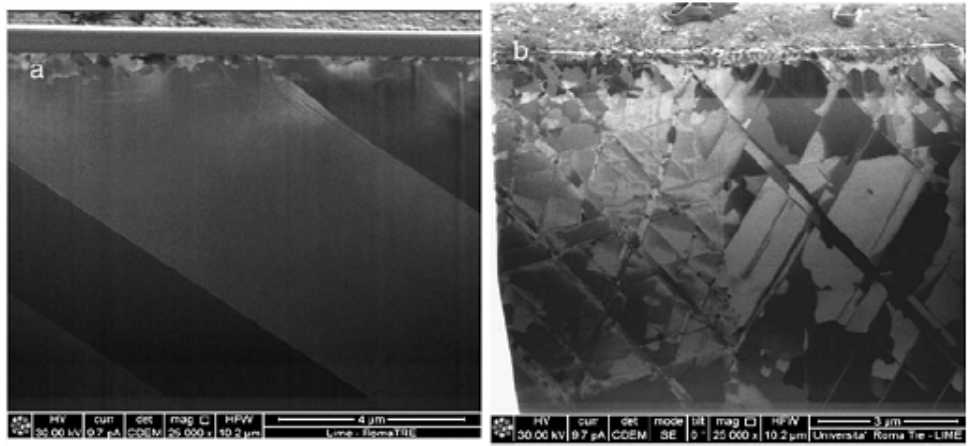

Figura 7.9: Imágenes FIB de la sección transversal del acero austenítico AISI 316L no cementado (parte a) y de la probeta C50-030 (parte b). La parte a muestra la existencia de maclas dentro del grano de austenita, mientras que en la parte b aparecen dos granos con estructuras irregulares en su interior.

Tenemos que reseñar que esta región no se podía observar mediante Microscopía Óptica al ser la capa cementada resistente a la corrosión electroquímica mediante ácido oxálico. Es importante señalar que mientras que en la parte a de la figura (7.9) la técnica FIB solo analiza un único grano, en la parte b de la figura (7.9) aparecen dos granos con el borde casi en el centro de la imagen. También se puede asegurar que no aparecen precipitados de carburos ni estructura en forma de agujas que se asocian a la fase de Hägg (fase $\chi$), resultado confirmado por otros estudios [Ernst, 2007].

## 7.2. Expansión de la austenita

La determinación de las fases resultantes (austenita expandida en nuestro caso) después del proceso de cementación se logró con la técnica de difracción de rayos X en su modo rasante (GIXRD) con distintos ángulos de incidencia. Para la evaluación del grado de expansión de la red en particular, se utilizó un ángulo de 10 grados. Con este valor de incidencia el espesor crítico bajo análisis es de $\approx 3$ $\mu$m. Esto se puede estimar a partir de la ecuación de atenuación de la radiación:

$I = I_0 \exp(-\mu/\rho\, \rho x)$,

que también puede expresarse como

$I = I_0 \exp(-\mu x)$,

con I la intensidad del haz tras atravesar la capa, $I_0$ la intensidad del haz incidente, $\mu/\rho$ es el coeficiente de atenuación másico (que consta del coeficiente de atenuación lineal, $\mu$, y la densidad del material), $\rho$ la densidad y x la distancia recorrida en el material [Cullity, 1956].

Si la radiación posee una energía de $\approx 8$ keV y utilizando el coeficiente de atenuación lineal del hierro $\mu_{Fe} = 2480$ cm$^{-1}$ podemos despejar el valor de x de la expresión si consideramos que el límite de apreciación de la intensidad de atenuación es del 1 %. Con este valor aplicamos una relación trigonométrica para determinar la profundidad en la que el recorrido del rayo cumple todas las condiciones mencionadas.

Se justifica la elección de $\mu_{Fe}$ al ser el acero inoxidable AISI 316L $\approx 70$ % de hierro y $\approx 17$ % de cromo en su composición, con solo $\approx 10$ % de níquel. Por otro lado, los coeficientes correspondientes al Cr y al Ni difieren mucho del correspondiente al Fe.

Los difractogramas posteriores muestran que se trata de una estructura similar a la austenita [Nosei, 2004] pero mostrando una dilatación de la red [Feugeas, 2002] a través de un desplazamiento de los picos de difracción hacia ángulos $2\theta$ menores [Blawert, 2001]. Se puede notar además la aparición de un pico satélite al (111) correspondiente a un ángulo 2θ ≈ 44,5º que puede ser asociado tanto a carburo de hierro del acero sin tratar como a la austenita sin expandir [Czerwiec, 2009]. Como se vio que el material base ya lo poseía es muy probable que sea a causa de un carburo ya presente.

Los cálculos de los parámetros de red se realizan mediante la ecuación de Bragg (5.1). Con esto ya podemos conocer a a partir de cada uno de los planos y ángulos conocidos. Para tener un valor más fiel se realiza la media de estos tres valores calculados y llevando a cabo la aplicación de la función de extrapolación de Nelson-Riley [Czerwiec, 2006], consistente en representar el parámetro de red en función de cotg(θ)·cos(θ) y determinar la intersección con el eje de ordenadas de la recta que atraviesa esos puntos mediante el software *OriginPro 8*. Es necesario mencionar que el valor del parámetro de red obtenido al operar el plano (200) nos indica una expansión mayor que en los cálculos que involucran los planos (111) y (202).

Algunos autores señalan [Lo, 2009] que esto es un indicio de la existencia de un proceso denominado *tetragonalización*, que indica un cambio de estructura cristalina: de cúbica a tetragonal. Sin embargo, este valor ligeramente diferente puede achacarse a las tensiones residuales y a fallos de apilamiento en este plano.

El análisis de los tres difractogramas correspondientes a las tres probetas C50-muestran los picos de la estructura, que vendrán identificados por el plano en cuestión al que representan y el valor del ángulo en el que se midieron, identificando además el pico satélite mencionado. Además, en el caso de la presencia de otros picos, indicaremos su posición angular y la composición correspondiente (obtenida a partir de las tablas de PCPDFWIN).

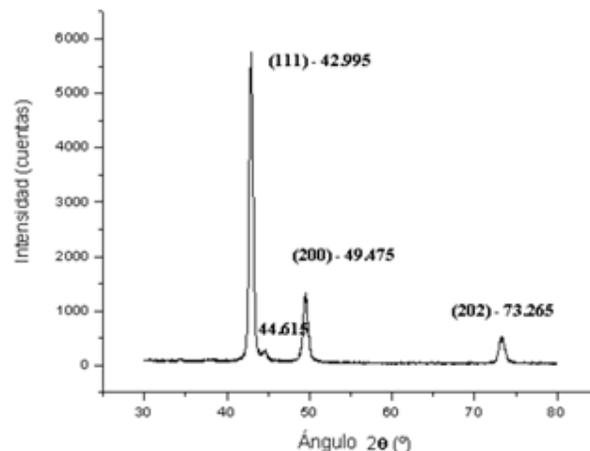

Figura 7.10: Difractograma de la muestra C50-030 en la que se localizan los picos de austenita expandida y la aparición de un pico satélite asociado al carburo de hierro del material base.

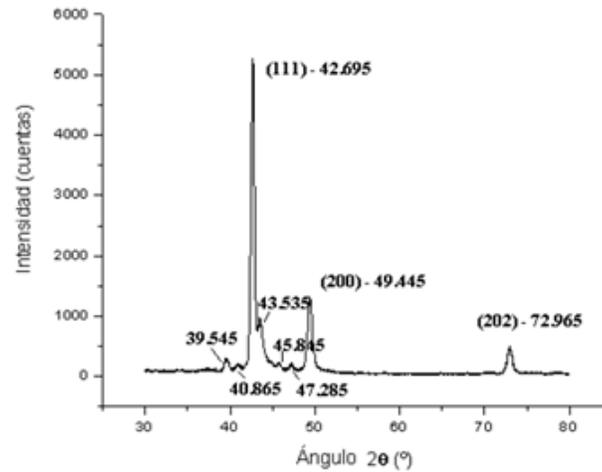

Figura 7.11: Difractograma de la muestra C50-060 en la que se indican los planos y respectivos ángulos de los picos de austenita. Aparecen cinco picos satélites: aparte del pico solitario asociado al carburo del material base (43,535º), se localizan cuatro más en esta sesión, achacados a la presencia de carburo de hierro.

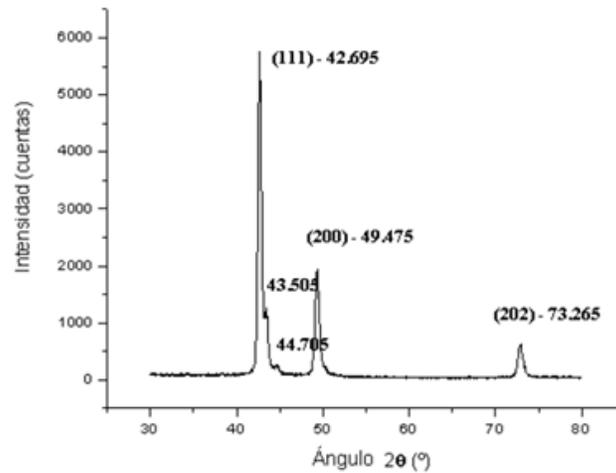

Figura 7.12: Difractograma obtenido por GIXRD de C50-120 que muestra los picos de austenita expandida y los planos asociados. Se muestran dos picos satélites: uno remanente del material base (43,505º) y otro nuevo, ambos asociados a la presencia de carburo de hierro.

Tras esto se indica el parámetro de red que se puede calcular para cada caso. El proceso operativo para calcular los parámetros de red ya ha sido descrito con anterioridad. De esta manera, entonces, sabremos qué expansión provoca cada tratamiento tras realizar posteriormente la extrapolación de Nelson-Riley. El porcentaje de expansión lo calcularemos a partir de la diferencia entre los valores de a entre cada muestra tratada y el material base y a su vez dividiremos por el valor de a del material base para realizar la comparación respecto a este. Por último, multiplicaremos por 100 para conocer el valor porcentual.

| Probeta | Parámetro de red (Å) | Expansión (%) |
|---------|----------------------|---------------|
| C50-030 | 3,660 | 2,12 |
| C50-060 | 3,661 | 2,15 |
| C50-120 | 3,666 | 2,29 |

Tabla 7.3: Valores de los parámetros de red de las diferentes muestras C50- y su porcentaje de expansión relativo al material base.

Una vez calculados los parámetros de red para las probetas C50- haremos lo mismo para el grupo de probetas C80-: a partir de las posiciones angulares de los picos, y aplicando la extrapolación de Nelson-Riley, conoceremos los valores.

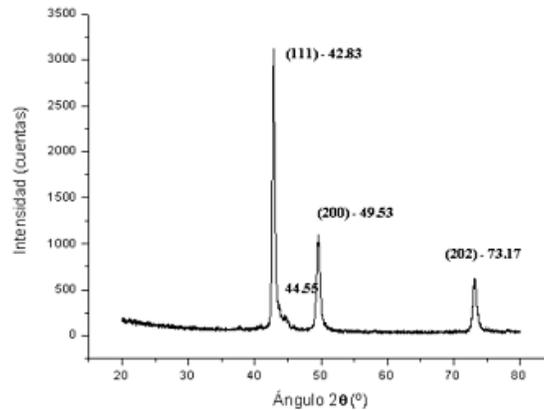

Figura 7.13: Difractograma de la probeta C80-030 donde se muestran los picos de austenita expandida y el pico satélite de carburo de hierro del material base.

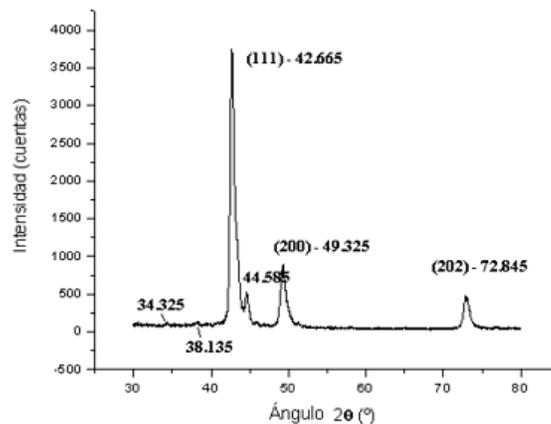

Figura 7.14: Difractograma de la muestra C80-060 que presenta los picos de austenita y tres picos satélites. Hay que mencionar que uno de ellos corresponde al carburo del material base (44,585º); mientras que el primero

(34,325º) se podría asociar a la presencia de un óxido (dando a entender pues las propiedades anómalas de esta sesión), el que queda es de carburo de hierro.

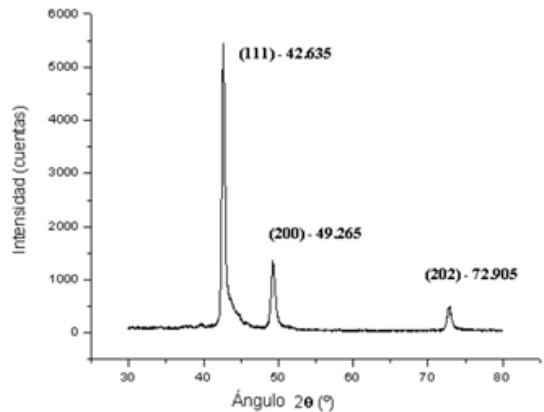

Figura 7.15: Difractograma de la sesión C80-120 en donde únicamente aparecen los picos de austenita expandida. Posiblemente la distribución de la intensidad del pico (111) solape al satélite.

Conociendo el valor del parámetro de red es necesario calcular el porcentaje de expansión respecto del material base. Este porcentaje se calcula de la misma manera que indicamos más arriba.

| Probeta | Parámetro de red (Å) | Expansión (%) |
|---------|---------------------|---------------|
| C80-030 | 3,654 | 1,95 |
| C80-060 | 3,669 | 2,37 |
| C80-120 | 3,659 | 2,09 |

Tabla 7.4: Parámetros de red de las probetas C80- y expansión relativa de estos respecto al valor del material base.

Se puede ver de manera general que la expansión de la austenita es proporcional al tiempo de tratamiento. Esto debe estar causado por la mayor cantidad de átomos de carbono que llegan a la superficie de la probeta y comienzan el proceso de difusión, ya que ingresa el metano a la cámara a la misma tasa durante todo el tiempo que dura la sesión. El desarrollo de la austenita expandida es mucho mayor en los primeros treinta minutos de la cementación, periodo en el que el valor del material base (3,584 Å) cambia a 3,660 y 3,654 Å para las probetas C50-030 y C80-030, respectivamente. Vemos además que para tiempos de cementación mayores los parámetros de red siguen aumentando, pero a una tasa mucho menor, llegando a 3,666 y 3,659 Å para C50-120 y C80-120, respectivamente. Como el tratamiento es a baja temperatura y de poca duración no se observan precipitados. La ausencia de trazas de carburos (aunque aparecen señales de ellos en los difractogramas, contradicción que se aclarará en la siguiente sección) concuerda con otros resultados experimentales [Ernst, 2004], donde el acero austenítico recibe un tratamiento en fase de gas a baja temperatura (470 ºC) durante lapsos de ≤ 38 horas. Además, aunque la expansión total observada para el caso de menor concentración de argón es un ≈10 % mayor que para el otro caso, para ambas concentraciones se determina un comportamiento muy similar, alcanzando en los primeros 30 minutos el 90 % de la expansión total. Esta rapidez de cementación se ha logrado explicar, tanto para plasmas como para gases [Williamson, 1994], en función de la temperatura del tratamiento [Blawert, 2001], suficientemente alta para fomentar la difusión del carbono hacia el interior de la muestra y suficientemente baja para no permitir iniciar la movilidad del cromo [Ernst, 2007] (que son los átomos que poseen mayor afinidad química con los átomos de C).

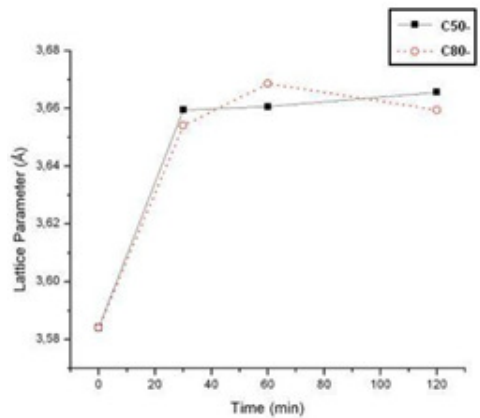

Figura 7.16: Comparación de los parámetros de red, calculados mediante la técnica de Nelson-Riley, para cada sesión y ambas atmósferas.

También podemos observar en los difractogramas un sostenido aumento de la intensidad de los picos (200), acompañada por una reducción de los picos (111). Este cambio de intensidades está provocado porque la cementación fomenta la texturación en la dirección <200>, por lo que el apilamiento se dará en los (200) y en menor grado en los (111).

## 7.3. Perfil elemental mediante Auger

Para analizar el perfil de concentración de los elementos en la parte más superficial de la capa cementada debemos recurrir a la Espectroscopia Electrónica de Auger [Feldman, 1986]. Los resultados que obtendremos tras realizar la sesión mostrarán los espectros diferenciales. Estos espectros han de ser convertidos a concentración de elementos. Para ello debemos identificar el pico correspondiente a cada elemento que se vaya a analizar y medir su altura. Este valor ha de dividirse por un factor de corrección para obtener un valor más fiel al real, ya que la función de respuesta del medidor no nos da la lectura real, absoluta, sino la relativa [Bubert, 2002]. Una vez corregidos todos los valores de la señal se calculará su cantidad porcentual respecto al resto de materiales que también hemos estudiado.

Este proceso se repite a diferentes profundidades, lo que permite establecer el per_l de concentraciones. Para ello, luego de un espectro Auger, se realiza una perforación controlada haciendo incidir haces de iones de argón durante un cierto tiempo.

En nuestro caso particular tenemos que la máxima profundidad que exploramos es de ≈156 nm, que corresponden a 65 minutos de sputtering de argón. Más profundo que este valor no es aconsejable, ya que la determinación de la profundidad a partir del tiempo de barrido con $Ar^+$ se vuelve altamente inexacta. En ciertas ocasiones detuvimos el análisis antes de alcanzar este valor de profundidad, ya que los valores de concentración elemental demostraron ser constantes durante mucho tiempo.

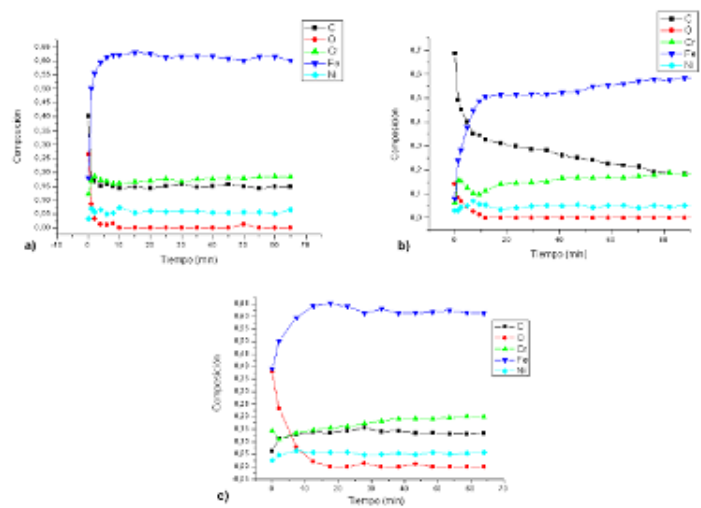

Figura 7.17: a) Perfil de Auger de la sesión C50-030. b) Perfil de Auger de la sesión C50-060. c) Perfil de Auger de la sesión C50-120.

Se observa que el oxígeno solo aparece en la superficie de las muestras, por lo que se identifica como contaminante que desaparece rápidamente con la profundidad. Analizando las muestras C50-030 y C50-120 (los resultados de C50-060 se omiten al fracasar el proceso de cementación), se observa que la concentración de hierro es casi constante (≈62 % y ≈63 %, para las probetas C50-030 y C50-120, respectivamente) más allá de 10 minutos de sputtering, lo que corresponde a una profundidad de ≈24 nm. Dichas concentraciones decaen drásticamente conforme nos vamos acercando más a la superficie.

Contrariamente, el cromo apenas presenta variaciones de concentración en la región que va desde la superficie hasta la mayor profundidad que podemos lograr (65 minutos de sputtering). En la región interna tiene una concentración de ≈18 %. También el carbono presenta unas concentraciones constantes (≈15 y ≈14 % para las probetas C50-030 y C50-120, respectivamente) menos para los primeros nanómetros de profundidad, alcanzando valores mucho mayores que pueden adjudicarse a contaminación superficial durante el análisis de AES[6].

Se identifica que a medida que el tiempo de tratamiento aumenta el carbono disminuye su concentración más lentamente, debido al mayor número de átomos de carbono presentes en la red. Inversamente, el hierro va aumentando su concentración lentamente debido al efecto de desplazamiento al que es sometido por parte del carbono. La señal del carbono medida debe considerarse como una composición del C grafítico (es decir, del disuelto en la red) y del C de carburo (enlazado con otros átomos). La llegada del cromo a la superficie permite un endurecimiento del acero, así como la introducción de una mejora en la resistencia al desgaste y a la corrosión. En el caso de la sesión de 120 minutos podemos observar que la concentración de carbono es baja desde el inicio del estudio y no llega tan abruptamente hasta su valor constante, por lo que se da a entender una falta de carbono superficial.

En cuanto a las probetas C80- también se han llevado a cabo análisis elementales mediante AES. Hay que tener en cuenta que el sputtering del argón necesario para analizar la capa cementada en función de la profundidad lleva consigo el efecto de *interdifusión*, que lo provoca el proceso de colisión en cascada iniciado por el ión que entra en el material a altas energías y se va frenando por colisiones inelásticas, las cuales hacen empujar al resto de átomos de la red hacia arriba (que causan la eyección del material, *sputtering*) y hacia abajo. También el proceso de *sputtering preferencial* (la eyección del material depende de los enlaces químicos y del elemento, así que unos serán más fáciles de arrancar que otros, por lo que transcurrido un tiempo la superficie será rica en el material poco arrancado) puede hacer que la lectura difiera algo con respecto a la concentración real a una profundidad dada.

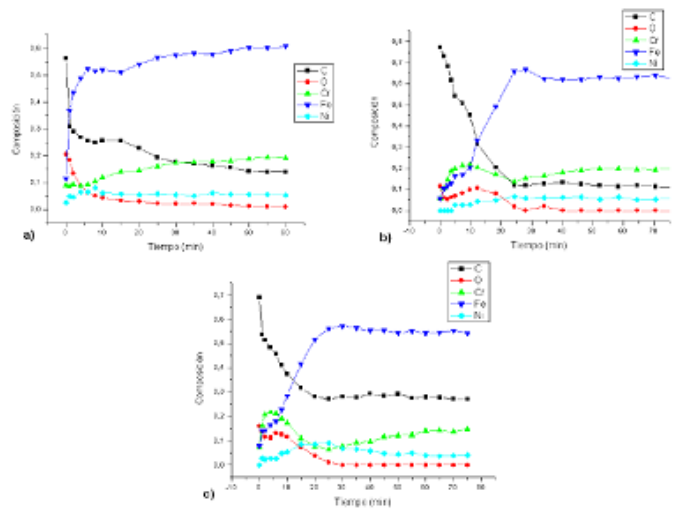

<hr />

[6] C50-120 no parece mostrar ese comportamiento, ya que estos datos fueron obtenidos analizando dos veces la misma probeta, por lo que la primera sesión barrió el carbono superficial por el sputtering de argón. En el primer estudio sí aparecían altos valores de C, pero la medición fue interrumpida por fallos técnicos.

Figura 7.18: a) Perfil de Auger de la sesión C80-030. b) Perfil de Auger de la sesión C80-060. c) Perfil de Auger de la sesión C80-120.

Las muestras C80- dan resultados muy diferentes a los que se obtuvieron de las probetas C50-. Existe ahora una evidente relación entre el tiempo de cementación y el perfil de concentración. Para C80-030 la concentración del hierro es de ≈35 % en la superficie (ignorando el primer valor por contaminación en la superficie que puede desvirtuar los resultados verdaderos), valor que aumenta al ≈52 % tras 5 minutos de sputtering (≈12 nm). Tras esto el aumento de concentración es algo menor, llegando al 60 % tras una hora de bombardeo (≈144 nm). Por su parte, el carbono posee una concentración de ≈30 % en la superficie, valor que se reduce a los 5 minutos de sputtering a un ≈25 %. Después decrece lentamente hasta la profundidad máxima alcanzada, llegando a una concentración de ≈15 %. El cromo posee una concentración en la superficie de ≈9 % y crece hasta el ≈19 % tras 60 minutos de análisis.

Un análisis similar se hace para la muestra C80-060, pudiendo determinar que el Fe posee una concentración en la superficie de ≈10 % (puesto que ahora también eliminamos el primer dato por contaminación superficial) que aumenta de manera rápida hasta ≈66 % tras 22 minutos de sputtering (≈53 nm), llegando a la profundidad máxima con un valor constante de ≈63 %. El C posee una concentración de ≈72 % en su superficie, decrece a un valor de ≈14 % tras 22 minutos de sputtering y vuelve a sufrir un decrecimiento continuo hasta el ≈11 %, valor que se mantiene hasta llegar a la profundidad máxima de estudio. En este análisis se puede observar que el Cr posee un comportamiento diferente. Mientras que la concentración aumenta a una baja tasa por debajo de los primeros ≈53 nm (correspondientes a 22 minutos de bombardeo de Ar), por encima de esta profundidad se aprecia una forma de cúpula que tiene un máximo a los 7 minutos de sputtering (≈22 %).

En cuanto a la probeta C80-120 se observan comportamientos similares a los estudiados en C80-060. El hierro posee una concentración superficial de ≈14 % y aumenta hasta ≈58 % tras 22 minutos de sputtering momento en el cual se logra apreciar un leve descenso de concentración a ≈55 % que llega hasta la profundidad máxima sondeada. El carbono, por su parte, posee un valor de concentración de ≈54 % en la misma superficie, decreciendo gradualmente hasta ≈28 % tras 22 minutos de bombardeo con iones de argón. Este valor se mantiene constante hasta el final del análisis. El cromo posee una concentración de ≈6 % tras 22 minutos de sputtering, creciendo hasta ≈16 % en la profundidad máxima. En esta muestra se repite el comportamiento del Cr observado en C80-060, presentando una cúpula de concentración con máximo de ≈21 % tras los primeros 5 minutos de bombardeo.

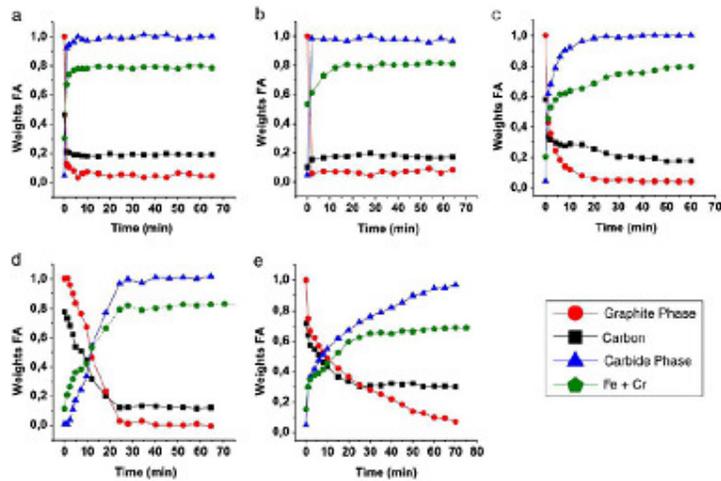

Figura 7.19: Perfiles de concentración de carbono para las muestras C50-030, C50-120, C80-030, C80-060 y C80-120 (partes a, b, c, d y e, respectivamente). Cuadrados: concentración total de carbono. Triángulos: carbono ligado químicamente. Círculos: carbono no ligado. Pentágonos: suma de hierro y cromo.

Para realizar un análisis más detallado del comportamiento del carbono se representa su concentración en función del tiempo de sputtering de las muestras C50-030, C50-120 y todas las C80- en la figura (7.19). Como ya habíamos mencionado anteriormente, las dos muestras C50- presentan una concentración de C casi constante (≈14 %) desde los 2 minutos de sputtering hasta la máxima profundidad analizada. Por otro lado, las probetas C80- indican un comportamiento diferente. Todas las concentraciones de carbono alcanzan valores constantes tras ≈25 minutos de sputtering (≈15 %, ≈11 % y ≈28 % para C80-030, C80-060 y C80-120, respectivamente) y alcanzan un máximo en la superficie (≈30 %, ≈72 % y ≈54 % para C80-030, C80-060 y C80-120, respectivamente).

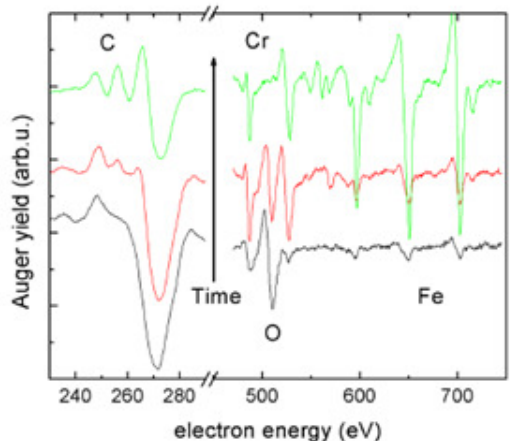

Figura 7.20: Evolución de la forma de las líneas de Auger para dos instantáneas tomadas en el perfil de profundidad de la muestra C80-030. En la parte inferior aparece un análisis de la superficie, donde se vislumbra el carbono *grafítico* típico. En la parte superior se refleja el análisis que se identifica normalmente como *carburo*. El espectro central corresponde a una mezcla de ambos.

Llevando a cabo un análisis más detallado de la evolución de la forma de las líneas de Auger se puede obtener información sobre el estado químico de los elementos analizados mediante AES. En la figura (7.20) quedan representados varios espectros de Auger en función de la profundidad para la probeta C80-030, desde la zona más superficial (espectro inferior) a la más interna (espectro superior). En este proceso se ve claramente cómo va desapareciendo el pico del oxígeno, mientras van ganando relevancia los de Cr y Fe para dar finalmente una aleación pura de Fe-Cr-C. Sin embargo, el punto a resaltar es el cambio de forma de la línea del carbono, donde la primera (más superficial) corresponde a una forma de *grafito* puro (carbono ligado de manera no química) y la última (más profunda) a *carburo* (carbono ligado químicamente) [Davis, 1978].

AES se convierte en una técnica importante a la hora de realizar un análisis químico en profundidad, cosa que no se consigue mediante XPS (Espectroscopia Fotoelectrónica de rayos X, por sus siglas en inglés). La forma de las líneas de Auger presentan información química [Steren, 1987] que se puede extraer mediante cálculos del Análisis de Factor (FA, siglas en inglés de Factor Analysis) [Malinowski, 1980]. Este método está incluido actualmente como paquete en muchos software de tratamiento de datos de AES [Vidal, 1988]. En la figura (7.19) estudiada anteriormente, donde se refleja el comportamiento del carbono total en función de la profundidad, se representa también el porcentaje de carbono libre y químicamente ligado en función del tiempo de sputtering. Tras decenas de minutos de bombardeo se comprueba que el carbono está en su estado de ligadura química, pero mientras más cerca de la superficie analicemos, se apreciará un aumento importante de la cantidad de carbono libre, tanto que en la superficie el 100 % del carbono está bajo esta forma [Passeggi Jr., 2002].

A pesar de que todas las probetas poseen este comportamiento, existen algunas diferencias entre ellas. Para el caso C50-030 y C50-120 el carbono está totalmente ligado de manera química más allá de los dos minutos de bombardeo iónico. En la probeta C80-030 se observa una rápida reducción del carbono libre, que desaparece más allá de los 15 minutos de sputtering. En cambio, para C80-060 el carbono libre tiene una reducción lenta y llega a desaparecer a los 25 minutos de bombardeo. Por último, para la probeta C80-120, la reducción gradual del carbono libre es muy lenta, desapareciendo cerca de los 60 minutos de sputtering.

Estos análisis mediante AES, que llegan a unos 156 nm de profundidad, han dado resultados que aparentemente difieren de los obtenidos mediante GIXRD, ya que esta técnica analiza zonas aún más profundas. Se ha determinado la existencia de carbono ligado químicamente. Esta ligadura del carbono tiene lugar en los primeros 30 minutos de tratamiento de cementación, que es el intervalo en que la austenita sufre su mayor grado de expansión. Ello deviene del análisis de las probetas C50-030 y C80-030, que a los ≈15 minutos de sputtering en AES se observa que el carbono está completamente combinado. El carbono libre únicamente está presente en la parte más superficial del acero. Para las probetas C50- no se aprecian variaciones importantes con el aumento del tiempo de cementación, al contrario de lo que ocurre con las C80-, donde la llegada de más carbono hace que este se aloje de manera libre, aumentando su concentración y la profundidad final que alcanza.

También se aprecia una diferencia en el perfil de concentración del cromo para ambas atmósferas utilizadas en el experimento de cementación. En las probetas C50- no se da una diferencia sustancial, cosa que no ocurre en las C80-, donde aparece una concentración que crece hasta los ≈22 minutos de sputtering (manteniéndose constante tras esta profundidad) y alcanzando un máximo de concentración a los ≈5 minutos de sputtering. Esta acumulación de cromo en la superficie con forma de cúpula se achaca al sputtering selectivo de hierro y cromo [Blawert, 1999] debido al bombardeo de los iones de argón y carbono que se dieron durante el proceso de cementación. Este fenómeno ya se ha comprobado en otros experimentos [Feugeas, 2003b], aunque utilizando como probetas acero AISI 304L. Como se puede observar en la figura (7.19) se representa en función de la profundidad la concentración de cromo sumada con la de hierro. Se puede ver que la forma es altamente parecida a la que presenta el carbono en forma de carburo, confirmando la suposición de la existencia de una fracción del carbono ligada químicamente al hierro y al cromo presentes en la estructura austenítica.

Ambas atmósferas de cementación tenían la misma cantidad de metano, luego la fuente de carbono era idéntica para los dos casos. La diferencia de composición elemental de las probetas C50- y C80- entonces puede ser adjudicada por los valores diferentes de densidad de corriente eléctrica que hubo durante el tratamiento: 8,1 mA/cm$^2$ para las C80- y 7,0 mA/cm$^2$ para las C50-. La densidad de corriente se identifica con el flujo de iones de Ar y C que llegan al cátodo y, en especial, a la superficie de las probetas. Estos resultados analizados dan a entender que existe un fuerte efecto entre la inclusión de carbono en las capas superficiales (además de la acumulación de cromo con el tiempo de cementación) y los flujos iónicos que resultan de valores de densidad de corriente superiores a 7,0 mA/cm$^2$ [García Molleja, 2010].

Por otro lado se tiene que a mayores profundidades no se da la formación de carburos, originados por carbono en solución sólida y presumiblemente localizados en los sitios intersticiales de la red fcc de la austenita. Los resultados de GIXRD demuestran estos hechos, al igual que otros experimentos [Blawert, 2001] en donde, mediante difracción de rayos X, se analiza la cementación iónica del acero X5CrNi189 a partir de PIII en una atmósfera de metano a 400 ºC. No obstante, otros autores [Ernst, 2007], utilizando la técnica PIII para nitro-carburizar (mediante mezcla de N$_2$ y C$_2$H$_2$ a 350 ºC) sobre el acero austenítico AISI 304L, encuentran la presencia de $Fe_3C$ junto a la fase $\gamma_N$.

Esta ausencia de precipitados también puede ser fomentada por la presencia de molibdeno en la estructura [Lee, 2009], ya que tiene la propiedad de reducir la probabilidad de precipitación y la promoción de la difusión de átomos intersticiales. Este fenómeno también se aprecia en procesos de nitro-carburización en vez de únicamente cementación. Experimentalmente [Lee, 2009], se comprueba la existencia de una temperatura límite para la precipitación del CrN en un proceso de nitro-carburización por el que el nitrógeno y el carbono difunden intersticialmente en procesos con plasma a baja temperatura: 480 ºC para el acero AISI 316L y 430 ºC para el acero AISI 304L, aunque no aparecieron precipitados de carbono. Análisis detallados de la difusión del carbono en aceros inoxidables AISI 316L [Ernst, 2007] mediante la cementación en fase gas a 475 ºC y una eliminación previa de la capa pasivante de cromo demuestran que en tiempos de cementación de 26-38 horas el carbono se difunde en el acero, constituyendo una capa de austenita expandida de 25 µm de espesor con una concentración de carbono del 12 %. A mayores tiempos aumenta la concentración de carbono en la capa cementada, pero aparecen precipitados con forma de aguja casi perpendiculares a la superficie, distorsionando en gran manera la matriz austenítica. Ciertos estudios identifican estas agujas con precipitados de $Fe_5C_2$ (que es la llamada fase $\chi$ o carburo de "Hägg"). Estas agujas están a una profundidad de 7 µm, justo donde el carbono está a una alta concentración ($\approx$10 %) conformando una solución sólida [Cao, 2003].

Por consiguiente, al examinar nuestras probetas mediante AES, se observa que aparece una capa cementada en la que el carbono configura una estructura en la que se encuentra químicamente ligado con el cromo y el hierro, formando carburos [Feugeas, 2003b]. Esto se da cuando la concentración de aquel es superior al $\approx$12 % [Cao, 2003]. No se observa la aparición de precipitados de tamaño macroscópico.

## 7.4. Dureza de las probetas cementadas

Los perfiles de dureza superficial se van a estudiar en una amplia zona: desde la misma superficie hasta profundidades de 6 milímetros. Por ello, es necesario combinar varias técnicas de indentación para tener resultados fiables en toda la zona de estudio. La región más cercana a la superficie se analiza a partir de pruebas de cargas de valor diferente (donde la máxima a emplear será 700 mN) aplicadas en la misma superficie. Utilizamos un indentador Berkovich[7] y estudiamos la profundidad de penetración. La contribución del material base no tendrá peso en estas medidas de dureza superficial.

---

[7] La indentación Berkovich usa una punta de diamante con forma piramidal, cuya base es un triángulo equilátero.

Antes de llevar a cabo estos experimentos, todas las probetas se limpian mediante un baño de metanol con ultrasonido, dejándolas secar tres minutos bajo radiación infrarroja. Como las técnicas de indentación son muy dependientes a la estructura del material (plano sobre el que se realizará la indentación, bordes de grano, presencia de inclusiones) para bajas cargas vamos a realizar 10 indentaciones para cada valor, sacando entonces un valor medio y su error estadístico correspondiente. Los cálculos de dureza ($H_{ns}$) para las probetas C50-030 y C80-030 se muestran en la figura (7.21) demostrando que a una profundidad de ≈200 nm la primera muestra posee un valor de 11,8 GPa y la segunda de 11,0 GPa. La dureza llega a reducirse hasta un valor de ≈8 GPa para ambas cuando se analiza una profundidad de ≈2000 nm.

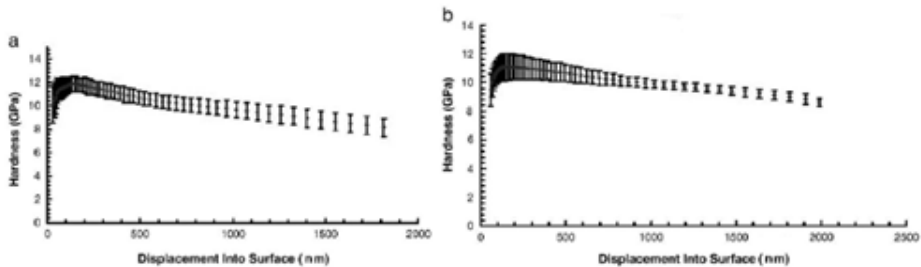

Figura 7.21: Perfiles de nanodureza determinados a través de nanoindentaciones sobre la superficie con diferentes cargas. La parte a corresponde a la probeta C50-030 y la parte b a la muestra C80-030.

Más allá de una profundidad de 2 μm la técnica deja de ser fiable, por lo que se reemplaza por la medida de nanoindentaciones ($H_{nc}$) en la sección transversal de las muestras. Para ello, dichas muestras se cortan, puliéndose la sección transversal siguiendo el protocolo mencionado en un apartado anterior. Los datos de dureza en función de profundidad se presentan en la figura (7.22) indicando que la dureza máxima se alcanza a una profundidad de ≈10 μm para las muestras de 30 minutos de tratamiento: C50-030 posee un valor de 7,5 GPa y C80-030 tiene una dureza de 8,8 GPa. Estos valores decrecen hasta una profundidad de 15 μm y se mantienen constantes hasta la máxima profundidad analizada, 140 μm, presentando C50-030 un valor de ≈5,4 GPa y C80-030 un valor de ≈5,2 GPa.

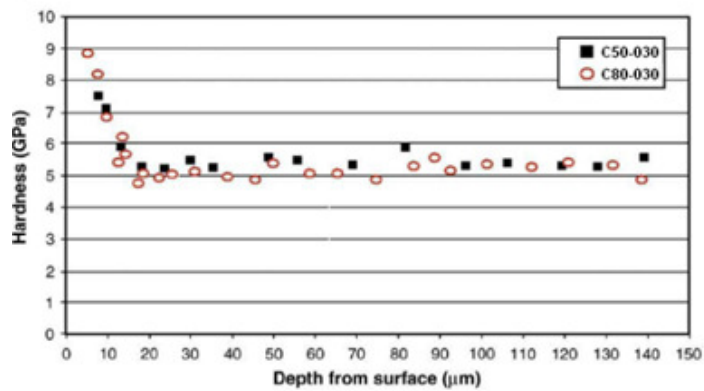

Figura 7.22: Perfiles de nanodureza de las muestras C50-030 y C80-030 determinados mediante nanoindentaciones. Se siguió el protocolo ISO 14577-1-2-3 (cargas que provoquen una profundidad de indentación ≈200 nm).

Para mayores profundidades, más allá de la frontera entre austenita expandida y austenita, se hace necesario determinar la dureza mediante el uso de un microindentador Vickers (HV). Los análisis se hacen en la sección transversal de las probetas, utilizando una carga de 300 g. Para lograr valores fiables se sigue repitiendo la medición varias veces en la misma profundidad, aunque esta vez serán 12 veces para eliminar el error introducido por el método de medida, ya que cada impronta depende de muchos factores, tales como la orientación del grano sobre el que estamos incidiendo y la posibilidad de que la pirámide quede entre dos planos de deslizamiento o dentro de la misma banda. En este caso vamos a recurrir a la *desviación estándar* para indicar el error, ya que no solo influye el error de la medida y su propagación al considerar tantas indentaciones, sino que hemos de tener en cuenta la gran variación de dureza en función del grano sobre el que indentemos (el cual, como acabamos de decir, puede que presente planos blandos o duros). La desviación se calcurará a través de

$$\sqrt{\sigma^2} = \sqrt{(\Sigma^N_{i=1}(x_m - x_i)^2/N)}, \tag{7.1}$$

con N = 12 el número de indentaciones realizadas y $x_m = 1/N \ \Sigma^N_{i=1} \ x_i$ el valor medio del conjunto de medidas. A partir de la desviación estándar conoceremos el error si dividimos el resultado obtenido por $\sqrt{12}$.

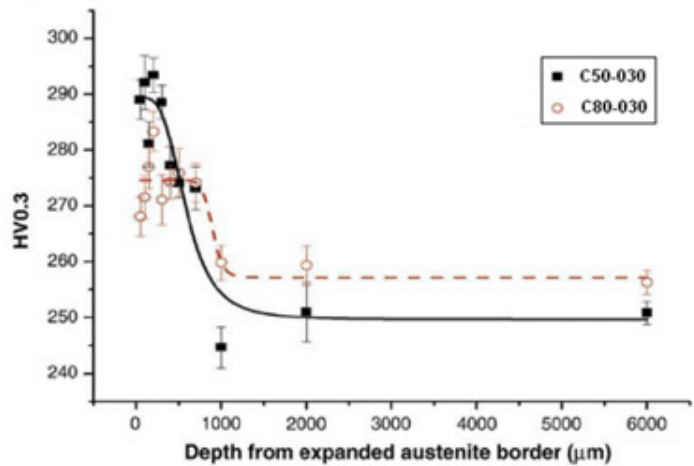

Figura 7.23: Perfiles de microdureza Vickers con carga de 300 g de las muestras C50-030 y C80-030.

Los resultados se muestran en la figura (7.23) y se ve claro que dentro de los primeros 250 µm la dureza de C50-030 es de 290 HV (=2,84 GPa) y la de C80-030 es de 275 HV (=2,70 GPa), decreciendo a mayores profundidades hasta alcanzar un valor de ≈255 HV (=2,50 GPa) en ambos casos. Este valor se mantiene constante a partir de ≈1 mm de profundidad y es muy semejante al valor de dureza del material base (264 HV, que equivale a 2,589 GPa).

La entrada de carbono dentro de la estructura austenítica se demuestra como útil a la hora de aumentar la dureza del material base. En ambas sesiones estudiadas se comprueba que los valores son muy superiores a la dureza del material base y que esta dureza disminuye con la profundidad. Como se puede ver claramente en las figuras, la dureza es independiente de las condiciones de cementación utilizadas, dando en ambos casos valores superficiales de ≈11 GPa y disminuyendo hasta ≈2,50 GPa a profundidades de ≈1 mm.

## 7.5. Resultados tribológicos

Para conocer si el proceso de cementación mejora las propiedades de resistencia al desgaste de la superficie debemos recurrir al uso del tribómetro. Como dijimos anteriormente, realizaremos las sesiones con una bola de alúmina ($Al_2O_3$) de 10 mm de diámetro, aplicando una carga de 10 N entre dicha bola y la probeta que está girando para completar una distancia de 500 m a un radio de 9 mm respecto el centro de las probetas. En la figura (7.24) se puede ver la forma de sujeción de la probeta en los ensayos de desgaste.

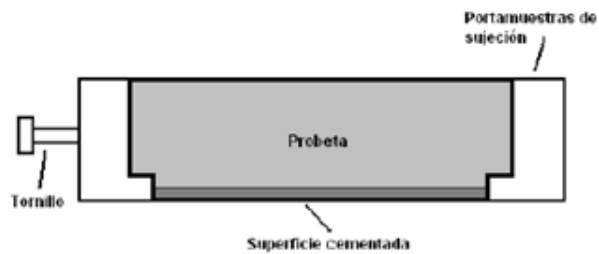

Figura 7.24: Esquema del montaje de la probeta en el tribómetro.

Tras el proceso mediremos con una lupa con regla incorporada la anchura de las huellas marcadas en el acero austenítico. Se llevaron a cabo 5 mediciones independientes de la anchura para realizar la media e intentar así eliminar el error introducido por posibles saltos y oscilaciones del motor. Para conseguir la anchura correcta en cada medición debemos aplicar el factor de corrección de lente, cuyo valor es de 0,893. Una vez hecho todo esto determinaremos la cantidad de volumen por unidad de longitud eliminado a partir de la expresión (5.2). Los resultados obtenidos los mostraremos en la tabla 7.5:

| Probeta | Radio (m) | Distancia (m) | Vueltas | Anchura (m) | $S$ $(\mathrm{m^3/m})$ |
|---|---|---|---|---|---|
| C50-030 | $9 \cdot 10^{-3}$ | 500 | 8842 | $0,93 \cdot 10^{-3}$ | $1,34 \cdot 10^{-8}$ |
| C50-060 | $9 \cdot 10^{-3}$ | 500 | 8842 | $0,45 \cdot 10^{-3}$ | $0,15 \cdot 10^{-8}$ |
| C50-120 | $9 \cdot 10^{-3}$ | 500 | 8842 | $1,46 \cdot 10^{-3}$ | $5,22 \cdot 10^{-8}$ |

Tabla 7.5: Parámetros del ensayo tribológico de las probetas C50- y valores de anchura y volumen por unidad de longitud medidos.

En los tres casos los valores de S son del mismo orden, por lo que la cantidad de material eliminado fue muy parecida para cada sesión. Sin embargo, se observan leves diferencias entre cada probeta C50-, observándose que C50-120 tiene un valor de S casi cuatro veces mayor que C50-030. Para determinar la relación del material que se perdió mediante desgaste en las probetas tratadas respecto del que se eliminó en el material base se emplea la siguiente relación porcentual:

$W_\% = (S_i/S_n) \times 100,$

donde $S_i$ y $S_n$ son los volúmenes por unidad de longitud para las muestras cementada y no cementada, respectivamente. Los datos se consignan en la siguiente tabla:

| Probeta | $W_\%$ |
|---------|--------|
| C50-030 | 60 |
| C50-060 | 7 |
| C50-120 | 234 |

Tabla 7.6: Porcentaje de desgaste de las probetas C50- relativo al del material base.

Las sesiones C50-030 y C50-060 muestran un comportamiento predecible, es decir, la resistencia al desgaste aumenta cuanto mayor sea el tiempo de tratamiento. Esto está causado por la expansión de la austenita, cuya propiedad interesante es su aumento de resistencia respecto a los aceros austeníticos sin tratar. Pero al estudiar la probeta C50-120 vemos que muestra una resistencia al desgaste menor que la del propio material base, en contra de lo que se esperaba.

En el caso de estudiar ahora la resistencia al desgaste de cada una de las tres probetas C80- la carga aplicada fue de 10 N y se incidió a unos 9 mm de radio respecto al centro de la superficie de la probeta. Cabe mencionar que el ensayo de tribología de la probeta C80-030 se hizo con un radio de 11 mm y la C80-060 con uno de 5 mm. Para que las condiciones fuesen equivalentes entre sí, se modificó entonces la distancia a recorrer de manera que todas dieran el mismo número de vueltas.

Una vez terminada la sesión de tribología medimos la anchura de la huella creada en el acero tratado. Hemos de tener en cuenta que la lente con la que observamos las huellas posee un factor de corrección de 0,893, el cual deberemos aplicar para conocer exactamente el valor de anchura real. Los valores medios de 5 mediciones aparecen en la tabla 7.7.

| Probeta | Radio (m) | Distancia (m) | Vueltas | Anchura (m) | $S$ (m$^3$/m) |
|---------|-----------|---------------|---------|-------------|---------------|
| C80-030 | $11 \cdot 10^{-3}$ | 610 | 8842 | $0,60 \cdot 10^{-3}$ | $0,36 \cdot 10^{-8}$ |
| C80-060 | $5 \cdot 10^{-3}$ | 278 | 8842 | $0,59 \cdot 10^{-3}$ | $0,34 \cdot 10^{-8}$ |
| C80-120 | $9 \cdot 10^{-3}$ | 500 | 8842 | $0,46 \cdot 10^{-3}$ | $0,16 \cdot 10^{-8}$ |

Tabla 7.7: Datos experimentales del ensayo de tribología de las probetas C80-. Se consignan los valores medidos de la anchura del surco y el cálculo del volumen por unidad de longitud.

Aquí podemos observar que la sesión C80-120 fue la que menos volumen perdió, mientras que la sesión C80-060 perdió más y la C80-030 perdió dos veces más volumen que C80-120, aunque solo dos centésimas más que en el caso C80-060. Es de destacar que todas perdieron material en el mismo orden de magnitud, por lo que fue de manera progresiva. Para tener una comparación mejor entre las tres C80- es necesario calcular el porcentaje de desgaste con relación al material base.

| Probeta | $W_\%$ |
|---------|--------|
| C80-030 | 16 |
| C80-060 | 15 |
| C80-120 | 7 |

Tabla 7.8: Porcentajes de desgaste tribológico en las probetas C80- respecto al obtenido en el material base.

La resistencia al desgaste aumenta con el tiempo de tratamiento, hecho esperable, puesto que la expansión de la austenita es la responsable de la mejora a la fricción. Además, todas y cada una de las sesiones son más resistentes que el material base, como presumiblemente se esperaba.

Los resultados globales indican una reducción en el desgaste gracias al tratamiento de cementación, mejorando con el tiempo de cementación de las probetas. La diferencia de atmósferas en cada sesión parece no tener efecto en los resultados de tribología.

En el caso particular de las probetas C80- la reducción de la tasa de desgaste se puede achacar a la concentración del carbono libre en la red austenítica, que aumenta con el tiempo de tratamiento, al igual que la profundidad a la que dicho carbono libre puede encontrarse.

Todo esto nos lleva a sugerir que la cantidad de carbono libre presente en la zona más superficial de la probeta actúa como lubricante sólido, reduciendo por consiguiente la tasa de desgaste.

Hay que mencionar que el software de control podía indicar, por sus mediciones de fricción, cuándo se desgastaba la capa cementada y pasaba la contraparte a tener contacto con el material base. En ninguno de los casos se observó la transición.

## 7.6. Resistencia a la corrosión

Los ensayos de corrosión fueron descritos en la sección 5.5 y un primer acercamiento a sus efectos tuvo lugar en la sección 6.5, donde se analizó superficialmente el material base tras sesenta días de inmersión en un ambiente salino. Tras el ataque con solución salina (NaCl) las probetas se lavaron con agua destilada para eliminar la capa de grafito que presentaban todas las muestras tratadas y eliminar el esmalte que cubría el resto de las probetas.

El grafito presentaba la forma de manchas de varios colores (por efecto de lámina delgada, donde las interferencias entre interfases de cada capa que recorre el rayo de luz hace que algunas longitudes de onda se destruyan) que se produjeron por el proceso de corrosión y la liberación de carbono. Tras esto se realizó una limpieza más profunda al utilizar etanol y terminar todo con un secado mediante aire caliente.

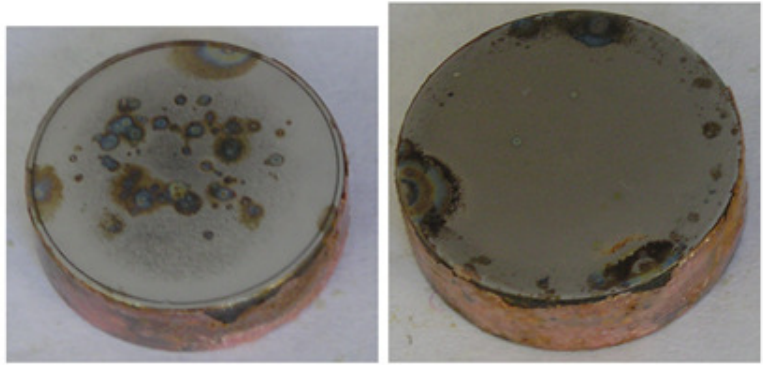

Figura 7.25: Izquierda: probeta C50-030 tras sufrir el tratamiento de corrosión. Derecha: muestra C80-030 antes de la limpieza con etanol.

Un estudio mediante Microscopía Óptica y Microscopía Electrónica de Barrido (SEM, por sus siglas en inglés, Scanning Electron Microscopy) muestra claramente que para las probetas C50- las superficies muestran regiones con presencia de manchas grisáceas, que bajo un análisis minucioso se revelan compuestas realmente por la agrupación de pequeñas picaduras. Estas manchas tienen un diámetro de ≈80 μm y las picaduras que las componen son de ≈5 μm de diámetro. Fuera de estos conglomerados de picaduras, la superficie presenta una densidad mucho menor de picaduras, repartidas uniformemente por todo el grano y con un diámetro de < 0,5 μm. La concentración aumenta en zonas próximas a los bordes de grano y en las bandas de deslizamiento, tal y como se puede apreciar mediante un estudio con SEM (parte b de la figura (7.26)). A la hora de analizar cuán profundas son las picaduras, los análisis realizados en la sección transversal de las probetas nos indican que la corrosión es un fenómeno principalmente superficial, ya que no aparecen canales de corrosión en toda la capa cementada.

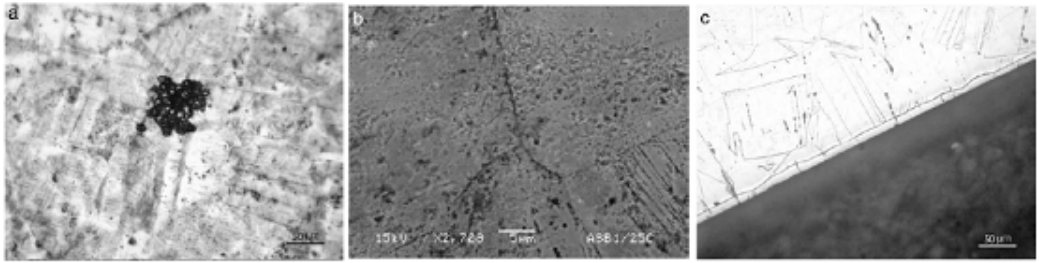

Figura 7.26: Imágenes de la muestra C50-030 donde aparecen los fenómenos de corrosión. a) Imagen SEM de una nube de picaduras. b) Imagen SEM de picaduras junto al borde de grano. c) Microscopía Óptica de la sección transversal de la muestra donde se verifica la ausencia de penetración por corrosión en la interfase.

En cuanto a las probetas C80- aparecen manchas generalizadas y desprendimientos de la capa cementada, indicando que con una alta concentración de argón se desmejora altamente la resistencia a la corrosión del acero inoxidable austenítico. Los estudios transversales muestran una corrosión generalizada que puede llegar incluso a varios milímetros de profundidad.

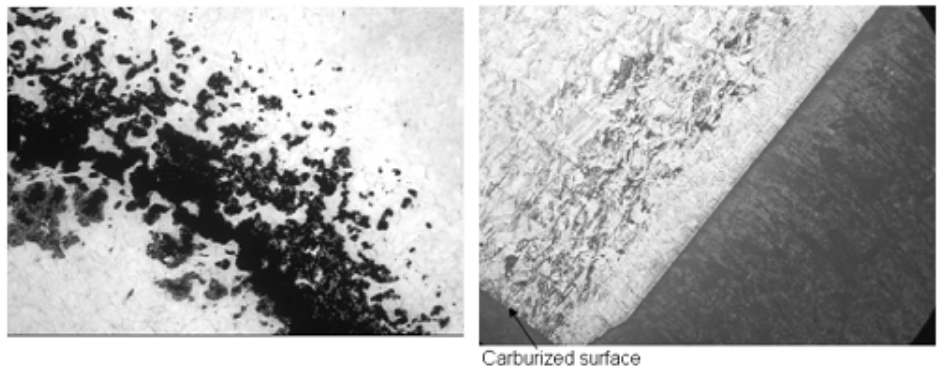

Figura 7.27: Imágenes de la muestra C80-030 donde se observa la corrosión. a) Imagen SEM donde se observa una región en la superficie afectada de corrosión generalizada. b) Observación mediante microscopio óptico de la sección transversal, apreciando la penetración de la corrosión (aproximadamente 2 mm de profundidad).

Si comparamos estas tres probetas con la muestra de material base observamos que el acero inoxidable sin tratar posee mejores propiedades anticorrosivas, mientras que las tratadas tienen en mayor o menor medida un ataque sobre su superficie.

Partiendo de que la corrosión da comienzo en los bordes de grano y las bandas de deslizamiento, tal y como se puede apreciar gracias a un análisis de la probeta bajo SEM (parte b de la figura (7.26)), se puede determinar que la corrosión será más intensa cuando el número de centros de alta energía sea más elevado. Experimentos con PIII (Implantación Iónica mediante Inmersión en Plasma, por sus siglas en inglés) en una atmósfera de metano determinan que el acero X5CrNi189 cementado iónicamente desarrolla una capa $\gamma_C$ que posee un comportamiento de corrosión similar al del material base bajo el ataque de una solución ácida ($H_2SO_4$) [Blawert, 2001]. Bajo esta condición de cementación, se produce una baja supersaturación de carbono que induce por consiguiente una baja densidad de defectos [Tsujikawa, 2005b]. Sin considerar los diferentes procesos de cementación, el material base y el medio corrosivo, el proceso de ataque se explica de la misma manera: los mencionados centros de alta energía. Estos centros se originan por la precipitación del cromo, que queda químicamente ligado al C, resultándole imposible crear una capa pasivante en la superficie. Esto hace que fuera de los precipitados de cromo el acero quede expuesto a los elementos corrosivos. Esta corrosión observada se hace más importante que otras observadas en otros experimentos [Ernst, 2007].

También se tiene que las diferencias de proceso corrosivo entre las muestras C50- y C80- se hacen muy evidentes. Esto puede explicarse por la mencionada formación de carburos de cromo. Sin embargo, los análisis mediante Microscopía Óptica de la capa cementada y los datos obtenidos mediante GIXRD no indican la presencia de dichos precipitados. Esta aparente contradicción se explica mediante AES, que localiza estos precipitados en la zona más superficial de la capa cementada y como la profundidad de penetración de los rayos X es muchísimo mayor que la profundidad que analiza AES, la contribución de los precipitados al difractograma estará eclipsada por la señal más intensa que proviene de zonas más internas, donde no existen dichos precipitados.

# Capítulo 8

# Estudio de la resistencia de las probetas ante irradiación iónica y altas temperaturas

Hasta este punto hemos analizado las propiedades elementales, físicas y mecánicas de los aceros inoxidables sometidos a un tratamiento de cementación bajo diferentes atmósferas y tiempos de tratamiento. La caracterización nos ha permitido considerar qué parámetros experimentales son necesarios controlar para evitar la precipitación de carburos, tales como la densidad de corriente. Además, determinamos que a un porcentaje fijo de metano [Chen, 2003] en la mezcla atmosférica (con un porcentaje ≤ 5 % de este para no crear hollín [Suh, 1997]) y alta presencia de argón, junto a baja cantidad de hidrógeno, fomenta una buena resistencia tribológica [García Molleja, 2010].

A partir de este capítulo nos interesaremos en caracterizar el comportamiento de estos aceros tratados bajo diferentes condiciones, tales como el bombardeo iónico y los tratamientos de larga duración a altas temperaturas [Bass, 2005]. Así conocemos la fortaleza del tratamiento bajo condiciones estrictas y el campo de aplicabilidad de estos en la industria.

## 8.1. Resistencia al bombardeo de iones ligeros

El proceso se lleva a cabo en una cámara de vacío (figura (4.3)), diferente a la que utilizamos en el anterior caso (figura (4.2)), en la que colocamos un cátodo especialmente diseñado para permitir el encastre de probetas (figura (4.4)) y evitar que sufran efectos de borde [Corujeira Gallo, 2010]. En esta nueva cámara utilizamos una fuente de alimentación DC rectificada que anula el ciclo negativo y minimiza la permanencia en tensión nula. Su resistencia de carga es de 16,8 $\Omega$ por lo que puede entregar una alta tensión. El proceso de evacuación, limpieza y calentamiento es idéntico al llevado a cabo para el anterior juego de probetas, aunque ahora el tiempo de tratamiento queda fijo en 80 minutos (para así conseguir una capa tratada gruesa) y que el aumento de temperatura se hace con mezcla de Ar y $H_2$. Los parámetros experimentales quedan iguales a los anteriormente consignados, aunque esta vez solo cementamos con una única mezcla de gases: 50 % de Ar, 45 % de $H_2$ y 5 % de $CH_4$: Para hacer un análisis sobre la viabilidad del acero AISI 316L cementado en comparación a los aceros austeníticos nitrurados hemos recurrido también al tratamiento de probetas de AISI 316L bajo una atmósfera de 80 % de $H_2$ y 20

% de $N_2$: El tiempo de tratamiento también duró 80 minutos y la presión de trabajo fue la misma que para los casos de cementación. En la tabla 8.1 se consignan los valores experimentales empleados en ambas sesiones:

| Probeta | Presión (mbar) | Tensión (V) | Corriente (A) | Temperatura (°C) |
|---|---|---|---|---|
| Nitruración | 4,985 | 762 | 0,412 | 406 |
| Cementación | 5,012 | 564 | 0,553 | 403 |

Tabla 8.1: Valores experimentales durante el tratamiento de nitruración y cementación. Probetas de acero AISI 316L de 6 mm de espesor y 2 cm de diámetro.

Gracias a esta tabla podemos observar que la presión de trabajo estuvo en un rango cercano a la que predefinimos inicialmente (5,000 mbar) y que la temperatura estuvo siempre en el rango establecido. El valor de tensión en la nitruración fue muy parecido a los valores que se pueden encontrar en la literatura [Nosei, 2004] mientras que la corriente se mantuvo en valores estables. Para el caso de la cementación, la tensión eléctrica de trabajo fue muy parecida a las que se manejaron con las probetas C50- (tabla 7.1) aunque a una corriente menor por la nueva fuente de potencia empleada, que nos permitía calentar a mayor velocidad el cátodo mediante bombardeo iónico por efecto Joule (debido a la transferencia de momento en la colisión entre la partícula acelerada por el campo eléctrico impuesto entre electrodos y las partículas que componen el cátodo) y llegar a la temperatura de trabajo de 400 ºC.

Conociendo que el cátodo que empleamos en estas sesiones tiene una superficie de 265,26 cm$^2$ la densidad de corriente durante todo el tratamiento fue de 1,55 mA/cm$^2$ para la nitruración y de 2,08 mA/cm$^2$ para la cementación, valores muy por debajo de los obtenidos en las muestras C50- y C80-, evitando así la precipitación de nitruros o carburos.

Con el mencionado fin de estudiar el efecto de la radiación sobre la austenita expandida, las probetas tratadas se someterán a un bombardeo mediante iones ligeros a altas energías [Sartowska, 2007]. Para lograr esto se recurre a un dispositivo de *plasma focus* denso de tipo *Mather* (figura (8.1)) de configuración coaxial (ánodo de 40 mm de diámetro, aislante de *Pyrex* de 15 mm, 40 mm de longitud libre y barras como cátodo dispuestas en circunferencia de 110 mm de diámetro) [Bernard, 2002]. Este dispositivo de plasma focus está basado en un proceso físico conocido como *pinzamiento en Z* (Z-pinch) por el que el plasma sufrirá una compresión en el eje axial (identificado como z) y eyectará tanto iones como electrones [Hussain, 2009] en dicha dirección [Rico, 2007], aunque en sentidos contrarios. Para determinar la contribución del bombardeo de iones sobre la superficie con el inherente efecto térmico se cubre la mitad de la probeta con una lámina de acero. De esta manera podremos ver los efectos térmicos [Sartowska, 2007b] en la zona oculta y considerarlos en la parte descubierta, identificándolos y adjudicando los demás efectos al proceso de bombardeo [Sartowska, 2005].

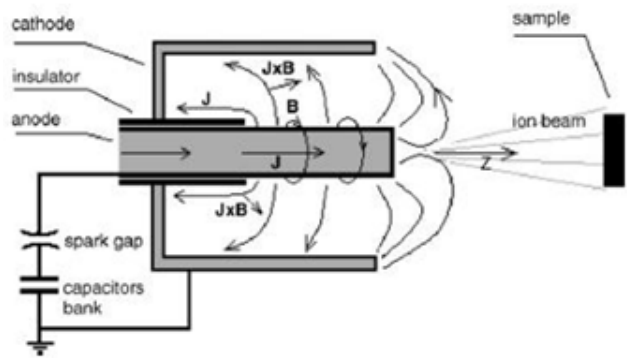

Figura 8.1: Dispositivo de plasma focus tipo Mather [Rico, 2007].

El dispositivo de plasma focus está dentro de una cámara a 1,600 mbar de presión de trabajo con el gas que utilicemos, que será deuterio ($^2$H o D) o helio ($^4$He). Esta cámara se evacuó previamente a 0,013 mbar para eliminar todo tipo de contaminante como puede ser el vapor de agua presente en la atmósfera. El cierre de un interruptor hace que se descargue el banco de cuatro condensadores de capacidad total de 4 µF, provocando la ruptura dieléctrica del gas en la zona colindante al aislante, puesto que la tensión de carga es de 31 kV y la inductancia parásita se minimizó a 47 nH. La corriente generada se combinará con el campo magnético originado por el paso de corriente a través del ánodo, dando lugar a una fuerza de Lorentz que desplaza al plasma [Milanese, 2005] por todo el recorrido coaxial y provocando un barrido de las moléculas (o átomos) del gas y dirigiéndolas al frente del dispositivo [Sigaut]. Cuando llegue al borde se dará inicio a una compresión radial hasta crear un cilindro de plasma frente al ánodo. Este plasma está a alta temperatura y posee una densidad elevada. El equilibrio al que llega es sumamente inestable [Schmidt, 1979], creándose rizos y torsiones del plasma [Haruki, 2006] hasta que se da el instante denominado *focalización* [Roth, 1995] en el que se rompe la columna. Esto provoca la eyección de iones a altísimas energías hacia el exterior de la configuración [Milanese, 2005]. Ahora bien, si a 82 mm de este punto colocamos la probeta podemos determinar cómo le afecta el bombardeo de diferentes tipos de gases, aplicados en un número creciente de veces (1, 5 y 10) [Feugeas, 2010]. Todo esto hace que la energía de operación del plasma focus sea de 2 kJ. Según esto, al colocar la probeta a 82 mm de distancia, la densidad de potencia que se entregue en la superficie es de 10 MW/cm$^2$, elevando entonces la temperatura hasta alrededor de 1500 ºC en las primeras centenas de nanosegundos. El choque térmico afectará incluso a capas que estén a varios micrómetros de profundidad. El proceso de plasma focus no crea haces iónicos uniformemente distribuidos, por lo que el cálculo de densidad de potencia ha de entenderse como un valor medio. Además, bajo los cráteres existen zonas altamente fundidas y con multitud de fallos pero no se analizaron perfiles de concentración de D y He con la profundidad.

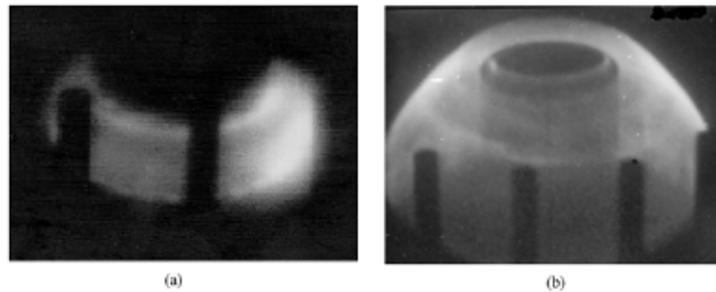

Figura 8.2: Imágenes con cámara ICC en un régimen de presión óptimo. a) Etapa inicial (exposición 100 ns a los 200 ns del inicio del proceso). b) Final de la etapa coaxial (tiempo de exposición de 5 ns a 800 ns del comienzo) [Milanese, 2005].

Una vez finalizado el proceso de bombardeo es necesario llevar a cabo una caracterización de las probetas. Las técnicas de estudio que realizaremos son:

- Caracterización morfológica mediante Microscopía Óptica y FIB

- Revelación de la estructura cristalina mediante GIXRD

- Estudio de la dureza superficial mediante microindentación Vickers

## 8.1.1. Caracterización óptica

Los análisis de Microscopía Óptica se han llevado a cabo a 500X para observar la morfología general en la superficie sin realizar ninguna limpieza previa [Nosei, 2004]. En el caso del acero AISI 316L cementado podemos observar la morfología final tras un disparo de deuterio y diez de helio.

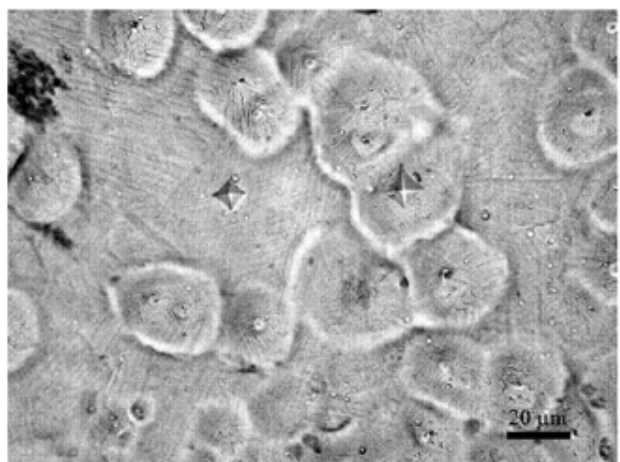

Figura 8.3: Morfología superficial del acero austenítico cementado con una descarga de D.

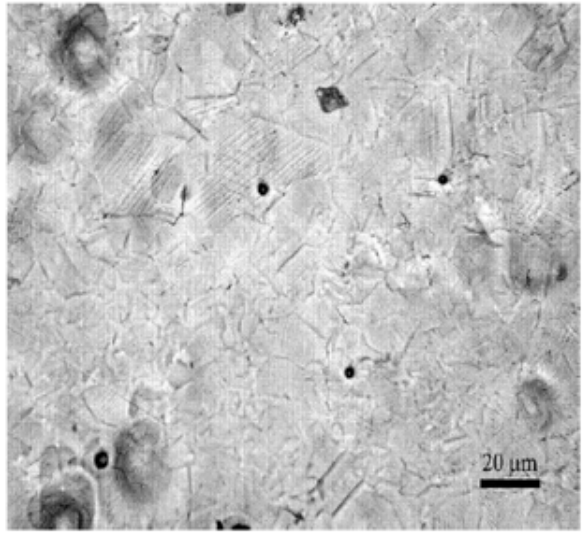

Figura 8.4: Imagen superficial del acero 316L cementado tras 10 pulsos de He.

Analizando la probeta que recibió una descarga de deuterio se puede comparar con la que recibió una única descarga de helio, demostrando que apenas hay diferencias entre ellas. Lo mismo ocurre cuando hacemos comparación entre las dos probetas cementadas que recibieron diez pulsos. En las figuras (8.3) y (8.4) se pueden identificar cráteres profundos y algunos puntos de eyección de material, típicos de los procesos de plasma focus [Feugeas, 2010]. Además, se localizan bandas de deslizamiento entrecruzadas debido a los elevados gradientes térmicos de calentamiento y enfriamiento. Estos gradientes por tanto provocan multitud de tensiones residuales que terminan desencadenando el deslizamiento de los planos (111) [García Molleja, 2010].

Según este análisis, la morfología superficial es independiente del gas empleado para irradiar la muestra, solo depende del número de descargas que sufra. Ahora analicemos además el acero AISI 316L nitrurado sometido a pulsos de gases ligeros.

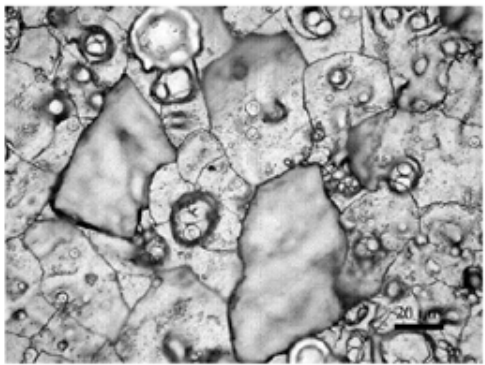

Figura 8.5: Superficie de acero austenítico nitrurado tras un pulso de D.

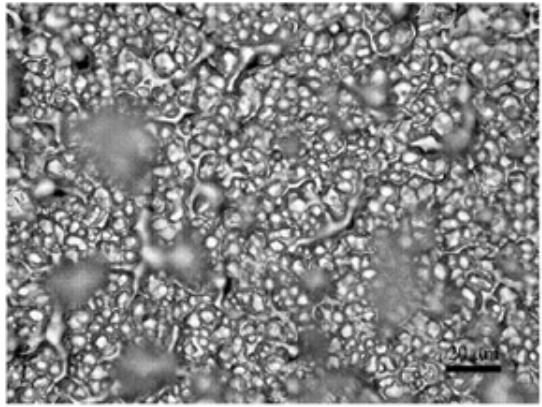

Figura 8.6: AISI 316L nitrurado analizado bajo microscopio tras 10 pulsos de He.

Al igual que en el anterior caso de cementación, en la nitruración también se confirma que la morfología es independiente del gas con el que se hagan las descargas y la dependencia es hacia el número de pulsos efectuados. Para el caso de una descarga, tanto con helio como con deuterio, se observan también cráteres, pero esta vez acompañados de descamado superficial. Este puede achacarse al rápido calentamiento y enfriamiento en la superficie, conllevando una diferencia de contracción entre las primeras monocapas superficiales y la zona nitrurada situada a más profundidad, lo que hace que al enfriarse se fracture y salte dicha zona superficial. En el caso de diez pulsos se aprecian claramente indicios de amorfización por los repetidos tratamientos térmicos que logran fundir las primeras capas. Esto nos ayuda a concluir que existe una mejor resistencia al bombardeo por parte de las probetas cementadas que de las nitruradas.

Mencionamos que la zona oculta por la lámina de acero únicamente presenta un ennegrecimiento apreciable a simple vista, pero la estructura morfológica es idéntica al material nitrurado o cementado antes de bombardear.

## 8.1.2. Microscopía de Haz Iónico Focalizado

Mediante la técnica FIB, combinada con SEM, es posible hacer un análisis exhaustivo sobre la morfología superficial y la estructura del acero tratado y bombardeado en los primeros micrómetros de profundidad [Söderberg, 2004]. Para ello recurrimos al uso de un haz de $Ga^+$ cuyo impacto en la superficie provocará un cráter de dimensiones controladas en el acero, pudiendo detectar simultáneamente los electrones secundarios que se eyectan por la colisión del galio con los átomos constituyentes del acero y obtener así una micrografía [Bemporad, 2008]. Cuando está ya formado el cráter en el material podemos también obtener una micrografía mediante SEM [Goldstein, 2003], que se consigue bombardeando la zona con electrones para arrancar los mencionados electrones secundarios [Tokaji, 2004].

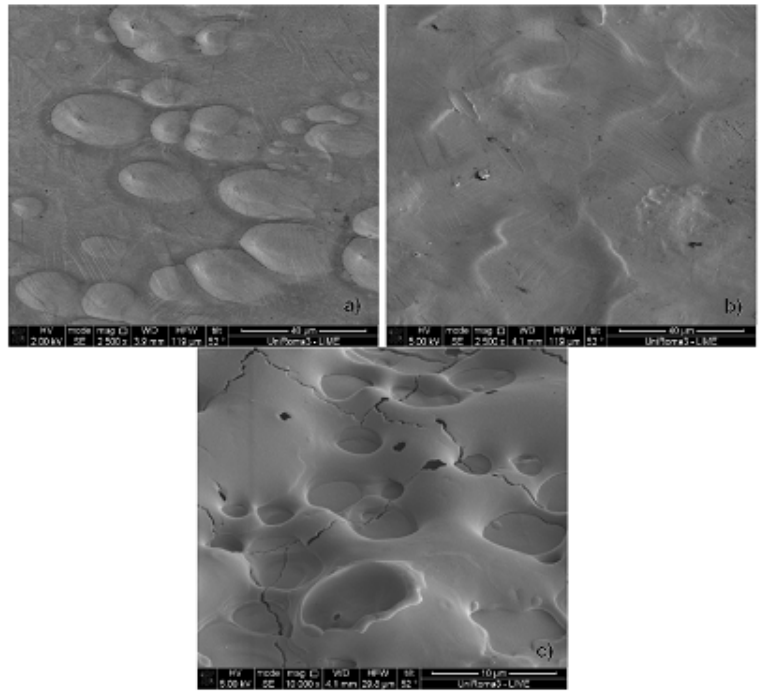

Figura 8.7: Imagen FIB/SEM del acero a) cementado, con un pulso de helio; b) cementado, con cinco pulsos de deuterio, y c) nitrurado, con cinco pulsos de deuterio.

Gracias al detalle de esta técnica y la capacidad de conseguir grandes aumentos se puede hacer un estudio de la morfología superficial y transversal de las probetas tratadas [Song, 2011]. De esta manera es posible hacer un análisis con mucho más detalle que con un microscopio óptico.

En la figura (8.7) analizamos la parte no oculta de las muestras de acero AISI 316L tratadas mediante cementación (partes a y b de la figura (8.7)) y nitruración (parte c de la figura (8.7)) que han sido sometidas a un bombardeo con deuterio y helio.

El análisis indica que ambos tipos de gases, deuterio y helio, al impactar sobre la superficie, provocan cráteres y eyección de material [Roshan, 2008], tal y como pudimos comprobar con el análisis de Microscopía Óptica. Se observan para las muestras cementadas (partes a y b de la figura (8.7)) bandas de deslizamiento entrecruzadas, provocadas por los elevados gradientes térmicos que se originan al llegar los iones ligeros a alta energía y transferir todo su momento en un pequeño intervalo de tiempo.

También observamos que para el caso c de la figura (8.7) se localizan los cráteres más grandes y profundos, además de una superficie irregular y abombada, resultado de la fusión y enfriamiento de la superficie, provocando para este caso una amorfización de la superficie. También se observan rebordes en los cráteres que son inmediatamente relacionados a la fusión del material y su rápido enfriamiento.

Además, se pueden ver grietas [Roshan, 2008] que recorren toda la superficie (quizás provocadas por el enfriamiento rápido) y zonas huecas. Estas zonas son debidas a la sucesión de pulsos que provocan, probablemente, el recubrimiento con material fundido de cráteres profundos provocados por pulsos anteriores.

En cambio, si analizamos la zona de la probeta tapada por la lámina de acero y que únicamente recibió el choque térmico podemos observar que la morfología superficial no ha cambiado en gran manera, observándose la presencia de granos y bandas de deslizamiento. En la probeta cementada aparecen gránulos achacables a grafito producidos por un inicio de exodifusión al activarse los procesos difusivos a altas temperaturas. La zona oculta solo sufre los choques térmicos, por lo que los procesos de difusión son los únicos que tienen relevancia en esta zona de la probeta. Normalmente, la difusión va hacia zonas más internas pero puede suceder que en las capas más superficiales, que están saturadas de carbono, los átomos no tengan lugares libres a los que acudir o que incluso la fuerza conducente no permita atravesar esta barrera, por lo que será mucho más fácil que difundan hacia el exterior de la probeta. La acumulación de estos átomos hará que se agreguen formando núcleos de grafito, que son los gránulos que observamos mediante FIB/SEM en la superficie.

A continuación, estudiamos las imágenes obtenidas en la sección transversal de las probetas analizadas. Iniciaremos el análisis sobre las probetas cementadas sometidas al bombardeo tanto de helio como de deuterio, finalizando con las nitruradas que han sido irradiadas con deuterio.

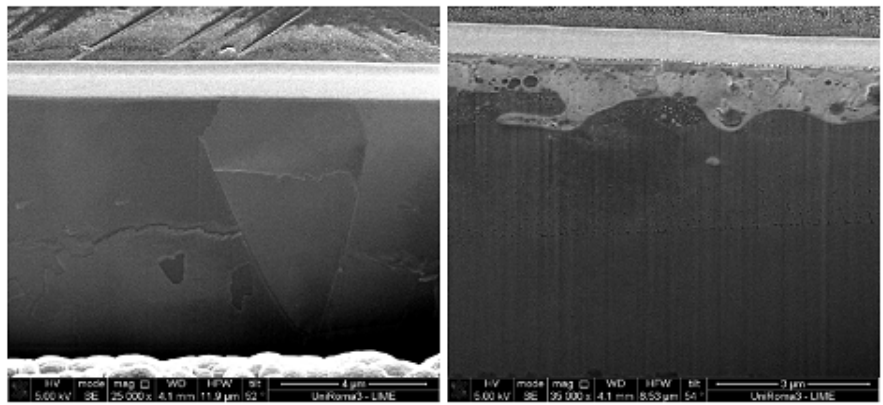

Figura 8.8: Sección transversal analizada con FIB/SEM del acero austenítico cementado bombardeado con un pulso de helio (Izquierda) y del acero cementado, con cinco descargas de deuterio (Derecha).

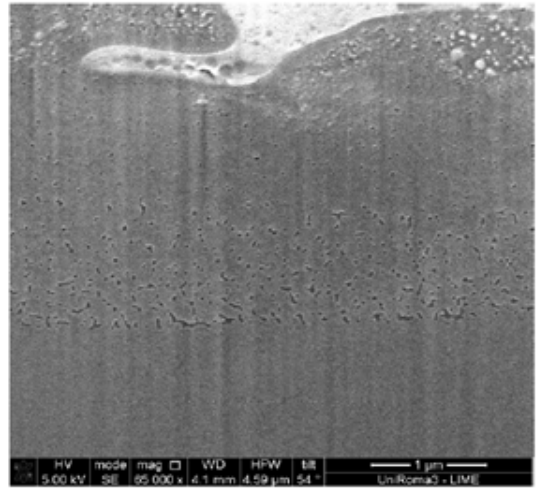

Figura 8.9: Aumento de la zona de cristalitas de la probeta cementada irradiada con cinco pulsos de deuterio.

Para los aceros austeníticos cementados observamos dos comportamientos diferentes. Para un pulso de helio no se aprecia severa amorfización de la zona más superficial, tal y como se puede observar en la parte izquierda de la figura (8.8). A lo sumo aparece una delgada línea en la parte superior de 107,9 nm de espesor. Debido a su aspecto regular es más probable que haya surgido por las tensiones que sufrió durante el pulido inicial de la probeta [García Molleja, 2010] y que no sea una zona amorfizada. Los bordes de grano son fácilmente identificables y dentro de ellos se pueden distinguir varias zonas, quizás provocadas por las tensiones y esfuerzos residuales que provoca la permanencia del carbono en la red austenítica expandida o incluso por el bombardeo directo de los iones de helio que provocan el rápido aumento de temperaturas asociado a todo el proceso. Esta apariencia de descamado (que se puede observar en la parte central de la figura (8.8), imagen izquierda) se puede identificar con la liberación de tensiones [Kim, 2005] entre planos de deslizamiento, que al provocar un cráter por el bombardeo de los iones de Ga se desprenden de la pared. En el caso de la probeta cementada bombardeada con cinco pulsos de deuterio, parte derecha de la figura (8.8), aparece en la superficie una zona amorfa caracterizada por un perfil de profundidad irregular (donde la región más delgada alcanza los 1,26 μm) y la apariencia granulosa provocada por el rápido enfriamiento, quedando incrustadas burbujas al solidificarse la matriz amorfa. Este fenómeno también puede apreciarse en la parte superficial de la misma probeta, parte b de la figura (8.7), puesto que los cráteres no están tan definidos como en el caso de la probeta cementada bombardeada con helio. Este fenómeno está debido a que la superficie queda fundida y es maleable, por lo que en un pequeño intervalo de tiempo el acero amorfo puede reacomodarse. Analizando la imagen podemos ver una región llena de pequeños granos a 3,00 μm de profundidad, que se puede observar mejor en la figura (8.9).

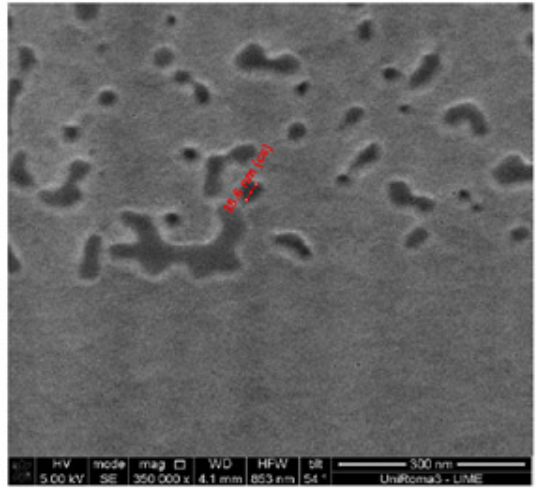

Figura 8.10: Detalle de las cristalitas, tanto aisladas como formando agregados, localizadas en la muestra cementada bombardeada con 5 descargas de deuterio. Imagen FIB/SEM tomada con 350000 aumentos.

Los tamaños de estas cristalitas son de 35,6 nm, aunque también se pueden discernir agrupaciones de estas con un contorno irregular de tamaño mucho mayor, tal y como puede verse en la figura (8.10). El proceso que desencadena estas cristalitas no está aún identificado del todo, por lo que es necesario un mayor estudio mediante FIB/SEM en experimentos posteriores. En un provisional primer intento de explicación podrían identificarse como los inicios de nucleación de la fase cristalina del acero austenítico, detenida por la rápida disminución de temperaturas. El crecimiento se da cuando se supera el tamaño crítico del núcleo y destaca el término volumétrico de g (que tiende a crecer, al contrario que cuando predomina el término superficial).

En el caso del acero inoxidable austenítico nitrurado y sometido a la irradiación de cinco descargas de deuterio, podemos ver en la figura (8.11) que la zona transversal presenta una gran burbuja hueca (de aproximadamente 1,61 μm de diámetro) en una zona de matriz amorfa de ≈3 μm de espesor, confirmando que el intenso bombardeo y los choques térmicos han provocado la fusión del acero. En este caso también a ≈3 μm de profundidad encontramos una región de anchura irregular (de 591,7 nm de espesor como se observa en la parte derecha de la figura (8.11), aunque en otras zonas alcanza un espesor de 1,27 μm, tal y como se observa en la imagen izquierda de la figura (8.11)) que contiene multitud de cristalitas, además de otros granos de mayor tamaño entre esta zona y la superficie. Parece ser que en los aceros previamente nitrurados la nucleación (y crecimiento) de cristales es mucho más rápida que para la cementación. El grosor irregular puede estar provocado por un recubrimiento con material fundido de un cráter formado previamente por un pulso. La zona de cristalitas siempre se detiene a una profundidad fija en torno a los tres micrómetros, esto quizás se achaca al golpe térmico de los iones y que únicamente elevan la temperatura por encima del valor de fusión del acero en la zona que va desde la super_cie hasta esta mencionada profundidad [Murtaza, 2009]. Finalmente, se pueden observar (izquierda de la figura (8.11)) grietas provocadas por los fuertes gradientes térmicos que alcanzan profundidades cercanas a los 7 μm.

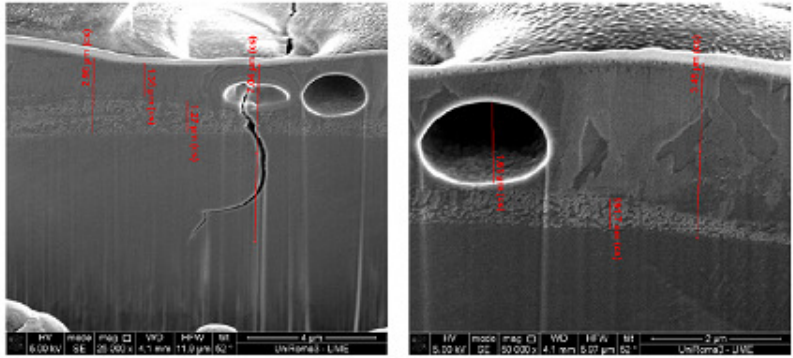

Figura 8.11: Imagen transversal FIB/SEM del acero austenítico nitrurado sometido a cinco descargas de deuterio. Las imágenes izquierda y derecha representan zonas contiguas de la misma probeta.

### 8.1.3. Difracción de Rayos X en Incidencia Rasante

A partir de un análisis con rayos X podemos realizar un estudio paralelo a FIB/SEM sobre la modificación de la estructura cristalina del acero austenítico sometido a bombardeo iónico. Las capas más superficiales son las más susceptibles de ser modificadas por las descargas [Murtaza, 2009], tal y como vimos en el apartado anterior, por lo que es necesario aplicar GIXRD para sondear los primeros micrómetros de material.

Como mencionamos en el capítulo dedicado a la descripción de los dispositivos experimentales, utilizamos una fuente de Cu K$\alpha$ a 40 kV de tensión y 30 mA de corriente. Gracias a un espejo de rayos X se puede conseguir un haz de rayos paralelos de tamaño 4x4 mm$^2$ debido a un ajuste de ranuras Soller horizontales y verticales.

Gracias al espejo se elimina la componente K$\beta$ de la radiación, cuya longitud de onda es 1,3922 Å. Hay que tener en cuenta que la radiación K$\alpha$ = 1,5418 Å es un valor medio de dos contribuciones [Cullity, 1956]: la perteneciente a la intensa K$\alpha_1$ cuya longitud de onda es 1,54056 Å y otra K$\alpha_2$ de menor intensidad y de longitud de onda 1,54433 Å. En principio se espera que los rayos X que lleguen a la probeta sean únicamente los de K$\alpha_1$, pero hay que tomar precauciones al elegir el valor de la longitud de onda usada a la hora de hacer cálculos.

Para analizar los primeros micrómetros incidiremos a 2 y 10º, por lo que sondearemos una profundidad entre 0,65 y 3,22 μm (bajo suposiciones que la red está compuesta únicamente de Fe y los fotones poseen una energía de 8 keV, además de que el límite de apreciación en la atenuación de intensidad es del 1 %).

Mediante un detector de centelleo con ranuras plano-paralelas recibiremos los fotones difractados que lo hagan bajo un ángulo 2θ entre 30 y 80º. Otros datos experimentales fueron el tamaño de paso (0,03º) y el tiempo de permanencia en cada punto del recorrido (1 s).

Los análisis de rayos X se realizarán sobre el material base AISI 316L y sobre la parte nitrurada o cementada y bombardeada con 1, 5 y 10 pulsos, estudiando el plano (111) de la estructura fcc. También se presentan junto a todos estos difractogramas los datos obtenidos para la zona oculta de la probeta (la que solo recibió los tratamientos térmicos, representada por un trazo de color celeste) la cual parece independiente del número de pulsos a la que fue sometida la muestra. El difractograma de esta zona es muy parecido (excepto quizás algo más de anchura en los picos a causa de liberación de tensiones y activación momentánea de la difusión al elevarse la temperatura [Valencia-Alvarado, 2007]) a los que se obtienen de las probetas tratadas antes de llevar al dispositivo de plasma focus.

Las dos siguientes figuras, (8.12) y (8.13), fueron bombardeadas mediante deuterio, la primera tuvo un tratamiento de cementación, mientras que la segunda fue sometida a una nitruración.

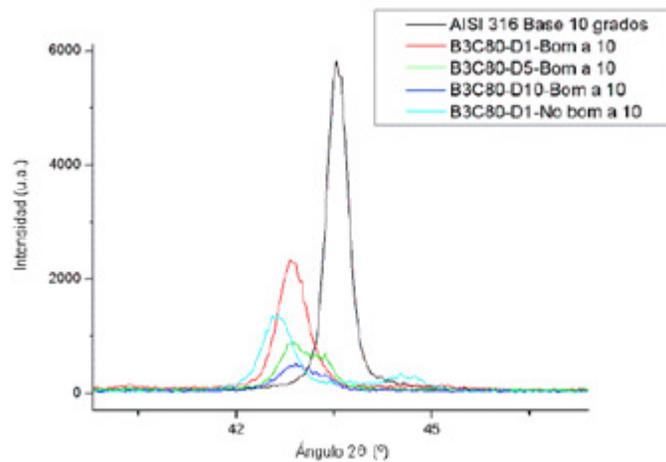

Figura 8.12: Difracción de rayos X realizada a 10º de incidencia en un acero austenítico cementado. Las descargas fueron hechas en atmósfera de deuterio. Material base: trazo negro. Zona no bombardeada: trazo celeste. Un disparo: trazo rojo. Cinco disparos: trazo verde. Diez disparos: trazo azul.

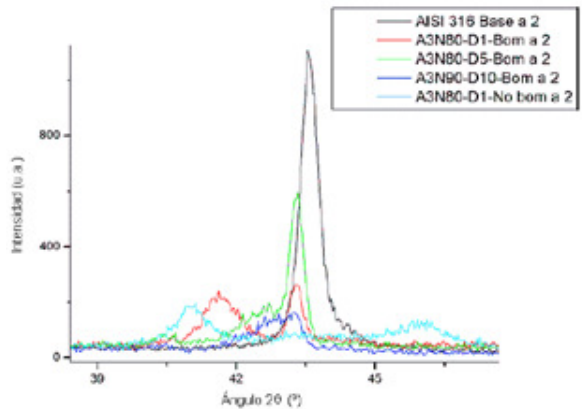

Figura 8.13: Difractogramas de muestras nitruradas de acero austenítico bombardeadas con iones de deuterio. El ángulo del haz incidente fue de 2º.

En ambos casos se observa que los picos (111) correspondientes a la austenita expandida se van desplazando a mayores valores de ángulos 2θ (traducido en menores parámetros de red) en función del número de pulsos realizados en el plasma focus. Se aprecia una tendencia hacia la posición del pico del material base (trazo negro), por lo que se puede intuir una pérdida de elementos expansores (C o N) con cada pulso [Rammo, 2006]. Esto se puede confirmar en la literatura a partir de un experimento inverso que consistió en laimplantación de nitrógeno dentro de acero X6CrNiTi1810 [Blawert, 1999b]. El pico (111) se desplazaba hacia valores angulares menores (correspondiendo a un mayor parámetro de red) cuantos más átomos de nitrógeno por centímetro cuadrado se implantasen.

Como es lógico, la zona que recibe el impacto de los iones tiene menos expansión que la zona oculta que solo recibió el choque térmico, que libera las tensiones residuales en mayor grado [Heuer, 2007]. Además de este corrimiento del pico se puede observar en la figura (8.13), correspondiente a las muestras nitruradas, un fenómeno interesante en el que sumado al pico de austenita expandida que va dirigiéndose a mayor número de disparos hacia mayores ángulos surge otro anclado en ≈43,3º y que al combinar estos resultados con los datos obtenidos mediante FIB/SEM, podemos adjudicar que el pico surge por las cristalitas (como podemos observar en la parte derecha de la figura (8.8), por ejemplo) que se forman en la matriz amorfa superficial al enfriarse la zona fundida de la probeta, dando lugar a la nucleación y efímero crecimiento de zonas con estructura cristalina fuertemente tensionada. Esta posición parece ser independiente del número de pulsos a los que se someta la superficie.

Un análisis minucioso de la figura (8.12) desvela que el pico en la posición 43,3º también aparece, pero está superpuesto con el pico de austenita expandida que pierde grado de expansión en función del número de descargas. Esto es lógico al considerar que la expansión de la austenita por nitruración es mayor que por cementación, desplazándose más hacia la izquierda y evitando la superposición.

Con los difractogramas asociados a cementación se puede indicar que en realidad la austenita expandida no va reduciendo su parámetro de red hasta el valor del material base, sino que lo hace hasta coincidir con el pico situado a 43,3º. Experimentos donde se llevaron a cabo veinte o más descargas de deuterio o helio sobre el acero cementado o nitrurado indican que llega un momento que se reduce en gran manera la pérdida de expansión a cada nuevo pulso, tendiendo el pico (111) hacia la posición de 43,3º y no hacia la posición angular del material base, por lo que se confirma así que el proceso tiene como límite dicha posición angular.

En el caso de llevar a cabo los bombardeos mediante helio en aceros AISI 316L sometidos previamente a nitruración (figura (8.15)) y cementación (figura (8.14)) obtenemos los mismos resultados: hay un desplazamiento progresivo del pico de austenita expandida hacia ángulos mayores cuanto mayor haya sido el número de pulsos recibidos, hasta alcanzar una saturación, en la que el pico deja de desplazarse y se posiciona en el mencionado valor de 43,3º. Este valor de saturación se da cuando sometemos a las probetas tratadas a un mínimo de veinte disparos.

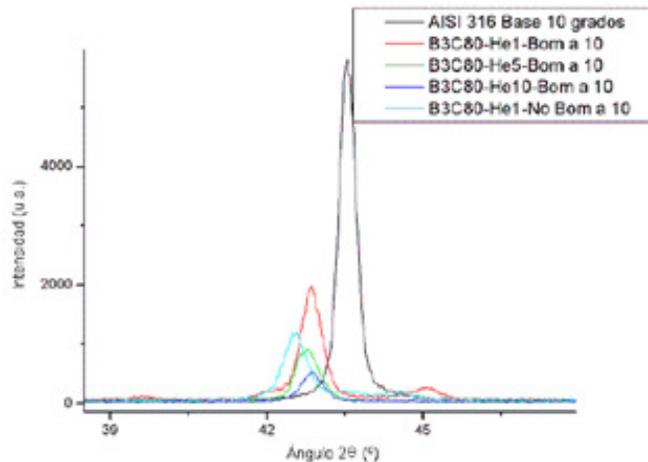

Figura 8.14: Muestra de acero austenítico cementada bombardeada con helio en la que se aplicó la técnica GIXRD. Incidencia rasante de 10º.

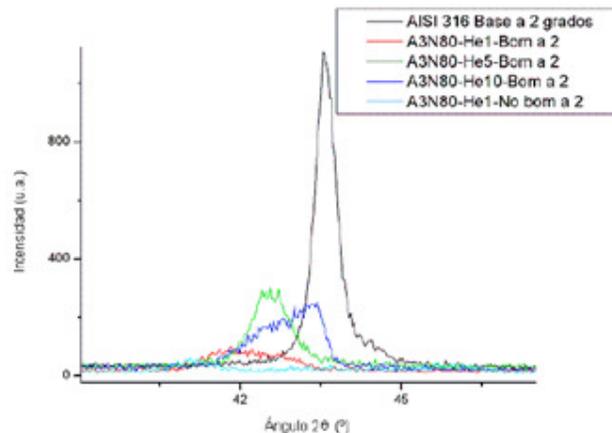

Figura 8.15: Acero AISI 316L nitrurado sometido a un bombardeo con iones de helio. El difractograma fue realizado con una incidencia a 2º.

Ambas figuras muestran resultados idénticos al anterior caso de bombardeo con deuterio. Para la cementación (figura (8.14)) se observa de nuevo el comportamiento de los anteriores casos: pérdida de expansión a un mayor número de descargas. En cuanto a la ausencia del pico de difracción situado en 43,3º, un análisis a 2º de incidencia rasante tampoco lo revela. Pudiera ser que esté superpuesto con los otros, pero los análisis mediante FIB/SEM, tal y como se observa en la parte izquierda de la figura (8.8), descartan esta posibilidad al no observar cristalitas en la sección transversal de la muestra, sumado a que la capa amorfa es inexistente. En su lugar aparece un pico situado en 45,08º, aunque solo para el caso de un pulso de helio. Este puede indicar un precipitado de $Fe_3C$ o de $Cr_3C_2$; quizás provocado por el sputtering preferencial sobre un elemento en particular que puede ocasionar el bombardeo de helio, que es más pesado que el deuterio. Para el caso de la nitruración, cuyo difractograma se muestra en la figura (8.15), también se da un desplazamiento del pico, representando una reducción de la expansión, a mayor número de descargas. Como es lógico, todos los picos están entre las posiciones del acero tratado sin bombardeo iónico y del material base. En este caso de nitruración otra vez aparece el pico en 43,3º, aunque está superpuesto con el pico (111) de los que se desplazan a mayores ángulos, determinado por el hombro de la distribución.

Todos estos análisis nos permiten comentar que a mayor número de disparos la austenita expandida va disminuyendo su parámetro de red hacia el valor de 3,6163 Å, que corresponde al parámetro de red de la estructura fcc cuyo pico (111) está situado en una posición $2\theta$ de 43,3º. Además, en la mayoría de casos aparece también este mencionado pico situado en 43,3º. Este comportamiento de los dos picos también se da para los planos (200) y (202).

- El primer caso, el que va acercándose a mayores valores angulares cuando se aumenta el número de pulsos, podría estar provocado por las elevadas temperaturas que alcanza la probeta por el impacto iónico.

Como la temperatura superficial, causada por la altísima densidad superficial de potencia que desarrolla el plasma focus (10 MW/cm$^2$), llega a 1500 ºC, puede asegurarse que a mayores profundidades se alcanzan valores que reactivan el proceso de difusión durante un corto periodo de tiempo [Ram Mohan Rao, 2005]. Este intervalo de tiempo, cuya duración puede estimarse en centenas de nanosegundos, es suficiente para que la red austenítica pierda átomos de nitrógeno [Manova, 2007] o carbono [Onink, 1995], que son los responsables de su expansión. Estos átomos posiblemente migrarán hacia zonas más profundas hasta que la temperatura de la probeta esté por debajo de 300 ºC [Sartowska, 2007b], momento en que quedarán fijos en el hueco intersticial al que hayan llegado [Farrell, 2005].

- El segundo, el localizado siempre a 43,3º, está provocado por el propio proceso de plasma focus: los iones que llegan, al impactar, crean tantos desórdenes estructurales y fallos en la red cristalina [Chalmers, 1959] que logran expulsar a casi todos los átomos de carbono o nitrógeno, unido al importante choque térmico que provoca la fusión de las primeras monocapas y la posterior amorfización de la estructura [Feugeas, 2010]. Al enfriarse rápidamente esta región no hay tiempo de comenzar un proceso a gran escala de nucleación y crecimiento de granos cristalizados [Chiang, 1997], por lo que quedará una matriz de acero amorfo con cristalitas de acero incluidas en ella, tal y como se ha apreciado en el apartado dedicado a la caracterización mediante FIB/SEM. Estas cristalitas pueden tener cantidades mínimas de nitrógeno o carbono que, junto a las elevadas tensiones residuales [Nosei, 2008] creadas durante el proceso de fusión y enfriamiento, provocan una expansión en la red (que siempre es la misma) y sitúa su pico (111) en 43,3º, correspondiente a un parámetro de red de 3,6163 Å.

Hay que mencionar finalmente que en los procesos de nitruración de aceros mediante plasma focus la austenita expandida desplaza el pico (111) a una posición máxima de ≈43,3º [Feugeas, 2010].

Una vez que determinamos el parámetro de red de la austenita a partir de la ley de Bragg podemos conocer el grado de expansión y su pérdida progresiva respecto al valor del parámetro de red del material base en función del tratamiento superficial realizado, el número de pulsos aplicados y el tipo de gas en el que se lleva a cabo la descarga. En la tabla 8.2 se consignan los datos referentes a la cementación del acero AISI 316L, cuyo parámetro de red ya fue determinado en 3,584 Å:

| Disparos | Deuterio (Å) | Porcentaje (%) | Helio (Å) | Porcentaje (%) |
|----------|--------------|----------------|-----------|----------------|
| 0 | 3,6696 | 2,39 | 3,6779 | 2,62 |
| 1 | 3,6545 | 1,97 | 3,6529 | 1,92 |
| 5 | 3,6491 | 1,82 | 3,6557 | 2,00 |
| 10 | 3,6429 | 1,64 | 3,6504 | 1,85 |

Tabla 8.2: Parámetros de red y expansión relativa respecto al material base de las probetas cementadas sometidas a 0, 1, 5 y 10 pulsos de deuterio y de helio.

Ahora si analizamos los parámetros de red para el proceso de nitruración del acero austenítico (tabla 8.3) tendremos los siguientes valores:

| Disparos | Deuterio (Å) | Porcentaje (%) | Helio (Å) | Porcentaje (%) |
|---|---|---|---|---|
| 0 | 3,8073 | 6,23 | 3,7984 | 5,98 |
| 1 | 3,7491 | 4,61 | 3,7415 | 4,39 |
| 5 | 3,6691 | 2,37 | 3,6743 | 2,52 |
| 10 | 3,6468 | 1,75 | 3,6556 | 2,00 |

Tabla 8.3: Valores calculados del parámetro de red y porcentaje de expansión relativa respecto al material base del acero austenítico nitrurado sometido a un número creciente de pulsos de deuterio y helio mediante plasma focus.

Los valores, en ambas tablas, que indican cero disparos fueron las mediciones de la estructura cristalina de la zona de la probeta que estaba oculta con una lámina de acero, por lo que sus valores son muy parecidos entre sí y también al valor del parámetro de red de la austenita expandida antes de llevar a cabo el bombardeo mediante plasma focus. Si comparamos estos valores con los parámetros de red calculados en la tabla 7.3 se observa un grado de expansión algo mayor. Esto puede deberse a las mejoras implementadas con la nueva cámara de vacío y el diseño especial del cátodo, que maximiza el proceso y evita la precipitación de carburos al no aplicar una intensidad de corriente elevada, lo que se traduce (también gracias a la gran superficie catódica) en una muy baja densidad de corriente eléctrica. Analizando las tablas puede determinarse que las probetas cementadas, al recibir un número de disparos fijo, presentan una pérdida de expansión austenítica de la misma magnitud, no importando por consiguiente el tipo de gas empleado para realizar el bombardeo. Por ejemplo, para las muestras de AISI 316L cementadas un disparo de deuterio conlleva una expansión del 1,97 % respecto al material base, mientras que si utilizamos helio la expansión queda reducida hasta un 1,92 %, ambos valores parecidos entre sí. El mismo fenómeno se da para las muestras nitruradas, por lo que bajo condiciones de bombardeo el factor más determinante es el proceso de modificación superficial previo.

En ambas tablas es posible observar que el porcentaje de expansión va decayendo de manera cada vez más lenta, por lo que se hace plausible postular que el menor parámetro de red que alcanzará el acero bombardeado será 3,6163 Å, el correspondiente al pico localizado en 43,3º, que equivale a una expansión de la red austenítica de un 0,90 % respecto del material base. Este valor de expansión se alcanzaría tras una sucesión de más de 20 descargas de plasma focus que, como hemos visto, es independiente del tratamiento previo y del gas empleado para irradiar. Esto se explica por el aumento de la capa amorfa al recibir sucesivos choques térmicos y bombardeo iónico intenso, que en última instancia destruiría toda la capa de austenita expandida y quedaría reemplazada por una matriz amorfa con pequeños núcleos cristalizados, pero sometidos a altas tensiones residuales internas [Nosei, 2008] que determinarían la posición angular detectada en los difractogramas de rayos X en el valor 43,3º.

## 8.1.4. Microdureza Vickers

En el caso de determinar la dureza superficial de las probetas tratadas y bombardeadas se vuelve a implementar microindentación Vickers, cuyos datos instrumentales ya quedaron consignados en la sección 5.3. Aplicamos en este caso cargas de 25 g en zonas superficiales alejadas entre sí y preferentemente en las regiones donde estén presentes las bandas de deslizamiento, tal y como se observa en la figura (8.3), donde se puede observar la marca de indentación realizada de manera correcta, bordes definidos y paralelos entre sí [Müller, 1973]. Para reducir el error estadístico realizamos un mínimo de tres indentaciones para que la dispersión siempre esté en torno al 10 %, calculada mediante la desviación estándar.

Conocemos por anteriores experimentos (tabla 6.3) que la dureza del material base AISI 316L es de 264 HV. Para el caso de la nitruración las durezas consignadas en la zona sujeta al bombardeo iónico son las siguientes:

| Disparos | Deuterio (HV) | Helio (HV) |
|----------|---------------|------------|
| 0 | $999 \pm 60$ | $965 \pm 70$ |
| 1 | $827 \pm 40$ | $815 \pm 40$ |
| 10 | — | — |

Tabla 8.4: Durezas Vickers sobre la superficie de acero nitrurado sometido a irradiación de iones ligeros.

En el caso de la cementación del acero austenítico los valores de dureza obtenidos son los que aparecen en la siguiente tabla.

| Disparos | Deuterio (HV) | Helio (HV) |
|----------|---------------|------------|
| 0 | $461 \pm 30$ | $443 \pm 40$ |
| 1 | $364 \pm 40$ | $293 \pm 6$ |
| 10 | $366 \pm 20$ | $367 \pm 20$ |

Tabla 8.5: Microindentaciones a 25 g sobre acero austenítico cementado bajo irradiación de iones de deuterio y helio.

Los cálculos de dureza para los valores de cero disparos se tomaron en todas las probetas sobre la zona oculta por la lámina de acero, haciendo una media y calculando la propagación del error. Generalmente, obtenemos un resultado conocido: la dureza en las probetas nitruradas es mayor que en las cementadas [Tsujikawa, 2005b]. Además, en la parte oculta que solo recibió el choque térmico el valor de dureza fue parecido a la que tenían las probetas antes de ser enviadas al plasma focus, indicando valores superiores a las zonas que sí recibieron el impacto iónico [Murtaza, 2009] y, por supuesto, al del material base tanto en el proceso de nitruración [Menthe, 2000] como en el de cementación [Qu, 2007]. Se puede identificar que para un tratamiento superficial dado, a igual número de descargas el valor de dureza es independiente al gas utilizado para bombardear. Podemos observar que en el caso de la nitruración diez disparos son suficientes para generar amorfización superficial y la destrucción de las capas más superiores (figura (8.6)), por lo que se hace imposible la medición de la dureza. En el caso de la cementación sí se puede comparar el proceso para cada tipo de gas en todas las muestras [Heuer, 2007]. Cuando el agente de bombardeo es deuterio no se observa cambio drástico en la dureza, pero sí para cuando se bombardea con helio. En este caso, a mayor número de disparos aumenta la dureza, aunque no se alcanza el valor del material tratado sin bombardear. Este aumento de dureza puede estar debido a un efecto de endurecimiento llevado a cabo por el tratamiento térmico y la creación de la capa amorfa fundida en la zona más superficial. De todas formas, a 10 descargas con plasma focus, el valor de la dureza es independiente al gas empleado para bombardear, siendo solo dependiente del número de descargas y al tratamiento previo.

## 8.2. Resistencia al tratamiento térmico

La observación del desplazamiento del pico (111) de austenita expandida por cada pulso de helio o deuterio fue adjudicado a la reactivación de los procesos de difusión de los átomos de carbono o nitrógeno dentro de la red austenítica que van a mayores profundidades, lo que conlleva una pérdida de concentración atómica de N o C en la capa tratada y la disminución del parámetro de red [Onink, 1995]. En un intento de comprender los procesos que se llevan a cabo, se tratarán térmicamente muestras de acero nitrurado y cementado a altas temperaturas y durante tiempos prolongados. También se va a estudiar el empleo del nitruro de aluminio (AlN) como barrera térmica, por lo que el tratamiento del acero constará de difusión iónica más deposición mediante sputtering por magnetrón, combinación denominada tratamiento dúplex. El nitruro de aluminio es un compuesto III-V [Venkataraj, 2006] de excelentes propiedades mecánicas, físicas y tribológicas [Xu, 2001], ampliamente utilizado en la industria por su ancho salto de banda energético [Olszyna, 1997], su óptimo coeficiente de acoplamiento electromecánico [Naik, 2000], su alta dureza como fase cúbica metaestable [Setoyama, 1996], su elevada velocidad de propagación del sonido [Clement, 2004] y las propiedades piezoeléctricas presentes en la dirección <0002> [Akiyama, 2006].

Las probetas cementadas y nitruradas que se utilizan en este apartado se obtuvieron de los procesos de nitruración y cementación realizados en el apartado 8.1, por lo que los valores experimentales son los mismos que se consignaron en la tabla 8.1. La mitad de ellas fueron sometidas a una deposición de AlN para poder hacer las comparaciones. Este proceso se lleva a cabo en una cámara de vacío diferente a las anteriormente utilizadas, de 94,03 L de capacidad y conectada a una bomba difusora respaldada por una bomba mecánica que permitieron lograr un vacío base de $9.60 \cdot 10^{-5}$ mbar. Previamente las probetas fueron limpiadas con un baño de alcohol y acetona para eliminar las impurezas de la superficie y se colocaron en un portamuestras a 3 cm de distancia del blanco del magnetrón (refrigerado por agua y conectado a una resistencia de 240 $\Omega$), que fue de aluminio puro y de 8 cm$^2$ de superficie. Una vez evacuada la cámara se realizan varias purgas con argón para reducir el vacío base y eliminar el vapor de agua adherido a las paredes de la cámara, asegurando que la presión medida constaba únicamente de átomos de argón. Tras este paso se coloca un obturador entre el magnetrón y la muestra para dejar a esta aislada del proceso de limpieza del magnetrón, realizado durante 10 minutos bajo una atmósfera de Ar puro a $5,33 \cdot 10^{-2}$ mbar. Los parámetros de descarga fueron 340 V y 125 mA. Una vez limpio el blanco se introduce una mezcla de 50 % de N$_2$ y 50 % de Ar a un flujo de 12,0 mL/min cada uno. La descarga se mantiene 15 minutos antes de retirar el obturador para asegurarnos haber llegado a condiciones de trabajo estables [Safi, 2000].

La deposición de nitruro de aluminio se realiza durante 30 minutos a una presión de trabajo de $6,65 \cdot 10^{-3}$ mbar, con 260 V de tensión DC de descarga e intensidad de corriente de 154 mA. Esto nos indica que la densidad de potencia que se desarrolla en el blanco es de 5,01 W/cm$^2$, valor que nos asegura una tasa de deposición bajo estas condiciones de 11 nm/min, por lo que estimamos que el espesor de la capa de AlN es de 330 nm. No pudimos medir la temperatura de las muestras que actuaron de sustrato pero se asegura que la única fuente de calentamiento fue el bombardeo de los átomos de nitrógeno y aluminio o los agregados formados de AlN [Smentkowski, 2000], por lo que se descarta una temperatura (a este rango de presiones de trabajo) suficiente para activar procesos de difusión.

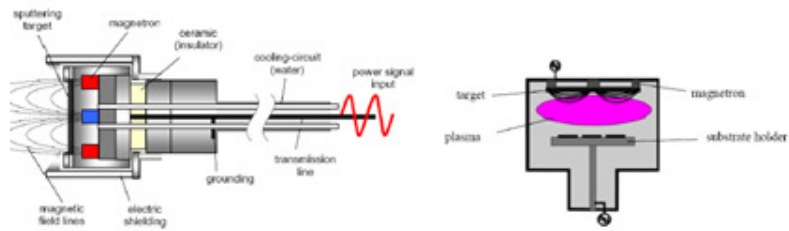

Figura 8.16: Izquierda: dibujo esquemático sobre la configuración de un magnetrón, con la colocación del blanco sobre el campo magnético. Derecha: esquema sobre la deposición mediante sputtering por magnetrón reactivo. Los iones de Ar chocarán con el blanco de Al, por lo que se eyectarán átomos de este material y se combinarán con el nitrógeno, llegando hasta el sustrato [Corbella Roca, 2005].

Los análisis GIXRD no indican que se haya logrado estructura cristalina en la capa de AlN [Kajikawa, 2003], debido a la fuerte rugosidad del sustrato de acero tratado, que en el apartado 4.2 determinamos que tenía un valor de 130 nm. Tras estos tratamientos se someterán las probetas que tienen AlN y las que no a tratamientos térmicos prolongados en un horno de marca VEFBEN de 15 A y 380 V de valores de corriente y tensión y que aplica calentamiento mediante una resistencia activa a intervalos para mantener un control estable de temperatura. Las temperaturas elegidas y los tiempos de tratamiento vienen indicados en la siguiente tabla:

| Temperatura (ºC) | 20 horas | 40 horas | 60 horas |
|---|---|---|---|
| 225 | Sí | Sí | No |
| 325 | Sí | No | No |
| 405 | Sí | Sí | Sí |
| 504 | Sí | No | No |

Tabla 8.6: Identificación de los tratamientos térmicos realizados a una temperatura y tiempo determinado.

Posteriormente a los tratamientos térmicos se realizaron análisis GIXRD a las probetas utilizando los valores experimentales ya utilizados en el desarrollo de esta tesis. En una primera inspección de los difractogramas no hay mucha diferencia entre una muestra cementada y nitrurada, ya que aparecen, junto a los picos adjudicados a la austenita, multitud de picos identificados con precipitados de cromo y óxidos de hierro. La intensidad de estos picos puede llegar a ser mayor que la de los picos austeníticos. En la figura (8.17) aparece el difractograma GIXRD tomado a 2º, donde se ven las diferencias de una probeta cementada con (trazo negro) y sin (trazo rojo) capa superficial de nitruro de aluminio tras sufrir un tratamiento térmico de 405 ºC durante 60 horas. En los picos de austenita no se muestran diferencias notables, aunque sí en los dos picos situados a menores ángulos, localizados a 33,008918 y 35,4641322º, identificados con la fase $\alpha$-$Fe_2O_3$. El resto de los picos se ajustan notablemente con el óxido de hierro identificado, aunque no se puede descartar la presencia de $Cr_3C_2$, puesto que ambos comparten muchas posiciones angulares. No se puede descartar tampoco la existencia de otros óxidos. Este fenómeno puede deberse al efecto de barrera protectora del AlN contra la oxidación del acero. Tampoco se observan diferencias entre los picos de precipitados, dando a entender que son procesos internos de la capa cementada y por tanto ajenos a la influencia de la capa de AlN.

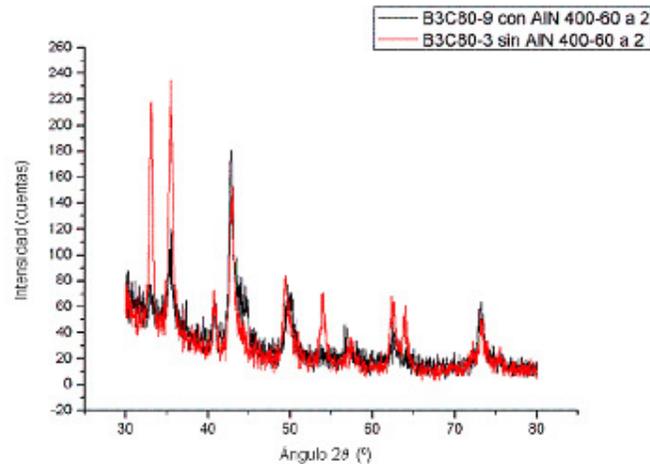

Figura 8.17: Difractogramas tomados a 2º de incidencia de dos probetas cementadas tratadas térmicamente durante 60 horas a 405 ºC. El trazo rojo representa a la probeta que no tiene una capa de AlN en su parte superior, mientras que el trazo negro representa a la probeta que recibió el tratamiento dúplex.

Fuera de este fenómeno de barrera contra la oxidación no aparecen efectos provocados por el AlN en los picos de austenita, por lo que nos centraremos en el estudio del pico (111) tal y como realizamos en el anterior apartado de irradiación mediante iones ligeros. De todas maneras incluiremos en el estudio tanto las probetas que tengan AlN como las que no tienen. En primer lugar se analizan las probetas de acero AISI 316L cementadas. El trazo para el material base es negro y el acero cementado se representa con un trazo rojo, mientras que para los tratamientos a 225 ºC son azul oscuro y verde para 20 y 40 horas de experimento, respectivamente. Para el tratamiento a 325 ºC se recurre al magenta y para los tratamientos a 405 ºC se utilizan trazos marrón claro, añil y marrón oscuro para los tratamientos de 20, 40 y 60 horas, respectivamente. Para finalizar, el experimento a 504 ºC se representa mediante un trazo verde oliva.

Analicemos en primer lugar, las probetas cementadas con una capa superior de nitruro de aluminio mediante GIXRD.

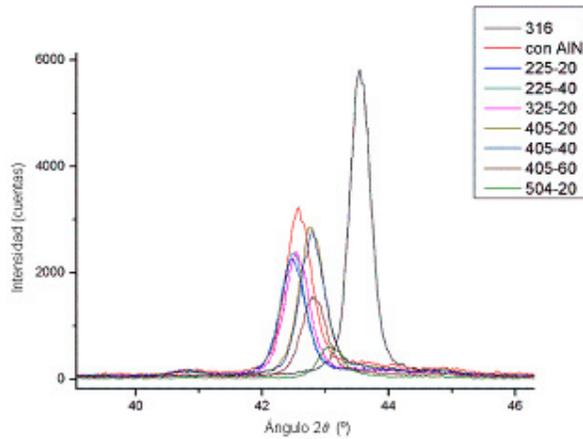

Figura 8.18: Difractograma a 10º de incidencia de muestra de acero AISI 316L cementada con capa superior de nitruro de aluminio. Se incluyen todos los tratamientos realizados al variar tiempo de tratamiento y temperatura.

El acero cementado, antes de llevarlo al horno, fue estudiado mediante GIXRD para revelar que su parámetro de red es de 3,6880 Å, que conlleva una expansión relativa respecto del material base de un 2,90 %. En la tablas siguientes se indican los parámetros de red y el grado de expansión relativa para cada una de las muestras tratadas térmicamente. En la tabla 8.7 aparecen los resultados de las probetas cementadas sometidas al tratamiento dúplex.

| Temperatura (ºC) | Tiempo (h) | Parámetro de red (Å) | Expansión (%) |
|---|---|---|---|
| 225 | 20 | 3,6796 | 2,67 |
| 225 | 40 | 3,6807 | 2,70 |
| 325 | 20 | 3,6773 | 2,60 |
| 405 | 20 | 3,6566 | 2,03 |
| 405 | 40 | 3,6557 | 2,00 |
| 405 | 60 | 3,6553 | 1,99 |
| 504 | 20 | 3,6323 | 1,35 |

Tabla 8.7: Valores del parámetro de red y la expansión relativa para el acero cementado con un recubrimiento superficial de AlN.

A continuación, se muestran los difractogramas de las probetas de acero austenítico únicamente cementadas.

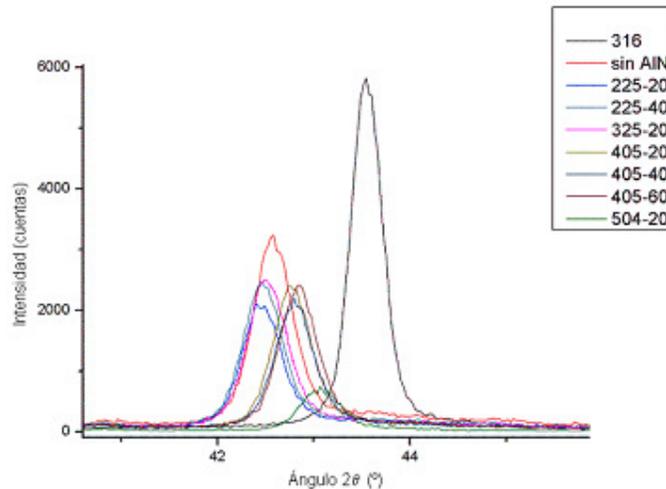

Figura 8.19: Difractograma a 10₀ de incidencia de muestra cementada sometida a procesos térmicos prolongados y de alta temperatura. Se incluyen todos los tratamientos realizados.

En la tabla 8.8 aparecen los datos correspondientes al parámetro de red y a la expansión relativa de la austenita expandida en función de la temperatura de tratamiento y el tiempo del experimento. En este caso no hay capa superior de AlN.

| Temperatura (°C) | Tiempo (h) | Parámetro de red (Å) | Expansión (%) |
|---|---|---|---|
| 225 | 20 | 3,6858 | 2,84 |
| 225 | 40 | 3,6841 | 2,79 |
| 325 | 20 | 3,6800 | 2,68 |
| 405 | 20 | 3,6594 | 2,10 |
| 405 | 40 | 3,6531 | 1,93 |
| 405 | 60 | 3,6521 | 1,90 |
| 504 | 20 | 3,6318 | 1,33 |

Tabla 8.8: Valores del parámetro de red y la expansión relativa para el acero cementado sin recubrimiento de AlN.

En ambas figuras se observa un comportamiento parecido: los tratamientos térmicos a 225 °C no presentan un desplazamiento apreciable del pico (111) respecto al de la muestra cementada a causa que dicha temperatura es insuficiente para activar los procesos difusivos en toda la capa cementada [Sartowska, 2007b]. Se observa también que a mayor tiempo de tratamiento menor intensidad, posiblemente a causa de una degradación de la estructura una vez que ha maximizado su eliminación de defectos estructurales.

El tratamiento a 325 °C sí presenta un leve desplazamiento del pico (111) hacia mayores valores angulares 2θ, es decir, a un menor parámetro de red. El desplazamiento aún es leve, puesto que la energía calorífica aportada, aunque superior al valor de activación de la difusión, no es elevada. No ocurre lo mismo a 405 y 504 °C, donde el desplazamiento es observable y acusado, identificando este con una disminución del parámetro de red al perder la austenita expandida átomos de carbono al difundir hacia el interior y quedar multitud de sitios intersticiales vacíos [Farrell, 2005].

En ambas tablas no se ve un predominio claro de un tipo de tratamiento u otro, por lo que se puede confirmar que la capa de nitruro de aluminio solo es relevante para evitar la oxidación superficial, no para frenar la difusión hacia el interior. Debido a que la capa de AlN es fina y la superficie sobre la que se deposita es altamente irregular, es posible que dicha capa actúe también como barrera contra la exodifusión hasta que al llegar a altos tiempos de tratamiento comience a despegarse del sustrato.

Una vez analizadas las muestras cementadas, haremos un tratamiento análogo con las probetas de AISI 316L nitruradas, donde algunas tienen nitruro de aluminio en la superficie tras un tratamiento dúplex y otras no. En la figura siguiente se muestran los difractogramas obtenidos con GIXRD bajo una incidencia de 10º de las muestras nitruradas con AlN superior. El código de colores se conserva, excepto el trazo rojo, que identifica ahora al acero nitrurado.

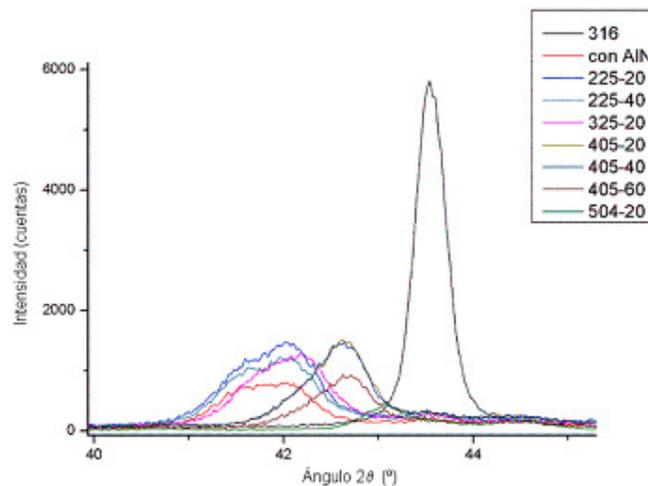

Figura 8.20: Difractograma a 10º de incidencia de muestra nitrurada con capa superior de AlN, incluyéndose todos los tratamientos realizados.

En este caso de nitruración, el parámetro de red de la austenita expandida por nitruración es de 3,7082 Å y una expansión relativa de 3,46 %. Esto da un valor de expansión menor que las primeras sesiones de tratamiento térmico y puede explicarse por una eliminación de defectos y leve reacomodamiento atómico que mejora la red cristalina.

En la tabla siguiente se dan los datos referentes a la expansión de la red austenítica y el parámetro de red obtenido tras los procesos de tratamiento térmico para el acero nitrurado con una capa superior de nitruro de aluminio.

| Temperatura (°C) | Tiempo (h) | Parámetro de red (Å) | Expansión (%) |
|---|---|---|---|
| 225 | 20 | 3,7308 | 4,09 |
| 225 | 40 | 3,7329 | 4,15 |
| 325 | 20 | 3,7176 | 3,73 |
| 405 | 20 | 3,6729 | 2,48 |
| 405 | 40 | 3,6753 | 2,55 |
| 405 | 60 | 3,6713 | 2,44 |
| 504 | 20 | 3,6313 | 1,32 |

Tabla 8.9: Parámetro de red y expansión relativa porcentual para la muestra nitrurada con deposición de nitruro de aluminio.

Ahora bien, en el caso de que no se deposite una capa superficial de AlN los difractogramas de las sesiones de tratamiento térmico presentarán el siguiente aspecto, indicado en la figura (8.21).

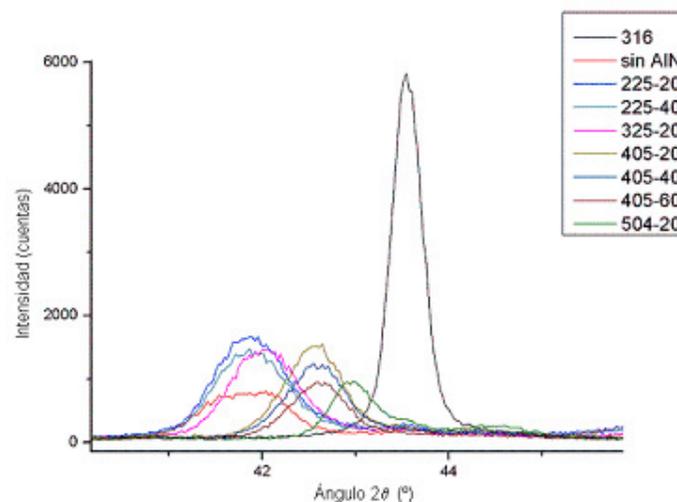

Figura 8.21: Difractograma a 10º de incidencia de muestra de acero AISI 316L nitrurada sometida a procesos térmicos prolongados y de alta temperatura.

Para el caso de los datos de parámetro de red y expansión relativa para cada uno de los tratamientos térmicos sobre acero únicamente nitrurado, en la tabla 8.10 se muestran todos los valores obtenidos.

| Temperatura (°C) | Tiempo (h) | Parámetro de red (Å) | Expansión (%) |
|---|---|---|---|
| 225 | 20 | 3,7308 | 4,10 |
| 225 | 40 | 3,7294 | 4,06 |
| 325 | 20 | 3,7173 | 3,72 |
| 405 | 20 | 3,6801 | 2,68 |
| 405 | 40 | 3,6756 | 2,56 |
| 405 | 60 | 3,6710 | 2,43 |
| 504 | 20 | 3,6398 | 1,56 |

Tabla 8.10: Parámetro de red y expansión relativa porcentual para la muestra nitrurada sin deposición de nitruro de aluminio.

Como en el caso de la cementación, la capa superficial de AlN solo tiene efectos contra la entrada de oxígeno atmosférico que produce la precipitación

de óxidos en los primeros micrómetros de la capa de austenita expandida [Lin, 2008]. No se aprecia ningún mecanismo por el que la capa de nitruro de aluminio frene o fomente los procesos difusivos y la pérdida de nitrógeno de la capa tratada. En ambos casos se observa que el tratamiento térmico solo tiene efectos importantes al aumentar de temperatura, puesto que se nota un desplazamiento de los picos hacia valores angulares mayores, es decir, a menores parámetros de red. El tiempo de tratamiento a un valor de temperatura fijo apenas tiene incidencia en el desplazamiento de estos picos (111), tal y como se observan en los difractogramas. Los tratamientos a 225 ºC no cambian en demasía el parámetro de red, al no haber suficiente energía para activar el proceso de difusión. También se observa que el desplazamiento es mínimo para el proceso a 325 ºC puesto que la energía térmica que tiene la red permite el paso de los átomos de N o C de una vacancia intersticial a otra, pero a poca velocidad [Sartowska, 2007b]. Para finalizar, los procesos a 405 y 504 ºC presentan una mayor difusión, al estar por encima de la energía umbral y con el remanente suficiente para acelerar el proceso.

# Bibliografía

# Parte V

# Conclusiones

                    Tesis presentada para optar al título de Doctor en Física

En esta tesis se trabajó en el campo de los tratamientos de superficies mediante plasmas en la variante de la difusión iónica. Las investigaciones se concentraron en particular sobre la modificación de la superficie de aceros inoxidables austeníticos mediante la cementación y la nitruración iónica.

Los procesos de tratamiento se realizaron utilizando descargas quiescentes DC pulsadas en mezcla de gases reactivos a presiones de ≈5,0 mbar. Para la cementación se utilizaron dos tipos de mezclas: 50 % Ar – 45 % $H_2$ – 5 % $CH_4$ (C50-) y 80 % Ar – 15 % $H_2$ – 5 % $CH_4$ (C80-) con tiempos de tratamiento de 30, 60 y 120 minutos. Para comparar posteriormente las propiedades del acero cementado con respecto al nitrurado se llevó a cabo una nitruración iónica con una mezcla de 80 % $H_2$ – 20 % $N_2$ a iguales condiciones experimentales. Las densidades de corriente durante los tratamientos variaron entre 7,0 y 8,1 $mA/cm^2$ para las muestras C50- y C80-, respectivamente, siendo estos valores los necesarios para que la superficie de las muestras alcancen temperaturas ≥ 400 ºC, temperatura necesaria para activar totalmente el proceso de difusión del carbono al interior del acero. En todos los casos se obtuvieron capas con una alta concentración de carbono, muy por encima de los porcentajes admitidos por los aceros y que resultan del diagrama Fe-C. Estas capas consisten en lo que se conoce como *austenita expandida* (también pudiéndose denominar como $\gamma_C$ o $\gamma_N$, según el tipo de gas que utilicemos para el tratamiento) debido a que la estructura es similar a la de la austenita, pero con un parámetro de red expandido a causa de la inserción (intersticial) dentro de la red cristalina del acero de átomos de carbono (austenita expandida tipo $\gamma_C$) o de los átomos de nitrógeno (austenita expandida $\gamma_N$). En los tiempos de tratamiento elegidos se obtuvieron capas de austenita expandida, por lo que se redujo considerablemente el tiempo para la creación de estas capas en comparación con otros tratamientos de hasta incluso días.

Los procesos atípicos que tienen lugar en el desarrollo de la austenita expandida se pueden resumir en los dos puntos siguientes:

- La brusca difusión de carbono (o nitrógeno) a partir de los 400 ºC (aunque el proceso difusivo se activa a partir de los 300 ºC) conduce a la supersaturación colosal de carbono (o nitrógeno) en la matriz de austenita.

- La generación de una frontera bien definida y recta, cuando atraviesa distintos granos con difusividades diferentes a causa de efectos diferentes de canalización, inclusiones, freno por bordes de grano. . .

Por otro lado, en nuestro proceso de cementación (o nitruración) se comprobó, mediante cálculos a partir de la configuración del plasma de tratamiento, que los iones no llegan a la superficie y simplemente se adsorben, sino que llegan con cierta energía cinética que les permite implantarse en los primeros nanómetros de profundidad, fomentando así el proceso difusivo dentro de la red cristalina, tras neutralizarse. Sin embargo, los mecanismos de alojamiento del carbono en los huecos intersticiales no están claros, habiendo diferentes hipótesis sobre si ocupan tanto los huecos octaédricos y tetraédricos o solo los primeros. Además, los defectos presentes en la red y la aparición de átomos metálicos diferentes al Fe complican una descripción teórica detallada, así como el valor de la temperatura con la que se realice el experimento. Por último, no está totalmente aclarado si el proceso difusivo sigue una única ley de Fick en todo el acero o van cambiando sus parámetros en función de la profundidad, así como los efectos de diferentes orientaciones cristalinas y los bordes de grano.

Los perfiles de concentración de los elementos con la profundidad en las superficies tratadas fueron estudiados mediante AES, abarcando en el análisis 156 nm de profundidad. Se observó que el carbono es el componente mayoritario en la superficie y que va decreciendo conforme se va yendo más hacia el interior de la probeta hasta los 48 nm de profundidad, punto a partir del cual se aprecia una concentración constante de carbono en el resto del sondeo de profundidad. La concentración de C, para las muestras C80-, aumenta en función del tiempo de tratamiento, alcanzando valores de ≈15 % para 30 minutos y ≈28 % para 120 minutos de tratamiento, respectivamente. En el caso de las muestras C50- se observa una concentración de ≈15 %, independientemente del tiempo de tratamiento. Se distinguen dos tipos de carbono: uno químicamente ligado (tipo carburo) y otro libre (tipo grafito). El segundo, en las probetas C80-, está en la parte más externa de la superficie y su proporción aumenta en función del tiempo de tratamiento. El primero se sitúa en zonas más internas de la superficie de las C80-, mientras que en las C50- está presente desde la superficie. Ambos tipos de carbono posiblemente se alojan en los huecos intersticiales octaédricos, aunque no se puede descartar que el carbono grafítico pueda localizarse también en agregados de tamaño nanométrico. El carbono tipo carburo aparece por las altas densidades de corriente empleadas en la cementación (por encima de 7,0 mA/cm$^2$) y está presente solo en las capas más superficiales. Esta es la explicación de por qué no se observan carburos en los estudios mediante GIXRD, ya que según la configuración experimental de esta técnica los datos obtenidos corresponden a profundidades en torno a los 3 μm. La observación de Cr en porcentajes muy superiores a los correspondientes al material base (17 %) se debe fundamentalmente al efecto de la incidencia de los iones de C y de Ar durante los procesos de tratamiento. Los estudios por GIXRD solo permiten la observación confiable a partir del primer micrómetro de profundidad. Estos estudios confirman la existencia de una estructura fcc (austenítica) pero con un parámetro de red que crece rápidamente del valor original de la austenita (3,584 Å) hasta ≈3,660 Å durante los primeros 30 minutos de tratamiento, continuando luego el crecimiento, pero mucho más lento, hasta valores de ≈3,666 Å para 120 minutos de tratamiento. Esta rápida dilatación de la red se debe a la rápida difusión del carbono al interior alojándose principalmente en los sitios octaédricos, tensionando a la misma y modificando las propiedades superficiales. Las imágenes FIB muestran también el desarrollo de maclas dentro de los granos originales de la austenita a causa del estado de compresión resultante, como también se puede apreciar en las imágenes ópticas superficiales por la proliferación de bandas de deslizamiento dentro de los granos de austenita. Esto se puede ver por el incremento de la dureza superficial desde los 2,589 GPa (equivalentes a 264 HV) típicos del acero base hasta valores de 12 GPa en las capas más superficiales, pero con valores no tan elevados, aunque siempre superiores a los del acero base hasta profundidades de 700 μm. Al mismo tiempo, los ensayos de desgaste ball-on-disc mostraron un aumento de la resistencia al desgaste, como así también una reducción de las fuerzas de fricción. Este último resultado puede explicarse debido al hecho de que el desarrollo del carbono libre (grafítico) en las capas externas funciona como lubricante natural.

Los resultados de corrosión pueden analizarse de la siguiente manera: para las probetas C50-, la resistencia al ataque de cloruros se mantiene comparándola con la correspondiente a la del material base. Sin embargo, la observación de algunas picaduras en la superficie de los aceros tratados que han desarrollado austenita expandida pueden considerarse inducidas por la activación de centros de corrosión en los bordes de grano y en las bandas de deslizamiento provocadas por el estado de tensiones. Este efecto podría reducirse mediante un tratamiento térmico posterior al desarrollo de austenita expandida, hecho que queda planteado como trabajo futuro. Pero al mismo tiempo, la segregación de Cr superficial observada implica una pérdida de ese elemento en la aleación base con la consiguiente pérdida de resistencia al ataque químico. Sin embargo, estos defectos que posibilitan la iniciación de las picaduras ocurren en las capas más externas, excesivamente delgadas (solo algunas decenas de nanómetros). Esto significa la posibilidad de ensayar procesos de remoción de estas capas eliminando las probables causas de iniciación de las picaduras de corrosión.

Con el fin de evaluar la estabilidad de la austenita expandida se sometieron muestras a los efectos de altas temperaturas y a procesos de irradiación superficial con haces de deuterio y de helio pulsados de altas energías.

Con respecto a las temperaturas, los resultados después de someter a la austenita expandida a temperaturas entre 225 y 504 ºC durante tiempos entre 20 y 60 horas en un horno en atmósfera normal, mostraron que no se observa degradación de la estructura de austenita expandida hasta que la temperatura alcanza un valor > 405 ºC. La degradación se puede observar a través de GIXRD como un gradual desplazamiento de los picos de difracción desde la posición correspondiente a la austenita expandida hacia los valores originales del material base. Esto muestra la importancia de la temperatura de trabajo (400 ºC) como temperatura umbral a partir de la cual los átomos de carbono (o de nitrógeno) adquieren suficiente movilidad dentro de la red de austenita. Un efecto adicional observado fue el hecho de que las superficies sometidas a altas temperaturas desarrollaron gradualmente capas delgadas de óxidos ($\alpha$-$Fe_2O_3$). Sin embargo, esta oxidación superficial fue eliminada por completo cuando previamente al ensayo se las recubrieron con capas delgadas (330 nm) de AlN depositadas mediante la técnica de sputtering por magnetrón en modo reactivo, mientras que la degradación de la austenita mantuvo el mismo comportamiento de las muestras sin recubrir.

La irradiación con haces pulsados y energéticos de iones ligeros fue llevada a cabo utilizando los haces generados en descargas plasma focus. Los iones creados en estos experimentos son polienergéticos con energías que varían entre 20 y 500 keV, de una duración temporal de ≈400 ns y fluencias de ≈$10^{15}$ $cm^{-2}$. Se usaron haces de deuterio y de helio[8] en experimentos que utilizaron la acumulación de 1, 5 y 10 pulsos individuales. En este caso solo se usaron probetas cementadas del tipo C50-, además de probetas nitruradas en atmósfera de 80 % de hidrógeno y 20 % de nitrógeno. Con un cátodo especialmente diseñado para evitar el efecto de borde en las probetas se alcanzaron densidades de corriente inferiores a ≈2 $mA/cm^2$, evitando así la formación de precipitados.

Mediante la observación de las probetas irradiadas con Microscopía Óptica se comprueba que superficialmente el bombardeo sobre muestras cementadas origina bandas de deslizamiento entrecruzadas (provocadas por los fuertes gradientes térmicos inducidos por los pulsos de iones) sobre las que aparecen cráteres debido a la localización de *clusters* de iones que forman parte de la estructura de los haces iónicos generados en los experimentos de plasma focus. Los resultados observados muestran que la irradiación con un solo pulso genera en la superficie una capa fuertemente alterada, observándose grietas y posible amorfización parcial de la misma, siendo más relevante en las muestras nitruradas. Sin embargo, los resultados parecen ser independientes del tipo de ión utilizado (D o He).

Los análisis FIB/SEM en la sección transversal muestran que el bombardeo iónico provoca la rápida fusión del material y su igualmente rápido enfriamiento, constituyendo una zona amorfa que llega hasta los 3 μm de profundidad. A esta profundidad, comienza a desarrollarse dentro de la matriz una gran cantidad de cristalitas (de aproximadamente 40 nm de diámetro) y agregados de ellas, dando la apariencia que son el inicio de la nucleación y crecimiento de estructuras cristalinas fuertemente tensionadas.

Los estudios mediante GIXRD a 2 y 10º muestran el desplazamiento del pico austenítico (111) hacia mayores valores angulares (o sea, menores parámetros de red). Esto sugiere que las elevadas temperaturas inducidas durante la incidencia de los haces de iones (choque térmico) generan las condiciones (elevación rápida de la temperatura) para que se produzca el movimiento difusivo de los átomos de carbono o de nitrógeno, dejando parcialmente los sitios intersticiales ocupados durante el proceso de cementación o nitruración iónica. Sin embargo, este calentamiento es seguido por un rápido enfriamiento (de solo microsegundos) que congelan los procesos difusivos de los átomos intersticiales. Este efecto se observa claramente por el desplazamiento gradual de los picos de difracción de la austenita expandida hacia ángulos mayores, es decir, una gradual contracción de la red. Pero, además de este efecto, es posible ver la gradual aparición de un pico de difracción a 43,3º (muy próximo al (111) correspondiente a la austenita) que resulta independiente del número de pulsos de irradiación y del tratamiento previo, y que equivale a un 0,90 % de expansión respecto de la red del material base. Observando el gradual corrimiento del pico (111) se puede ver que este también alcanzaría este valor de 43,3º después de sufrir la superficie la incidencia de ≈20 pulsos.

---

[8] Como el rango de energías involucradas en el plasma focus varía entre 20 y 500 keV se tiene que el rango de penetración de los iones de deuterio en la austenita expandida por cementación varía entre 1,65 y 2,43 μm, mientras que para los iones de helio el rango de penetración varía entre 0,85 y 1,02 μm. Es posible que se difundan a distancias más profundas cuando la temperatura por choque térmico es aún elevada.

Este pico podría surgir por las cristalitas observadas a través de los estudios FIB, que se trataría simplemente de la austenita (libre ya de los átomos de carbono o nitrógeno) levemente dilatada por desarrollarse en el seno de material amorfo tras ser fundido por la transferencia de energía de los haces de iones. Algunas observaciones FIB muestran el desarrollo de granos de austenita inmersos dentro de un cuerpo de material aparentemente amorfo.

La austenita expandida es una estructura claramente derivada de un proceso de difusión ultra rápido del C (o N) en la matriz de austenita, que se dispara a la temperatura de 400 ºC. Esta estructura es estable mientras la temperatura que le da origen no sea superada, ya que el proceso de difusión se volverá a desencadenar, provocando entonces la pérdida de carbono (o nitrógeno) y su gradual contracción hacia su estructura austenítica original.

La austenita expandida posee excelentes propiedades superficiales (desgaste, dureza, etc.) y frente a la corrosión. Sin embargo, los resultados del tratamiento de cementación (o nitruración) iónica inducen en la capa externa (solo algunos nanómetros de espesor) condiciones que favorecen el comienzo de corrosión a través de pequeños centros de picaduras. La posibilidad de su eliminación con posterioridad al proceso de cementación (o nitruración) posibilitaría llevar la superficie a las condiciones anticorrosivas originales del material base.